\documentclass[aps,onecolumn,preprint,showkeys,showpacs,amsmath,amssymb,pre,groupedaddress]{revtex4}

\usepackage{amsfonts}
\usepackage{graphicx}
\usepackage{natbib}
\usepackage{dcolumn}
\usepackage{bm}
\usepackage{rotating}
\usepackage[letterpaper,left=0.8in,right=0.8in,top=0.8in,bottom=0.8in]{geometry}

\begin{document}

\preprint{}

\title{The World-Trade Web:\\Topological Properties, Dynamics, and Evolution }

\author{Giorgio Fagiolo}
\affiliation{Sant'Anna School of Advanced Studies, Laboratory of
Economics and Management, Piazza Martiri della Libert\`{a} 33,
I-56127 Pisa, Italy. Tel: +39-050-883356 Fax: +39-050-883343.
E-mail: giorgio.fagiolo@sssup.it}

\author{Javier Reyes}
\affiliation{Department of Economics, Sam M. Walton College of
Business, University of Arkansas, USA. E-mail:
JReyes@walton.uark.edu}

\author{Stefano Schiavo}
\affiliation{Department of Economics and School of International
Studies, University of Trento, Italy and OFCE. E-mail:
stefano.schiavo@ofce.sciences-po.fr}

\date{November 2008}

\begin{abstract}
\noindent This paper studies the statistical properties of the web
of import-export relationships among world countries using a
weighted-network approach. We analyze how the distributions of the
most important network statistics measuring connectivity,
assortativity, clustering and centrality have co-evolved over time.
We show that all node-statistic distributions and their correlation
structure have remained surprisingly stable in the last 20 years --
and are likely to do so in the future. Conversely, the distribution
of (positive) link weights is slowly moving from a log-normal
density towards a power law. We also characterize the autoregressive
properties of network-statistics dynamics. We find that
network-statistics growth rates are well-proxied by fat-tailed
densities like the Laplace or the asymmetric exponential-power.
Finally, we find that all our results are reasonably robust to a few
alternative, economically-meaningful, weighting schemes.
\end{abstract}



\maketitle

\section{Introduction}
In the last decade, a lot of effort has been devoted to the
empirical exploration of the architecture of the World Trade Web
(WTW) from a complex-network perspective
\cite{LiC03,SeBo03,Garla2004,Garla2005,serrc07,Bhatta2007a,Bhatta2007b,
Garla2007,FagioloClustArxiv,FRS2007wp,Communities2007,fagic08}. The
WTW, also known as International Trade Network (ITN), is defined as
the network of import/export relationships between world countries
in a given year. Understanding the topological properties of the
WTW, and their evolution over time, acquires a fundamental
importance in explaining international-trade issues such as economic
globalization and internationalization \cite{StiglitzBook,GlobBook}.
Indeed, it is common wisdom that trade linkages are one of the most
important channels of interaction between world countries
\cite{Krugman1995}. For example, they can help to explain how
economic policies affect foreign markets
\cite{HelliwellPadmore1985}; how economic shocks are transmitted
among countries \cite{Artis2003}; and how economic crises spread
internationally \cite{Forbes2002}. However, direct bilateral-trade
relationships can only explain a small fraction of the impact that
an economic shock originating in a given country can have on another
one, which is not among its direct-trade partners \cite{AbeFor2005}.
Therefore, a complex-network analysis
\cite{AlbertBarabasi2002,DoroMendes2003,Newman2003,PastosVespignani2004}
of the WTW, by characterizing in detail the topological structure of
the network, can go far beyond the scope of standard
international-trade indicators, which instead only account for
bilateral-trade direct linkages \footnote{For example, ``openness to
trade'' of a given country is traditionally measured by the ratio of
exports plus imports to country's gross domestic product (GDP).}.

The first stream of contributions that have studied the properties
of the WTW has employed a \textit{binary-network analysis}, where a
(possibly directed) link between any two countries is either present
or not according to whether the trade flow that it carries is larger
than a given lower threshold \cite{SeBo03,Garla2004,Garla2005}.
According to these studies, the WTW turns out to be characterized by
a high density and a right-skewed (but not exactly power-law)
distribution for the number of partners of a given country (i.e.,
the node degree). Furthermore, there seems to be evidence of
bimodality in the node-degree distribution. While the majority of
countries entertain few trade partnerships, there exists a group of
countries trading with almost everyone else
\cite{FRS2007wp,fagic08}. Also, the binary WTW is a very
disassortative network (i.e., countries holding many trade partners
are on average connected with countries holding few partners) and is
characterized by some hierarchical arrangements (i.e., partners of
well connected countries are less interconnected among them than
those of poorly connected ones). Remarkably, these properties are
quite stable over time \cite{Garla2005}.

More recently, a few contributions have adopted a
\textit{weighted-network approach} \cite{Barr04,Barr05,Bart05} to
the study of the WTW, where each link is weighted by some proxy of
the trade intensity that it carries. The motivation is that a binary
approach cannot fully extract the wealth of information about the
trade intensity flowing through each link and therefore might
dramatically underestimate the role of heterogeneity in trade
linkages. Indeed, Refs. \cite{FagioloClustArxiv,FRS2007wp,fagic08}
show that the statistical properties of the WTW viewed as a weighted
network crucially differ from those exhibited by its unweighted
counterpart. For example, the weighted version of the WTW appears to
be weakly disassortative. Moreover, well-connected countries tend to
trade with partners that are strongly-connected between them.
Finally, the distribution of the total trade intensity carried by
each country (i.e., node strength) is right-skewed, indicating that
a few intense trade connections co-exist with a majority of
low-intensity ones. This is confirmed, at the link level, by Refs.
\cite{Bhatta2007b,Bhatta2007a} who find that the distribution of
link weights can be approximated by a log-normal density robustly
across the years \cite{LiC03}. The main insight coming from these
studies is that a weighted-network analysis is able to provide a
more complete and truthful picture of the WTW than a binary one
\cite{fagic08}.

Additional contributions have instead focused on specific features
of the structure and dynamics of the WTW. For example, Refs.
\cite{Garla2004,Garla2007} find evidence in favor of a
hidden-variable model, according to which the topological properties
of the WTW (in both the binary and weighted case) can be well
explained by a single node-characteristic (i.e., country
gross-domestic product) controlling for the potential ability of a
node to be connected. Furthermore, Ref. \cite{serrc07} studies the
weighted network of bilateral trade imbalances \footnote{That is,
they weight each bilateral trade relation by the difference between
exports and imports. Notice that, as happens also in Refs.
\cite{Bhatta2007a,Bhatta2007b}, their across-year comparison may be
biased by the fact that trade flows are expressed in current U.S.
dollars and do not appear to be properly deflated.}. The Authors
note that also the international trade-imbalance network is
characterized by a high level of heterogeneity: for each country,
the profile of trade fluxes is unevenly distributed across partners.
At the network level, this prompts to the presence of high-flux
backbones, i.e. sparse subnetworks of connected trade fluxes
carrying most of the overall trade in the network. The Authors then
develop a method to extract (for any significance level) the flux
backbone existing among countries and links. This turns out to be
extremely effective in sorting out the most relevant part of the
trade-imbalance network and can be conveniently used for
visualization purposes. Finally, Ref. \cite{Communities2007}
considers the formation of ``trade islands'', that is connected
components carrying a total trade flow larger than some given
thresholds. The analysis of the evolution of the WTW community
structure \cite{NewmanCommunities} finds mixed evidence for
globalization.

In this paper we present a more thorough study of the topological
properties of the WTW by focusing on distribution dynamics and
evolution. More specifically, following the insights of Ref.
\cite{fagic08}, we employ a weighted network approach to
characterize, for the period 1981-2000, the distribution of the most
important network statistics measuring node connectivity,
assortativity, clustering and centrality; as well as link weights.
We ask three main types of questions: (i) Have (and, if so, how) the
distributional properties of these statistics (and their correlation
structure) been changing within the sample period considered? (ii)
Can we make any prediction on the out-of-sample evolution of such
distributions? (iii) Do the answers to the previous questions change
if we play with a number of alternative, economically-meaningful
weighting schemes (i.e., if we allow for different rules to weight
existing links)?


The rest of the paper is organized as follows. Section
\ref{Sec:Data} presents the data sets and defines the statistics
studied in the paper. Section \ref{Sec:Results} introduces the main
results. Finally, Section \ref{Sec:Conclusions} concludes and
discusses future extensions.

\section{Data and Definitions} \label{Sec:Data}

We employ international-trade data provided by \cite{GledData2002}
to build a time-sequence of weighted directed networks. Our balanced
panel refers to $T=20$ years (1981-2000) and $N=159$ countries. For
each country and year, data report trade flows in current US
dollars. To build adjacency and weight matrices, we followed the
flow of goods. This means that rows represent exporting countries,
whereas columns stand for importing countries. We define a ``trade
relationship'' by setting the generic entry of the (binary)
adjacency matrix $\tilde{a}_{ij}^t=1$ if and only if exports from
country $i$ to country $j$ ($e_{ij}^t$) are strictly positive in
year $t$.

Following Refs. \cite{LiC03,Bhatta2007a,Bhatta2007b,Garla2007}, the
weight of a link from $i$ to $j$ in year $t$ is defined as
$\tilde{w}_{ij}^t=e_{ij}^t$ \footnote{In Section
\ref{SubSec:Robustness} we explore what happens if we employ a few
alternative definitions for link weights.}. Thus, the sequence of
$N\times N$ adjacency and weight matrices
$\{\tilde{A}^t,\tilde{W}^t\}$, $t=1981,...,2000$ fully describes the
within-sample dynamics of the WTW.

A preliminary statistical analysis of both binary and weighted
matrices suggests that $(\tilde{A}^t,\tilde{W}^t)$ are sufficiently
symmetric to justify an undirected analysis for all $t$. From a
binary perspective, on average about 93\% of WTW directed links are
reciprocated in each given year. This means that, almost always, if
country $i$'s exports to country $j$ are positive
($\tilde{a}_{ij}^t=1$), then $\tilde{a}_{ji}^t=1$, i.e. country
$j$'s exports to country $i$ are also positive. To check more
formally this evidence from a weighted perspective, we have computed
the weighted symmetry index defined in Ref.
\cite{FagioloSymmEcoBull}. The index ranges in the sample period
between 0.0017 and 0.0043, signalling a relatively strong and stable
symmetry of WTW weight matrices \footnote{The expected value of the
statistic in a random graph where link weights are uniformly and
independently distributed as a uniform in the unit interval is 0.5
\cite{FagioloSymmEcoBull}. Furthermore, the expected value computed
by randomly reshuffling in each year the empirically-observed
weights among existing links ranges in the same period from 0.0230
to 0.0410. Therefore, the empirical value is significantly smaller
than expected.}. We have therefore symmetrized the network by
defining the entries of the new adjacency matrix $A^t$ so that
$a_{ij}^t=1$ if and only if either $\tilde{a}_{ij}^t=1$ or
$\tilde{a}_{ji}^t=1$, and zero otherwise. Accordingly, the generic
entry of the new weight matrix $W^t$ is defined as
$w_{ij}^t=\frac{1}{2}(\tilde{w}_{ij}^t+\tilde{w}_{ji}^t)$. This
means that the symmetrized weight of link ${ij}$ is proportional to
the total trade (imports plus exports) flowing through that link in
a given year. Finally, in order to have $w_{ij}^t\in[0,1]$ for all
$(i,j)$ and $t$, we have re-normalized all entries in $W^t$ by their
maximum value $w^t_{\ast}=max_{i,j=1}^{N}w_{ij}^t$.

For each $(\tilde{A}^t,\tilde{W}^t)$, we study the distributions of
the following node statistics:

\begin{itemize}
  \item \textit{Node degree} \cite{AlbertBarabasi2002,Pastor2001}, defined as
  $ND_i^t=A_{(i)}^t\textbf{1}$, where $A_{(i)}^t$ is the $i$-th row of $A^t$
  and $\textbf{1}$ is a unary vector. ND is a measure of binary
  connectivity, as it counts the number of trade partners of any
  given node. Although we mainly focus here on a weighted-network
  approach, we study ND because of its natural interpretation in
  terms of number of trade partnerships and bilateral trade agreements.

  \item \textit{Node strength} \cite{DeMontis2005}, defined as $NS_i^t=W_{(i)}^t
  \textbf{1}$, where again $W_{(i)}^t$ is the $i$-th row of $W^t$. While ND tells
  us how many partners a node holds, NS is a measure of weighted connectivity, as
  it gives us an idea of how intense existing trade relationships of country $i$ are.

  \item \textit{Node average nearest-neighbor strength} \cite{DeMontis2005}, that is
  $ANNS_i^t=(A_{(i)}^t W^t\textbf{1})/(A_{(i)}^t\textbf{1})$. ANNS measures how intense are
  trade relationships maintained by the partners of a given node.
  Therefore, the correlation between ANNS and NS is a measure of
  network assortativity (if positive) or disassortativity (if
  negative). It is easy to see that ANNS boils down to average nearest-neighbor degree (ANND) if
  $W^t$ is replaced by $A^t$.

  \item \textit{Weighted clustering coefficient} \cite{Saramaki2006,FagioloClustArxiv},
  defined as $WCC_i^t=({[W^t]}^{\left[\frac{1}{3}\right]})_{ii}^{3}/(ND_i^t(ND_i^t-1))$. Here
  $Z_{ii}^3$ is the $i$-th entry on the main diagonal of $Z\cdot Z\cdot Z$ and
  $Z^{\left[\frac{1}{3}\right]}$ stands for the matrix obtained from $Z$ after raising each
  entry to $1/3$. WCC measures how much clustered a node $i$ is from a weighted perspective, i.e.
  how much intense are the linkages of trade triangles having country $i$ as a
  vertex \footnote{Cf. Ref. \cite{Saramaki2006} for alternative definitions of the clustering
  coefficient for weighted undirected networks. Here we employ the above formulation because it
  is the only one retaining two properties important to fully characterize clustering in trade networks, namely
  (i) $WCC_i^t$ takes into account the weight of all links in any given triangle; (ii) $WCC_i^t$  is invariant to
  weight permutations in one triangle.}. Again, replacing $W^t$ with $A^t$, one obtains the standard
binary clustering coefficient
  (BCC), which counts the fraction of triangles existing in the neighborhood of any give node
  \cite{WattsStrogatz1998}.

  \item \textit{Random-walk betweenness centrality}
  \cite{newm05,five06}, which is a measure of how much a given
  country is globally-central in the WTW. A node has a higher
  random-walk betweenness centrality (RWBC) the more it has a position of strategic significance
  in the overall structure of the network. In other words, RWBC is
  the extension of node betweenness centrality  \cite{Scott2000} to weighted
  networks and measures the probability that a random signal can find its way
  through the network and reach the target node where the links to follow are
  chosen with a probability proportional to their weights.

\end{itemize}

The above statistics allow one to address the study of node
characteristics in terms of four dimensions: connectivity (ND and
NS), assortativity (ANND and ANNS, when correlated with ND and NS),
clustering (BCC and WCC) and centrality (RWBC). In what follows, we
will mainly concentrate the analysis on ND and the other weighted
statistics (NS, ANNS, WCC, RWBC), but we occasionally discuss, when
necessary, also the behavior of ANND and BCC.

We further explore the network-connectivity dimension by studying
the time-evolution of the link-weight distribution $w^t=\{w_{ij}^t,
i\neq j=1,\dots,N\}$. In particular, we are interested in assessing
the fraction of links that are zero in a given year $t$ and becomes
positive in year $t+\tau$, $\tau=1,2,\dots$ and the percentage of
links that are strictly positive at $t$ and disappear in year
$t+\tau$. This allows one to keep track of trade relationships that
emerge or become extinct during the sample period \footnote{Notice
that the fraction of strictly positive links (over all possible
links) also defines network density. More on that below.}.

\bigskip

\section{Results} \label{Sec:Results}

\subsection{Shape, Moments, and Correlation Structure of Network
Statistics} \label{SubSec:Shape} 
\begin{figure}[h]
\begin{minipage}[t]{8cm}
\centering
{\includegraphics[width=8cm]{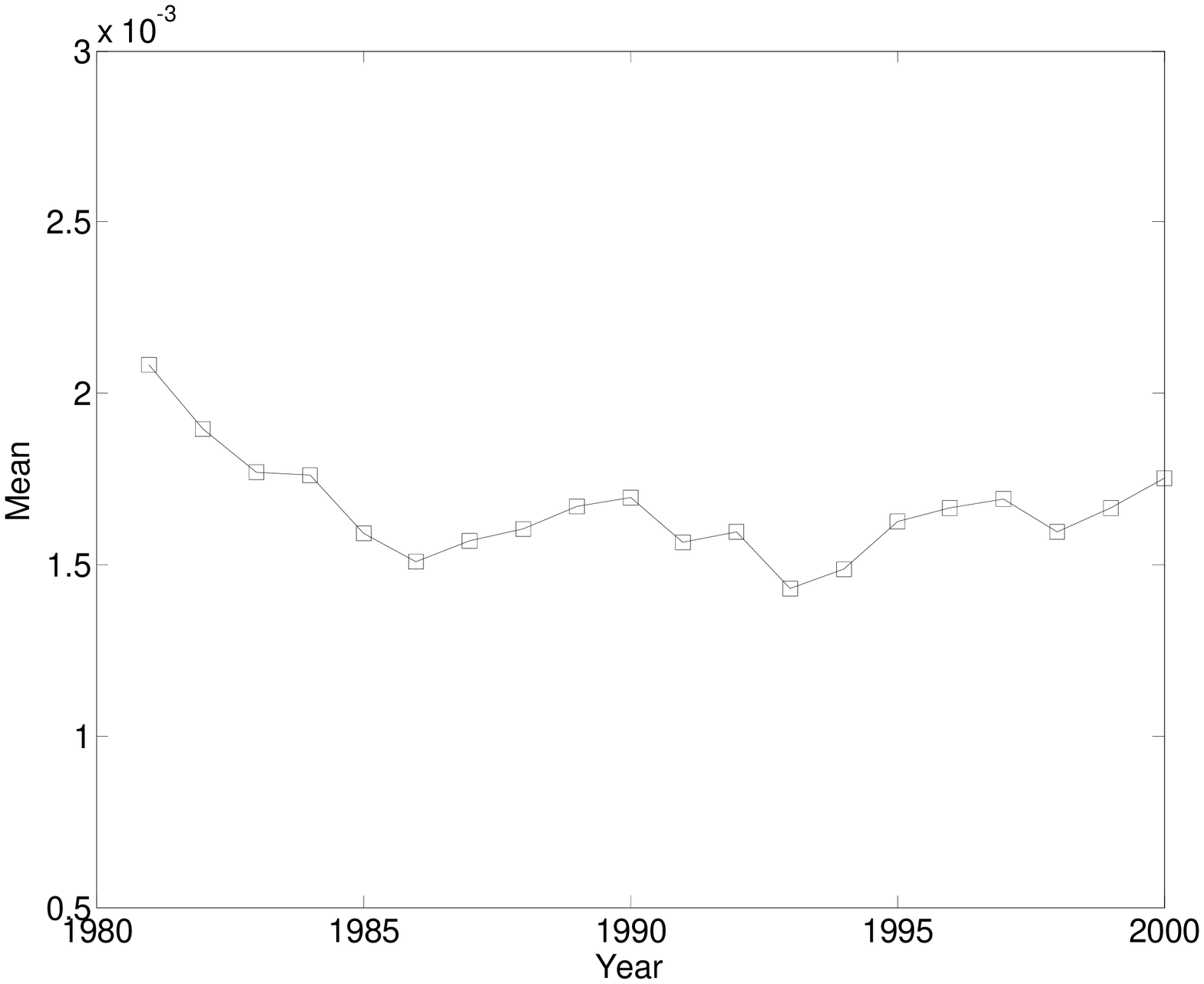}}
\end{minipage}\hfill
\begin{minipage}[t]{8cm}
\centering
{\includegraphics[width=8cm]{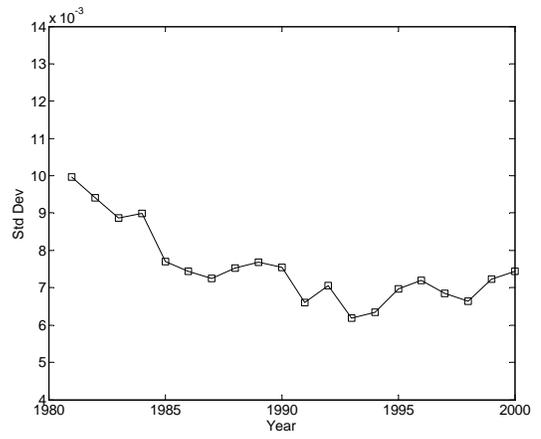}}
\end{minipage}
\begin{minipage}[t]{8cm}
\centering
{\includegraphics[width=8cm]{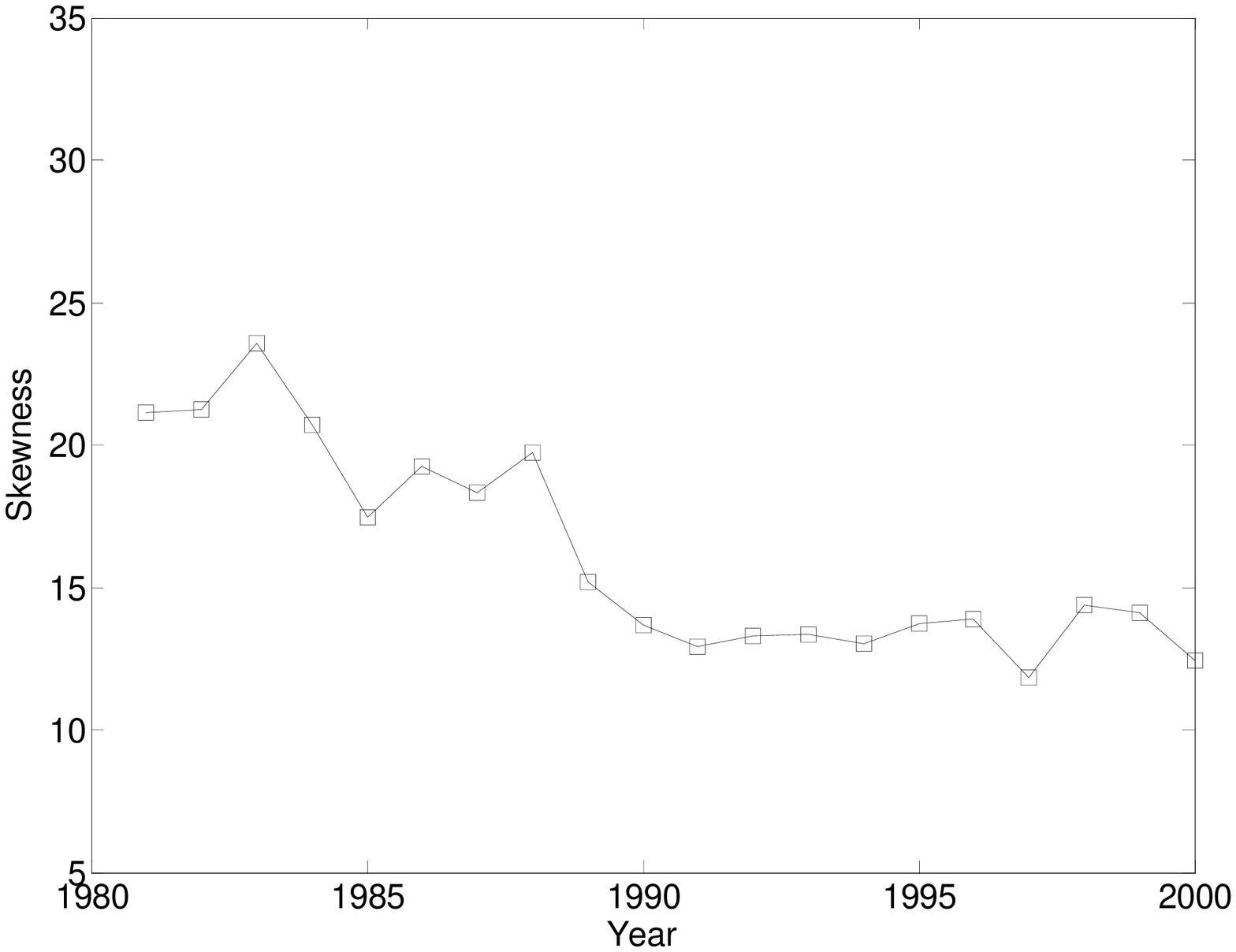}}
\end{minipage}\hfill
\begin{minipage}[t]{8cm}
\centering
{\includegraphics[width=8cm]{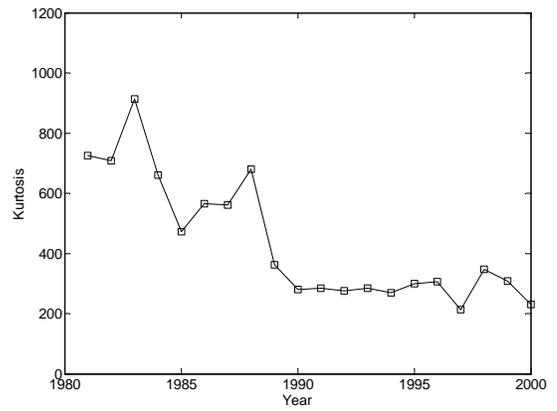}}
\end{minipage}
\caption{First four sample moments of the link-weight distribution
vs. years.} \label{Fig:linkweights_distr_stats}
\end{figure}

\begin{figure}[h]
\centering
{\includegraphics[width=7.5cm,height=6cm]{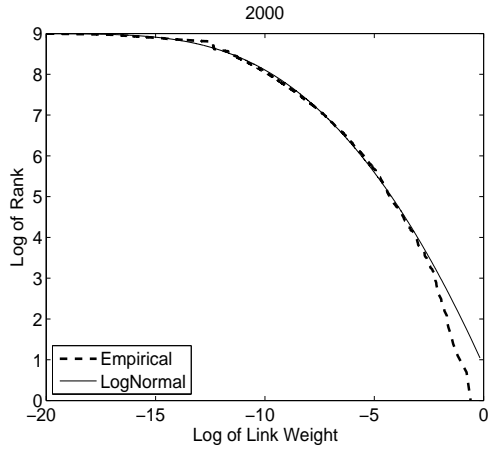}}
\caption{Size-rank (log-log) plot of the link-weight distribution in
year 2000.} \label{Fig:srp_linkweights}
\end{figure}

We begin by studying the shape of
the distributions of node and link statistics and their dynamics
within the sample period under analysis. As already found in Refs.
\cite{Bhatta2007a,Bhatta2007b}, link weight distributions display
relatively stable moments (see Figure
\ref{Fig:linkweights_distr_stats}) and are well proxied by
log-normal densities in each year (cf. Figure
\ref{Fig:srp_linkweights} for an example). This means that the
majority of trade linkages are relatively weak and coexist with few
high-intensity trade partnerships. The fact that the first four
moments of the distribution do not display remarkable structural
changes in the sample period hints to a relatively strong stability
of the underlying distributional shapes \footnote{If any, there
seems to be some evidence towards declining higher-than-one
moments.}. We shall study this issue in more details below.

\begin{figure}[h]
\begin{minipage}[t]{8cm}
\centering {\includegraphics[width=8cm]{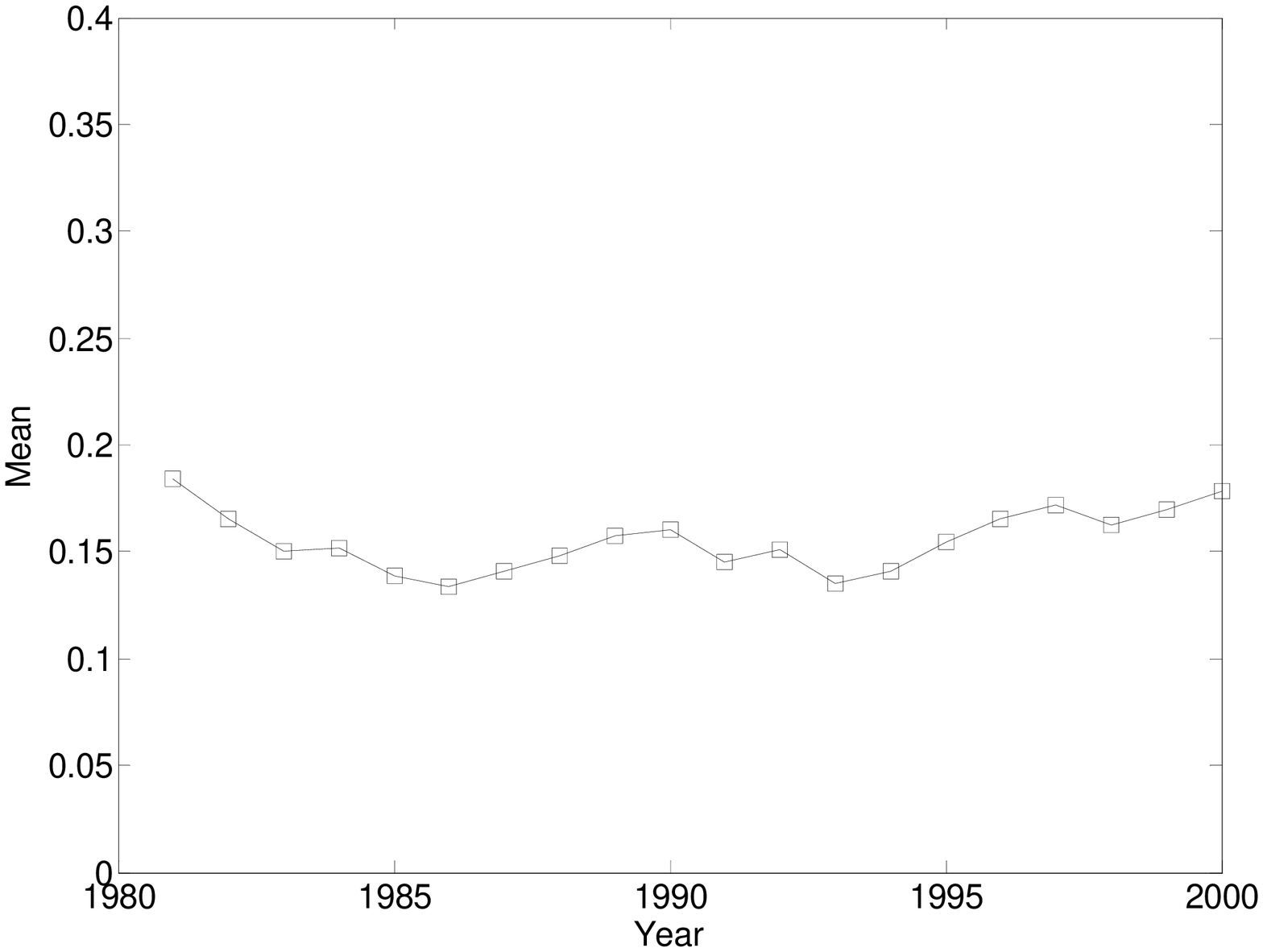}}
\end{minipage}\hfill
\begin{minipage}[t]{8cm}
\centering {\includegraphics[width=8cm]{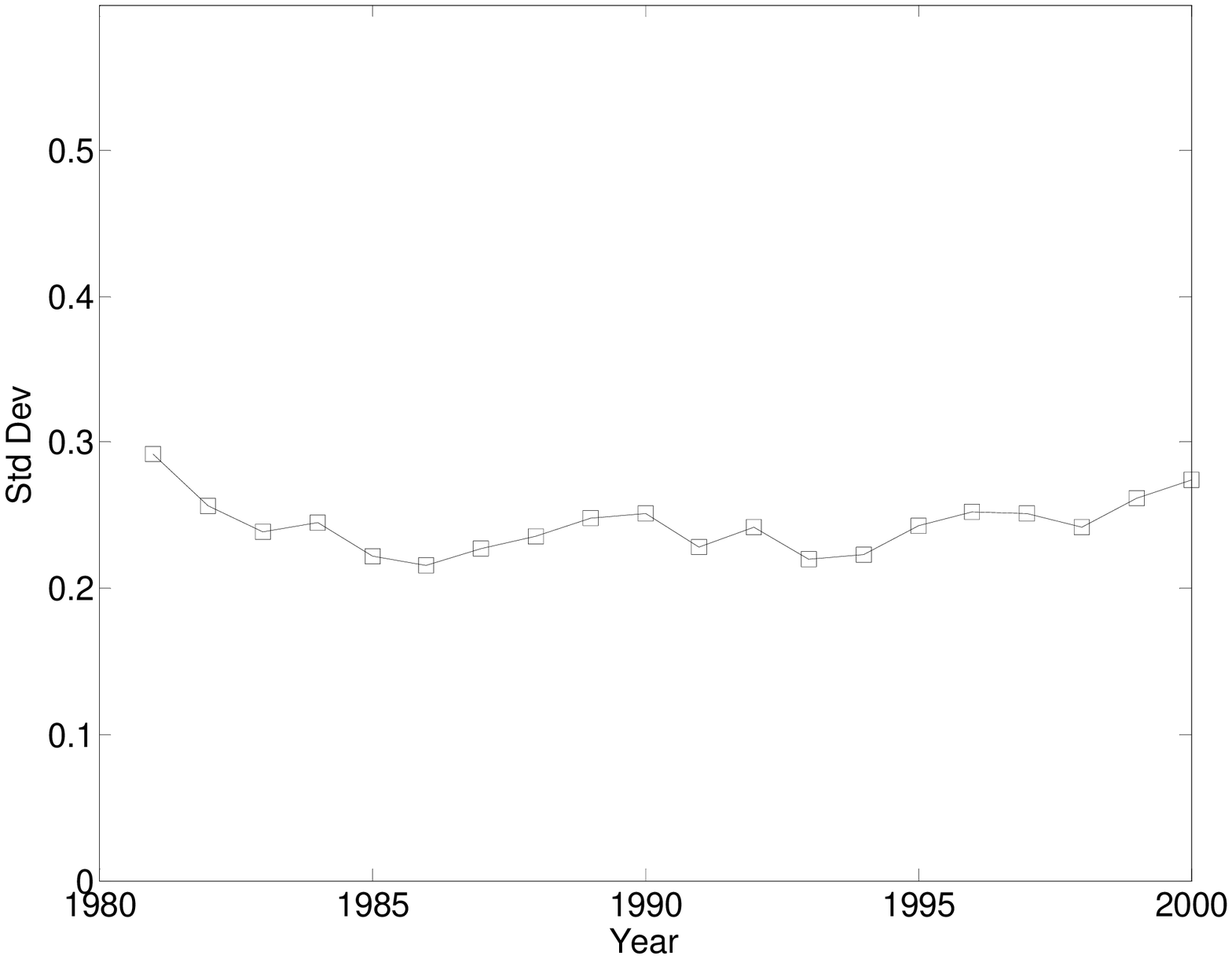}}
\end{minipage}
\begin{minipage}[t]{8cm}
\centering {\includegraphics[width=8cm]{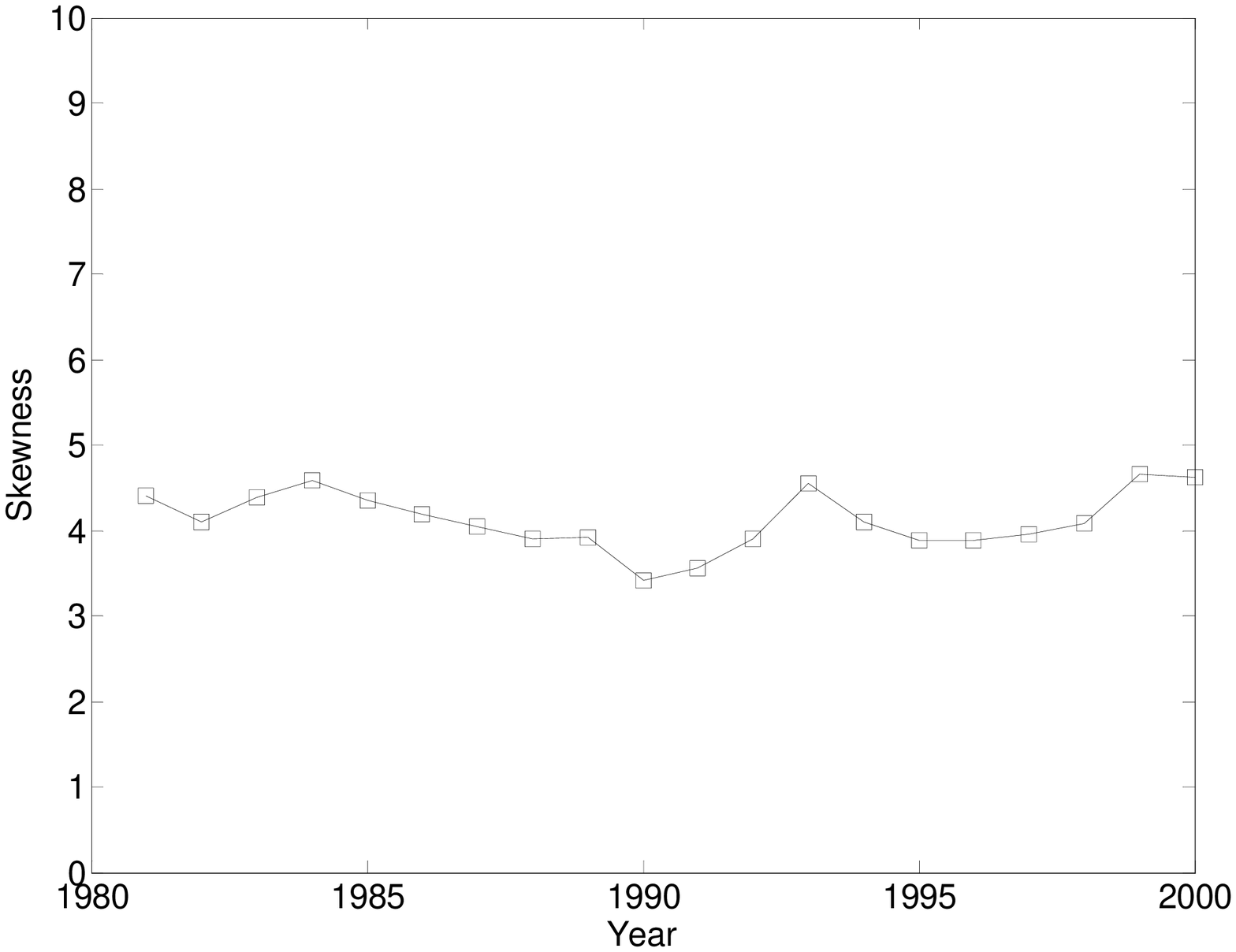}}
\end{minipage}\hfill
\begin{minipage}[t]{8cm}
\centering {\includegraphics[width=8cm]{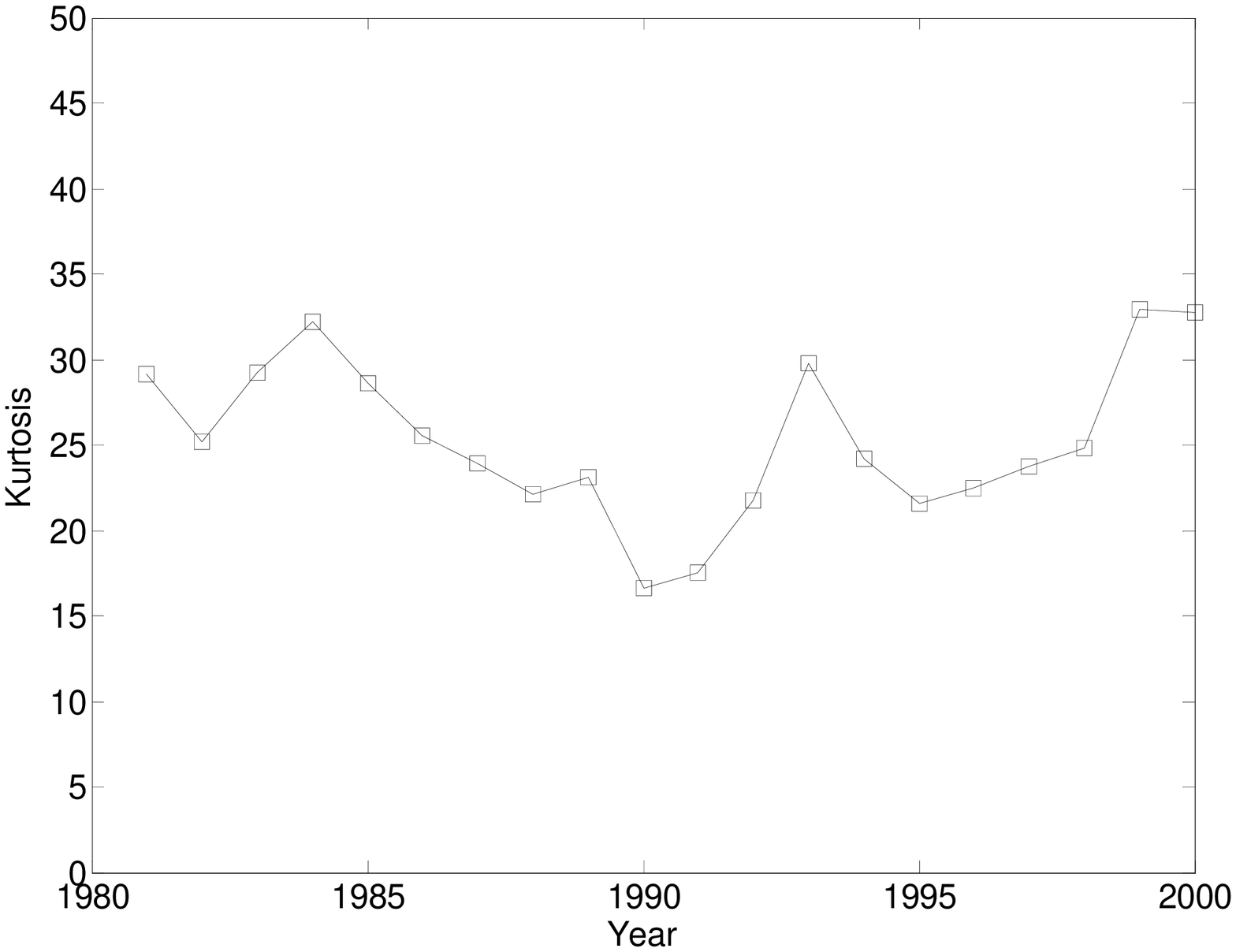}}
\end{minipage}
\caption{Sample moments of node strength (NS) distribution vs.
years.} \label{Fig:ns_distr_stats}
\end{figure}

A similar stable pattern is detected also for the moments of the
distributions of all node statistics under analysis, see Figure
\ref{Fig:ns_distr_stats} for the case of NS distributions. To see
that this applies in general for node statistics, we have computed
the time average (across 19 observations) of the absolute value of
1-year growth rates of the first four interesting moments of ND, NS,
ANNS, WCC and RWBC statistics, namely mean, standard deviation,
skewness and kurtosis \footnote{More formally, let $X_i^t$ be the
value of the node statistic X at time $t$ for country $i$ and
$M^k(\cdot)$ the moment operator that for $k=1,2,3,4$ returns
respectively the mean, standard deviation, skewness and kurtosis.
The time-average of absolute-valued 1-year growth rates of the
$k$-th moment-statistic is defined as
$\frac{1}{T-1}\sum_{t=2}^{T}{|(E^k(X_i^t)/E^k(X_i^{t-1})-1)|}$. }.
Table \ref{Tab:moments_gr} shows that these average absolute
growth-rates range in our sample between 0.0043 and 0.0615, thus
indicating that the shape of these distributions seem to be quite
stable over time.

\begin{figure}[h]
\begin{minipage}[t]{7.5cm}
\centering
{\includegraphics[width=7.5cm,height=6cm]{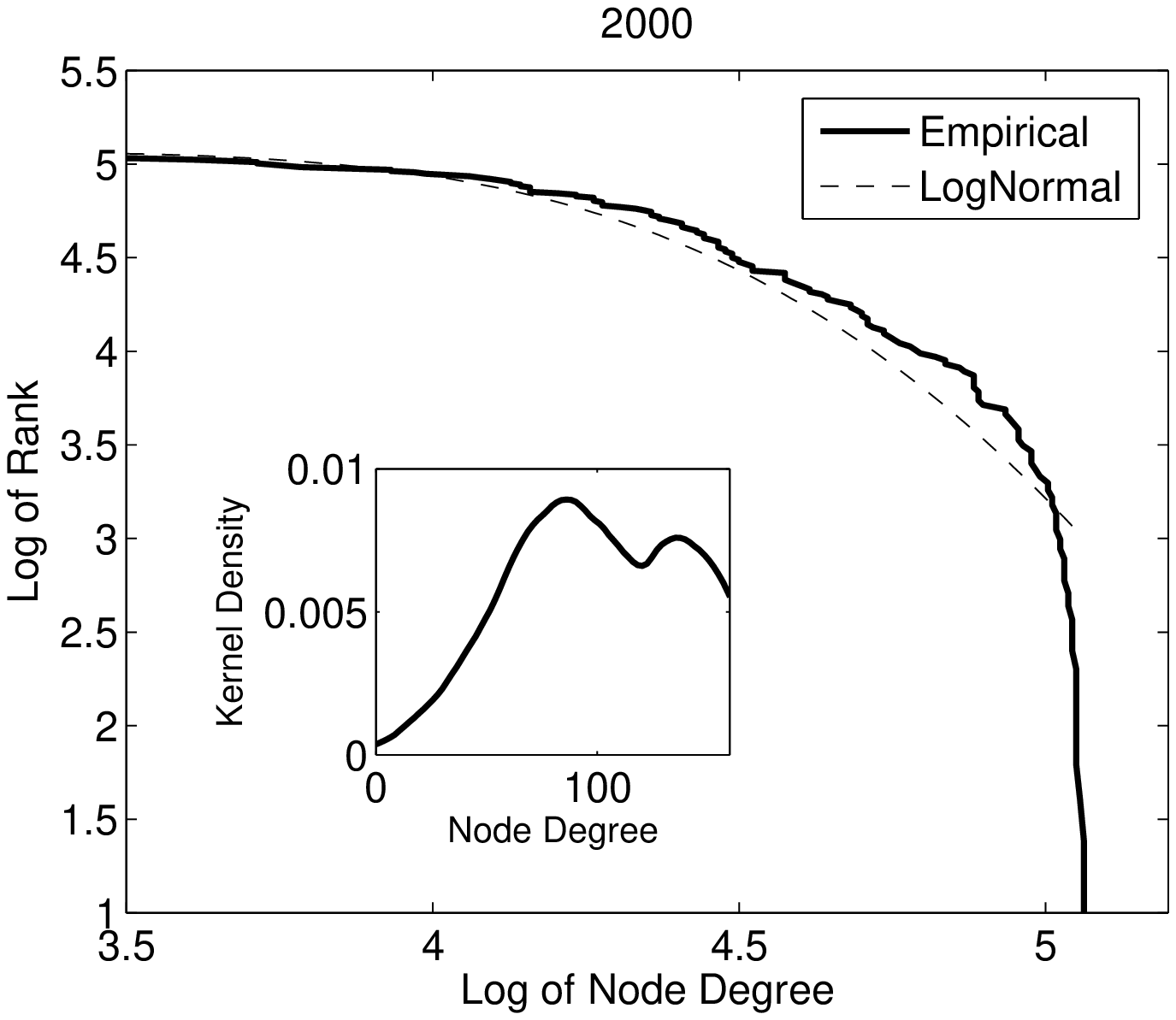}}
\caption{Size-rank (log-log) plot of node-degree distribution in
year 2000. Inset: Kernel density estimate.}
\label{Fig:nd_rsp_plus_kernel_2000}
\end{minipage}\hfill
\begin{minipage}[t]{7.5cm}
\centering
{\includegraphics[width=7.5cm,height=6cm]{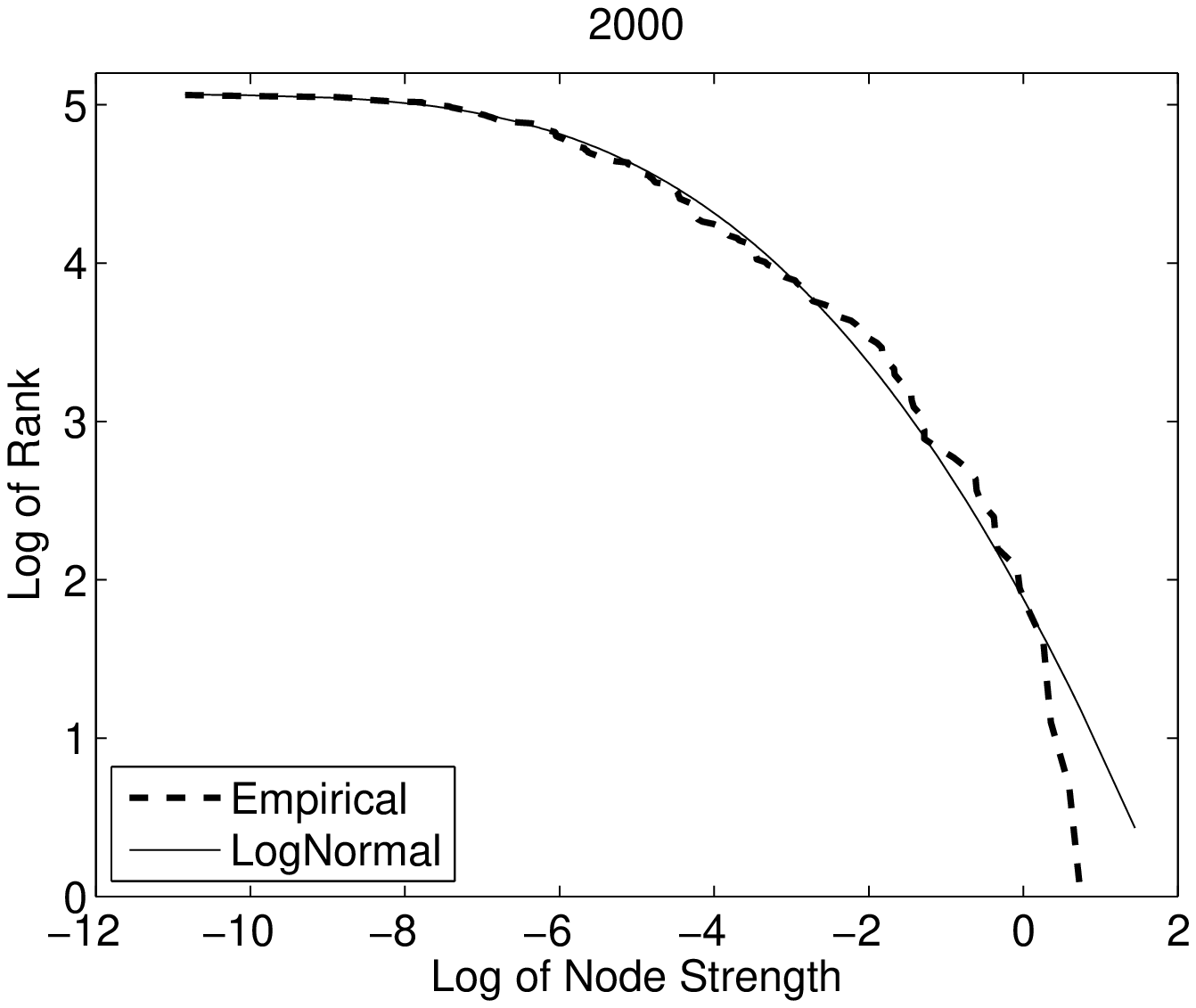}}
\caption{Size-rank (log-log) plot of node-strength distribution in
year 2000.} \label{Fig:srp_ns_2000}
\end{minipage}
\end{figure}

\begin{figure}[h]
\begin{minipage}[t]{7.5cm}
\centering
{\includegraphics[width=7.5cm,height=6cm]{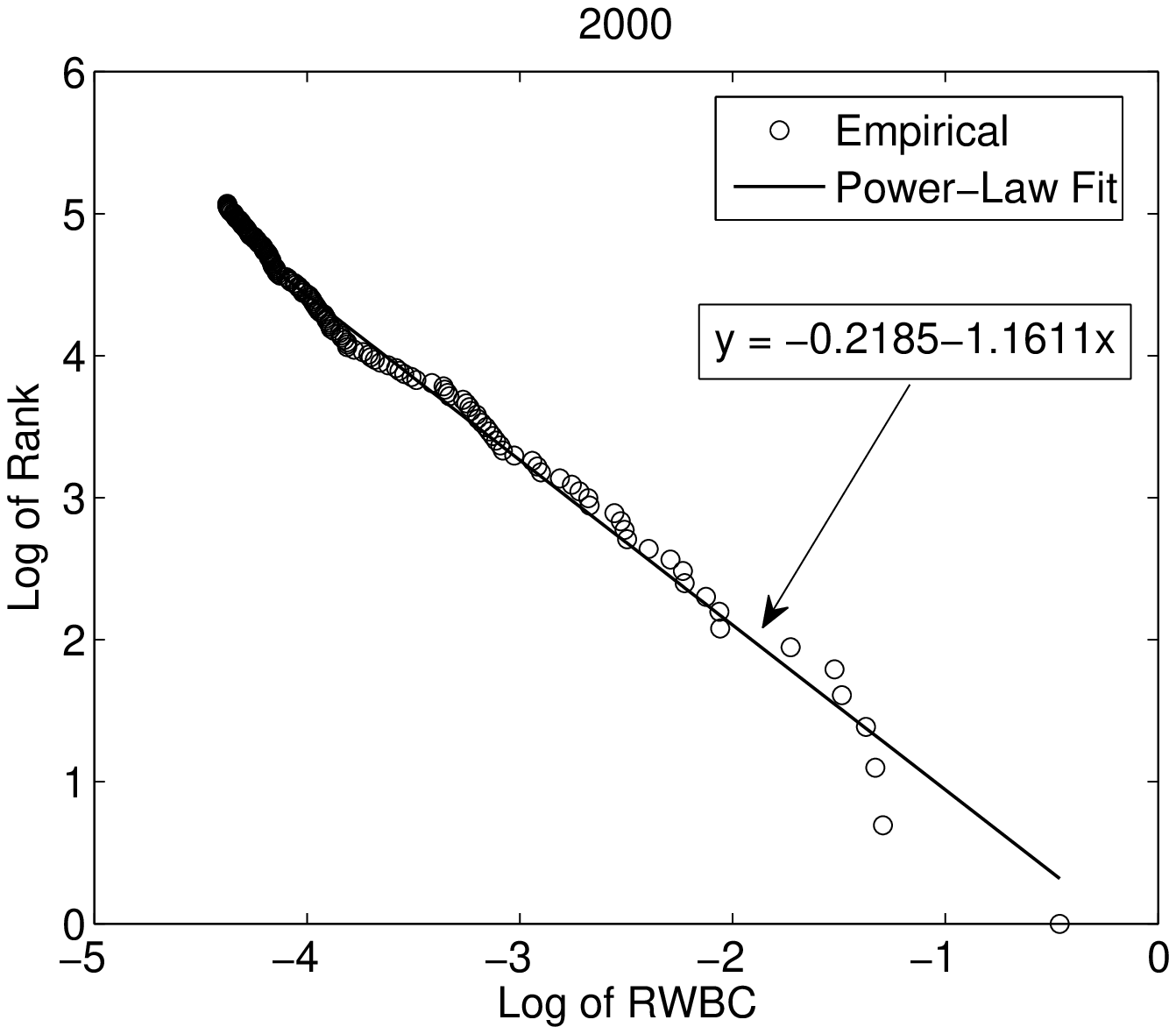}}
\caption{Size-rank (log-log) plot of node random-walk betweenness
centrality (RWBC) distribution in year 2000. Solid line: Power-law
fit (equation of the regression line in the inset).}
\label{Fig:srp_rwbc_2000}
\end{minipage}\hfill
\begin{minipage}[t]{7.5cm}
\centering
{\includegraphics[width=7.5cm,height=6cm]{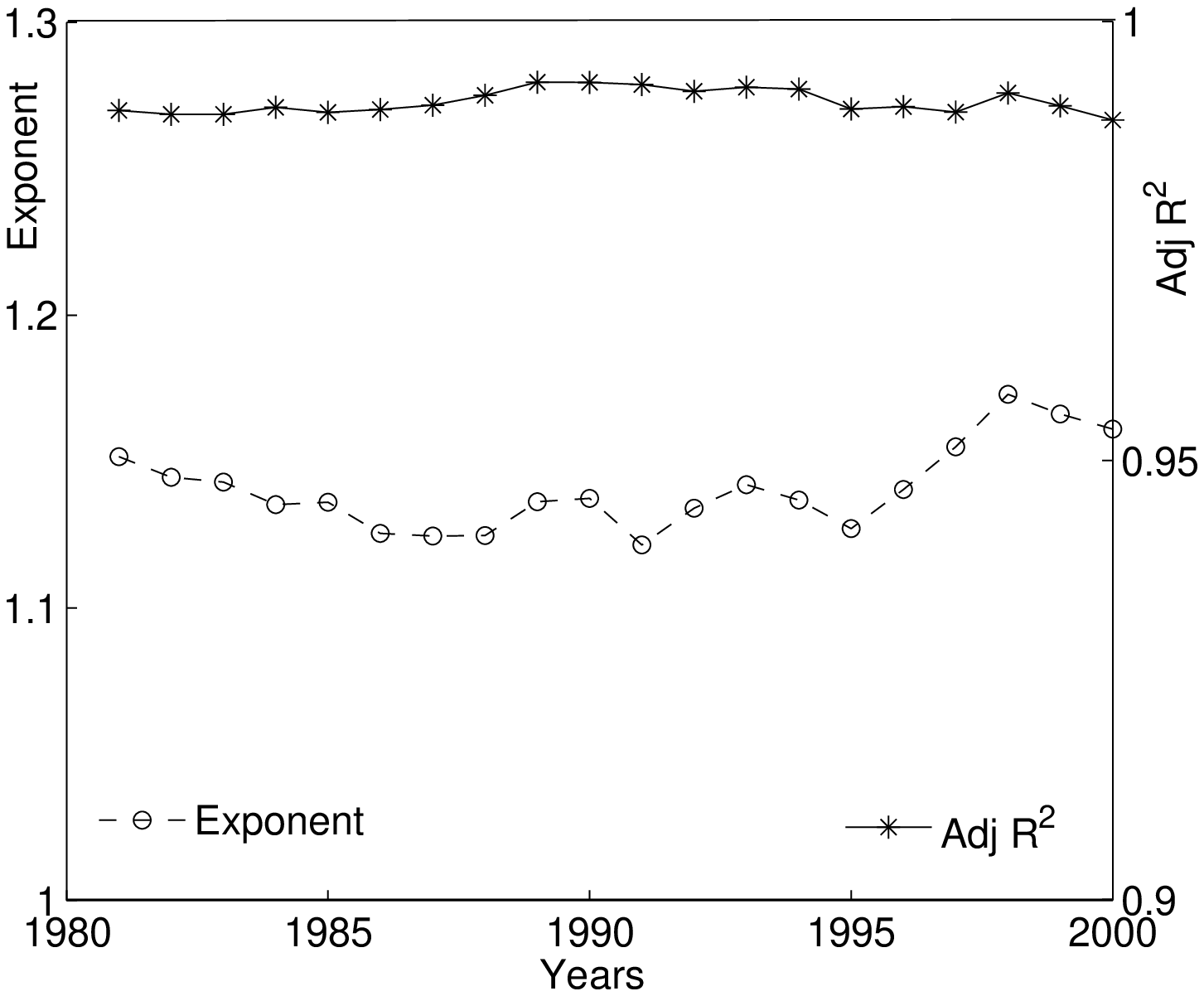}}
\caption{Left Y-axis scale: Estimated power-law exponent for node
random-walk betweenness centrality (RWBC) distributions vs. years.
Right Y-axis scale: Adjusted $R^2$ associated to the power-law fit.}
\label{Fig:power_law_rwbc}
\end{minipage}
\end{figure}

But how does the shape of node- and link-statistic distributions
look like? To investigate this issue we have begun by running
normality tests on the logs of (positive-valued) node and link
statistics. As Table \ref{Tab:normality} suggests \footnote{Table
\ref{Tab:normality} reports p-values for the Jarque-Bera test
\citep{jbtest1,jbtest2}, the null hypothesis being that the logs of
positive-valued statistics are normally distributed with unknown
parameters. Alternative normality tests (Lilliefors,
Anderson-Darling, etc.) yield similar results.}, binary-network
statistic distributions are never log-normal (i.e., their logs are
never normal), whereas all weighted-network statistics but RWBC seem
to be well-proxied by log-normal densities. To see why this happens,
Figure \ref{Fig:nd_rsp_plus_kernel_2000} shows the rank-size plot of
ND in 2000 (with a kernel density estimate in the inset) \footnote{A
rank-size plot is simply a transformation of a standard
cumulative-distribution function (CDF) plot, where the behavior of
the upper tail is accentuated. Indeed, suppose that
$(x_1,\dots,x_N)$ are the available empirical observations of a
random variable $X$, and sort the $N$ observations to obtain
$(x_{(1)},\dots,x_{(N)})$, where $x_{(1)}\geq x_{(2)}\geq \dots \geq
x_{(N)}$. A rank-size plot graphs $\log(r)$ against $\log(x_{(r)})$,
where $r$ is the rank. However, since $r/N=1-F(x_{(r)})$, then
$\log(r)=\log[1-F(x_{(r)})]+\log(N)$. Since the existing literature
on the WTW has discussed at length the upper-tail behavior of node
and link distributions, we have preferred to visualize the shape of
the distributions of interest using a rank-size plot instead of
employing standard (two-tailed) CDF plots.}. It is easy to see that
ND exhibits some bimodality, with the majority of countries
featuring low degrees and a bunch of countries trading with almost
everyone else. Figure \ref{Fig:srp_ns_2000} shows instead, for year
2000, how NS is nicely proxied by a log-normal distribution. This is
not so for RWBC, whose distribution seems instead power-law in all
years, with slopes oscillating around -1.15, see Figures
\ref{Fig:srp_rwbc_2000} and \ref{Fig:power_law_rwbc}. Therefore,
being more central is more likely than having high NS, ANNS, or WCC
(i.e., the latter distributions feature upper tails thinner than
that of RWBC distributions; we shall return to complexity issues
related to this point when discussing out-of-sample evolution of the
distributions of node and link statistics). The foregoing
qualitative statements can be made quantitative by running
comparative goodness-of-fit (GoF) tests to check whether the
distributions under study come from pre-defined density families. To
do so, we have run Kolmogorov-Smirnov GoF tests
\cite{Massey1951,Owen1962} against three null hypotheses, namely
that our data can be well described by log-normal, stretched
exponential, or power-law distributions. The stretched-exponential
distribution (SED) has been employed because of its ability to
satisfactorily describe the tail behavior of many real-world
variables and network-related measures \cite{SED1,SED2}. Table
\ref{Tab:kstest} reports results for year 2000 in order to
facilitate a comparison with Figures
\ref{Fig:nd_rsp_plus_kernel_2000}-\ref{Fig:srp_rwbc_2000}, but again
the main insights are confirmed in the entire sample. It is easy to
see that the SED does not successfully describe the distributions of
our main indicators. On the contrary, it clearly emerges that NS,
ANNS and RWBC seem to be well described by log-normal densities,
whereas the null of power-law RWBC cannot be rejected. For ND,
neither of the three null appears to be a satisfactory hypothesis
for the KS test.

\begin{figure}[h]
\centering {\includegraphics[width=7.5cm,height=6cm]{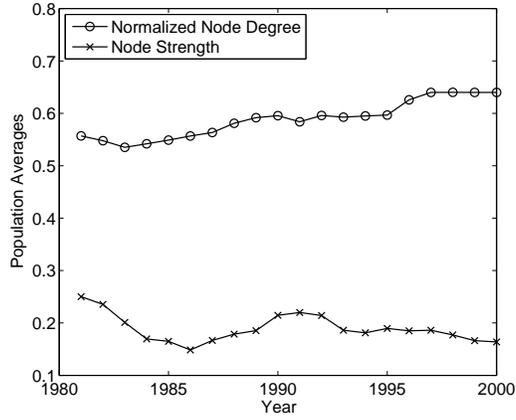}}
\caption{Population averages of node degree normalized by population
size $N=159$ and node strength vs. years.} \label{Fig:ave_nd_ns}
\end{figure}

\begin{figure}[h]
\begin{minipage}[t]{7.5cm}
\centering
{\includegraphics[width=7.5cm,height=6cm]{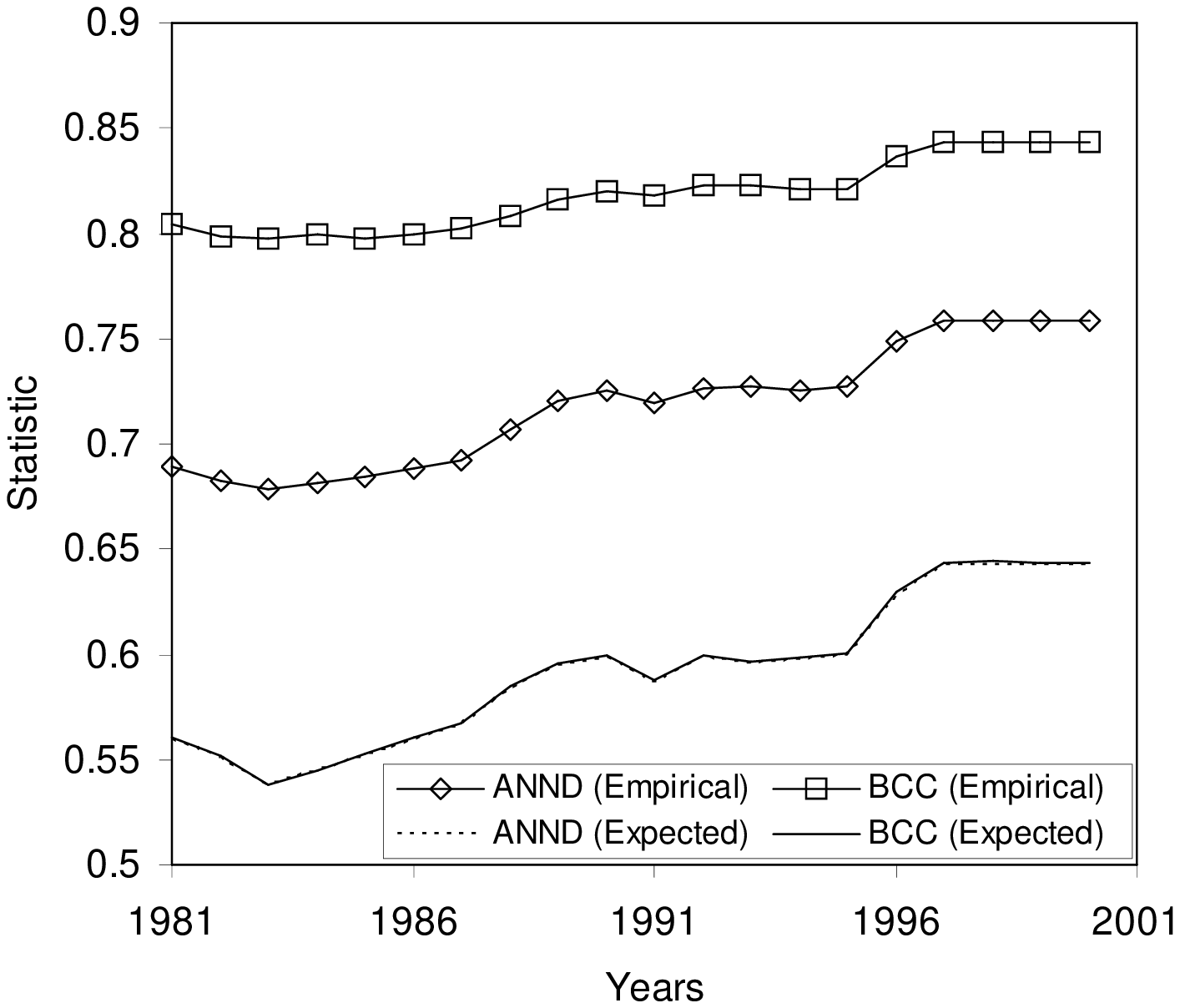}}
\end{minipage}\hfill
\centering
{\includegraphics[width=7.5cm,height=6cm]{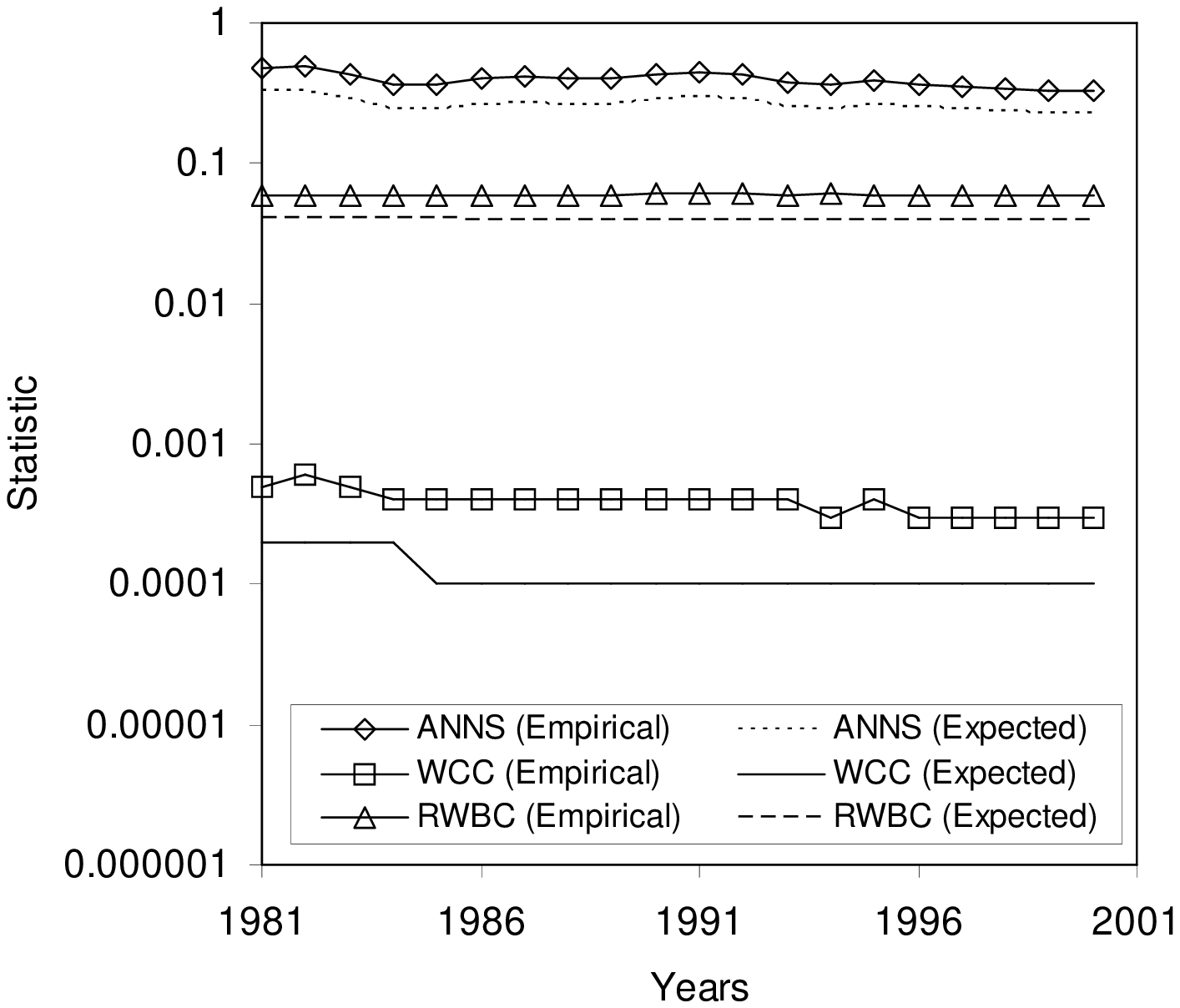}}
\caption{Population average vs. expected values of node statistics.
Expected values for binary-network statistics (left) are computed by
reshuffling binary links by keeping observed density fixed. Expected
values for weighted-network statistics (right) are computed by
reshuffling observed link weights while keeping the binary sequence
fixed (edge-crossing algorithm).} \label{Fig:expected}
\end{figure}

We now discuss in more detail the evolution over time of the moments
of the distributions of node statistics. As already noted in Refs.
\cite{Garla2004,Bhatta2007a,Bhatta2007b,fagic08}, the binary WTW is
characterized by an extremely high network density
$d^t=\frac{1}{N(N-1)}\sum_i \sum_j{a^t_{ij}}$, ranging from 0.5385
to 0.6441. Figure \ref{Fig:ave_nd_ns} plots the normalized (by $N$)
population-average of ND, which is equal to network density up to a
$N^{-1}(N-1)$ factor, together with population-average of NS. While
the average number of trade partnerships is very high and slightly
increases over the years, their average intensity is rather low (at
least as compared to NS conceivable range, i.e. $[0,N-1]$) and tends
to decline \footnote{At the extreme, if in every year $t$ the
network were an Erd\"{o}s-Renyi random-graph \cite{Bollo1985} with
link probability equal to network density $d^t$ and link weights
drawn from an i.i.d. uniform r.v. defined on the unit interval
(U[0,1]) -- uniformly weighted ER graph henceforth -- the expected
NS would have been $\frac{1}{2}(N-1)[d^t]^2$, that is a value
ranging over time between 22.9079 and 32.7699.}. As far as
ANND/ANNS, clustering and centrality are concerned, a more
meaningful statistical assessment of the actual magnitude of
empirical population-average statistics requires comparing them with
expected values computed after reshuffling links and/or weights. In
what follows, we consider two reshuffling schemes (RSs). For binary
statistics, we compute expected values after reshuffling existing
links by keeping fixed the observed density $d^t$ (hereafter, B-RS).
For weighted ones, we keep fixed the observed adjacency matrix $A^t$
and re-distribute weights at random by reshuffling the empirical
link-weight distribution $w^t=\{w_{ij}^t, i\neq j=1,\dots,N\}$
(hereafter, W-RS) \footnote{Notice that under if the network were an
uniformly weighted ER graph (see above), one would have obtained
$E(ANND_i^t)=1+(N-2)d^t$,
$E(ANNS_i^t)=E(NS_i^t)=\frac{N-2}{2}(d^t)^2$, $E(BCC_i^t)=d^t$ and
$E(WCC_i^t)=\frac{27}{64} d^t$; see \cite{FagioloClustArxiv}.}.
Figure \ref{Fig:expected} shows empirical averages vs. expected
values over time. Notice that empirical averages of ANND, BCC, ANNS,
WCC, and RWBC are larger than expected, meaning that the WTW is on
average more clustered; features a larger nearest-neighbor
connectivity, and countries are on average more central than
expected in comparable random graphs.

\begin{figure}[h]
\begin{minipage}[t]{7.5cm}
\centering
{\includegraphics[width=7.5cm,height=7cm]{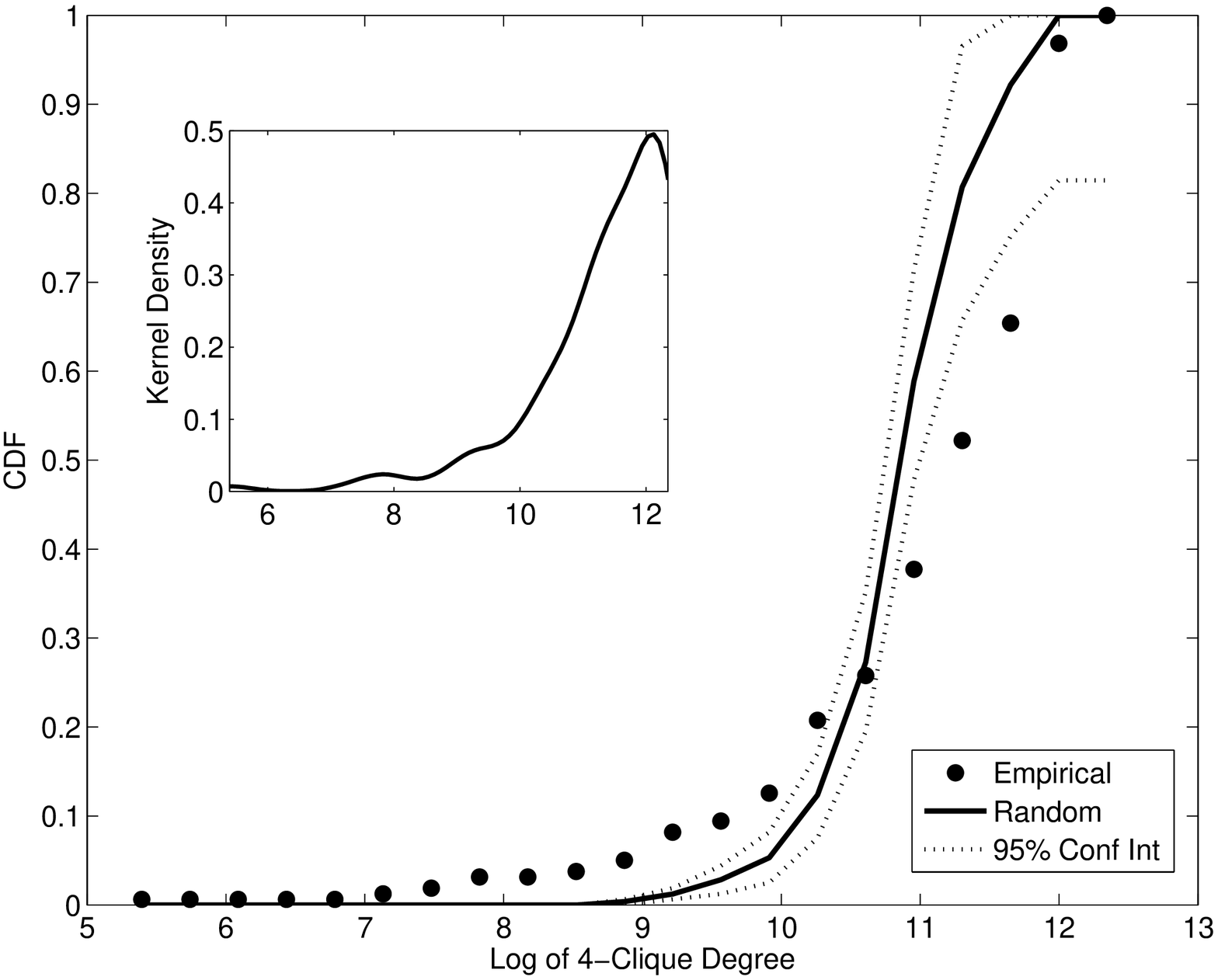}}
\end{minipage}\hfill
\begin{minipage}[t]{7.5cm}
\centering
{\includegraphics[width=7.5cm,height=7cm]{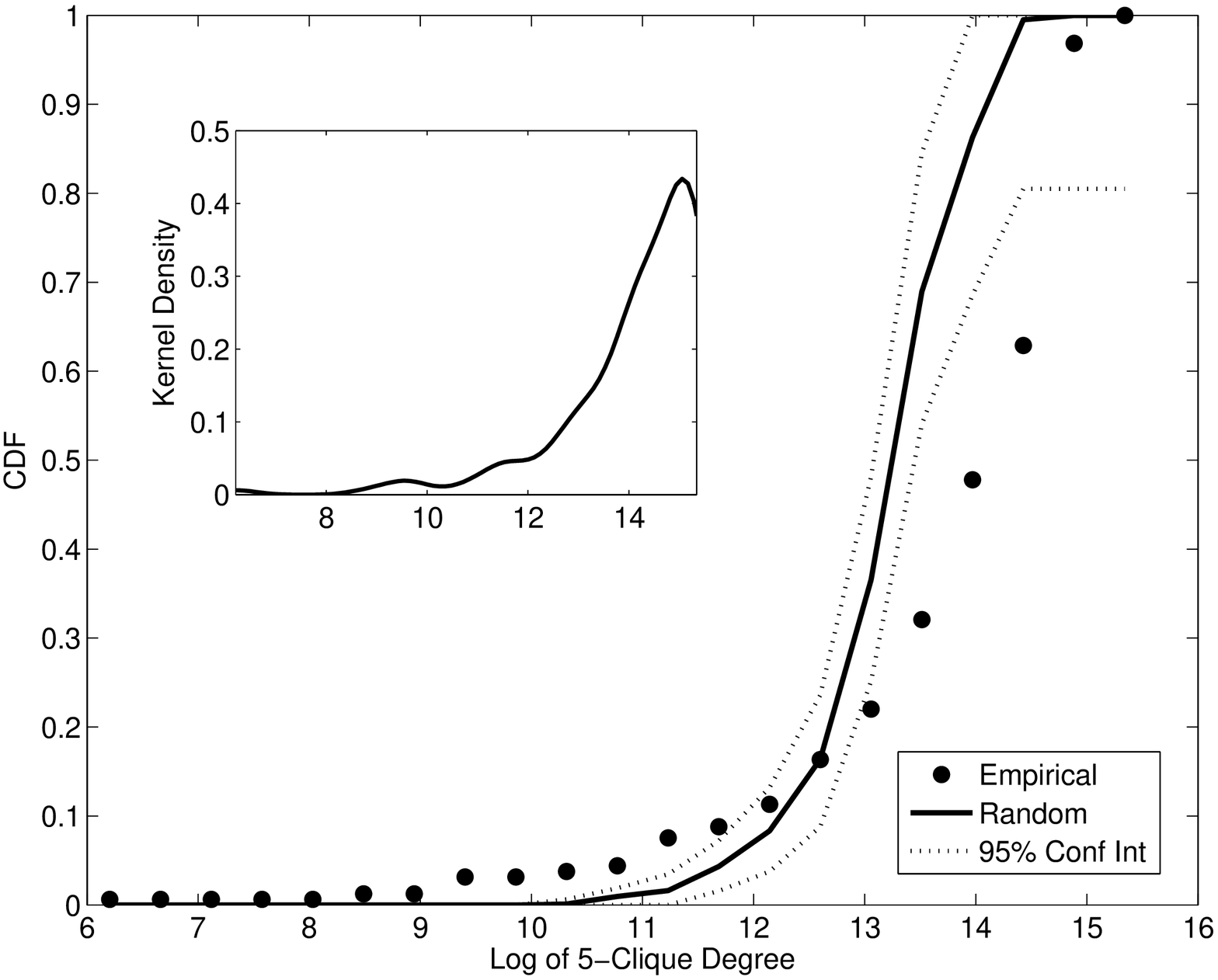}}
\end{minipage}
\caption{Clique-degree distributions in year 2000 of order $k=3$
(left) and $k=4$ (right). Main plots: Cumulative distribution
function (CDF, filled circles) together with expected shape of CDF
in randomly-reshuffled graphs with the same degree distributions as
the observed one. Dotted lines: 95\% confidence intervals for the
CDF estimate based on 10000 samples. X-axis: Log of the number of
nodes belonging to $k$-order cliques.} \label{Fig:CliqueDeg}
\end{figure}

\begin{figure}[h]
\begin{minipage}[t]{7.5cm}
\centering
{\includegraphics[width=7.5cm,height=7cm]{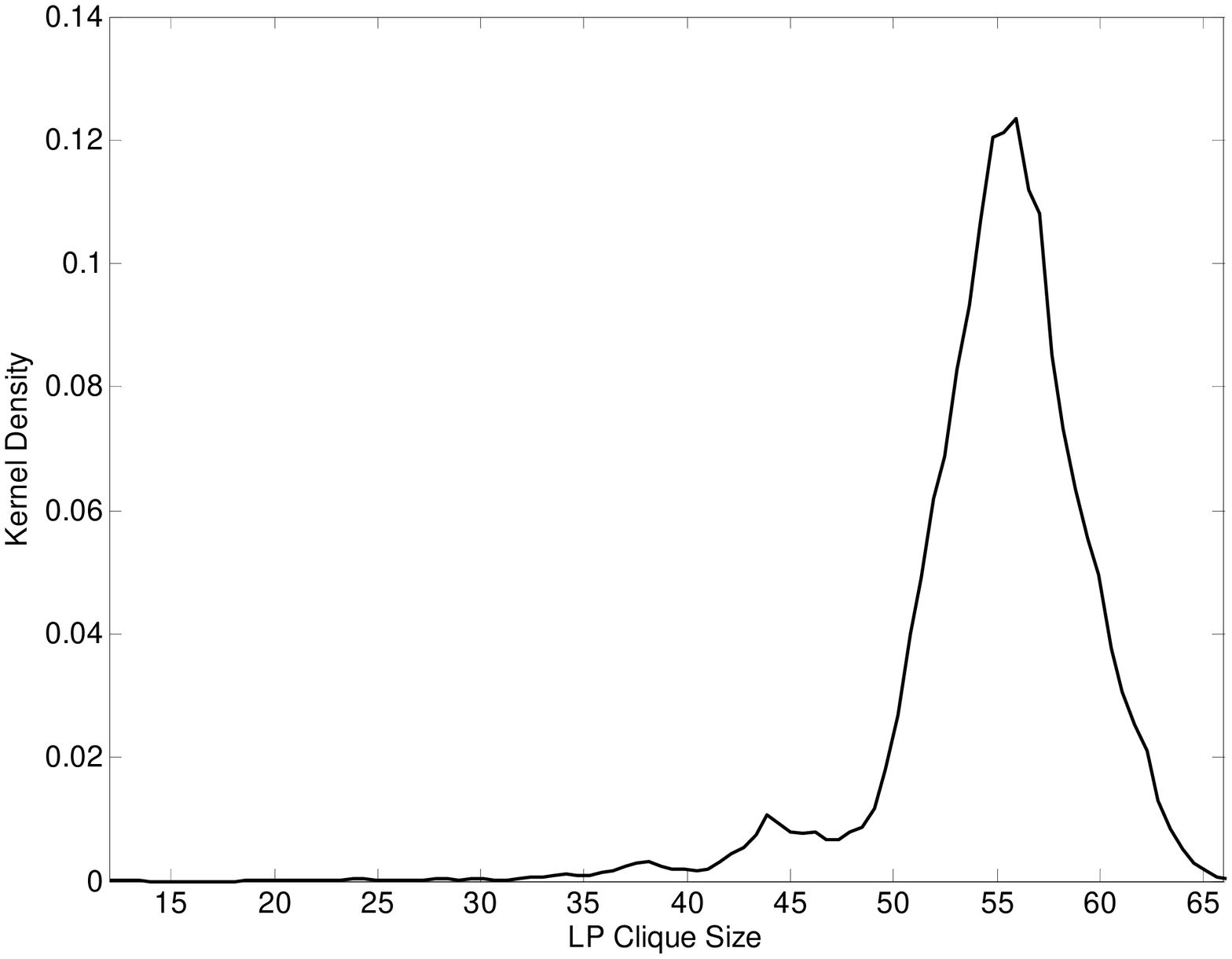}}
\end{minipage}\hfill
\begin{minipage}[t]{7.5cm}
\centering
{\includegraphics[width=7.5cm,height=7cm]{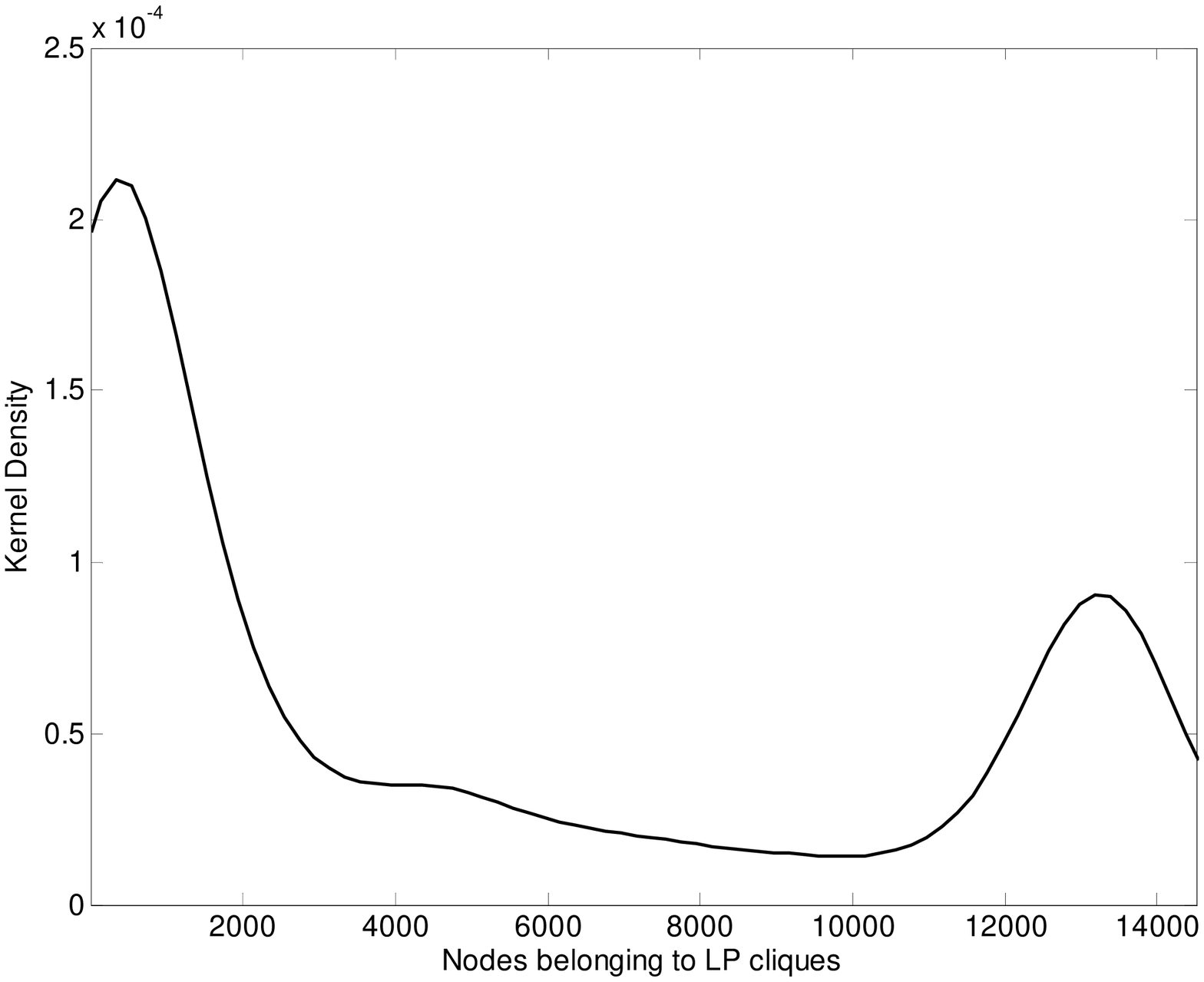}}
\end{minipage}
\caption{Luce and Perry \cite{CliqueDeg3} cliques (in year 2000).
Left: Kernel estimate of LP clique-size distribution. Right: Kernel
estimate of the distribution of number of nodes belonging to LP
cliques of any size (greater than or equal to the minimal one, i.e.
12).} \label{Fig:LP-Cliques}
\end{figure}

The relatively high clustering level detected in the WTW hints to a
network architecture that, especially in the binary case, features a
peculiar clique structure. To further explore the clique structure
of the WTW we computed, in the binary case, the node k-clique degree
(NkCD) statistic \citep{CliqueDeg1}. The NkCD for node $i$ is
defined as the number of $k$-size fully connected subgraphs
containing $i$. Since NkCD for $k=2$ equals node degree, and for
$k=3$ is closely related to the binary clustering coefficient,
exploring the properties of NkCD for $k>3$ can tell us something
about higher-order clique structure of the WTW. Figure
\ref{Fig:CliqueDeg} reports ---for year 2000--- an example of the
cumulative distribution function (CDF) of NkCDs for $k=4,5$ (very
similar results hold also for the case $k=6$). In the insets, a
kernel estimate of the corresponding probability distributions are
also provided. We also compare empirical CDFs with their expected
shape in random nets where, this time, links are reshuffled so as to
preserve the initial degree sequence (i.e., we employ the
edge-crossing algorithm, cf. Refs. \cite{CliqueDeg1,CliqueDeg2} for
details). The plots indicate that the majority of nodes in the WTW
are involved in a very large number of higher-degree cliques, but
that the observed pattern is not that far from what would have been
expected in networks with the same degree sequence (if any, observed
NkCDs place relatively more mass on smaller NkCDs and less on
medium-large values of NkCDs). Notice that these findings are at
odds with what commonly observed in many other real-world networks,
where NkCD distributions are typically power law. However, this
peculiar feature of the WTW does not come as surprise, given its
very high average degree, and the large number of countries trading
with almost everyone else in the sample. Note also that the high
connectivity of the binary WTW makes it very expensive to compute
NkCD distributions for $k>6$. In order to shed some light on the
clique structure for larger values of $k$, we have employed standard
search algorithms to find all Luce and Perry (LP) $k$-cliques
\cite{CliqueDeg3}, that is maximally connected k-size subgraphs
\footnote{A subgraph is defined to be maximal with respect to some
property whenever either it has the property or every pair of its
points has the property and, upon the addition of any point, either
it loses the property or there is some pair of its points which does
not have the property. Thus Luce and Perry's cliques are maximal
complete subgraph such that every pair of points in the clique is
adjacent, and the addition of any point to the clique makes it
incomplete. Note that an agent belonging to a LP clique of order $k$
will automatically belong to all the sub-cliques of that one of
order $h<k$ with that node as one vertex.}. Notice that in general
the WTW does not exhibit LP cliques with size smaller than 12 (this
is true in any year). Furthermore, a large number of LP cliques are
of a size between 50 and 60 (see Figure \ref{Fig:LP-Cliques}, left
panel, for a kernel-density estimate of LP clique size distribution
in 2000). Hence, the binary WTW, due to its extremely dense
connectivity pattern, seems to display a very intricate clique
structure. Additional support to this conclusion is provided by
Figure \ref{Fig:LP-Cliques} (right panel), where we plot a
kernel-density estimate of the distribution of the number of agents
belonging to LP cliques of any size. Although a large majority of
countries belong to less than 2000 LP cliques, a second peak in the
upper-tail of the distribution emerges, indicating that a
non-negligible number of nodes are actually involved in at least
10000 different LP cliques of any size (greater than 12).

\subsection{Correlation Structure and Node Characteristics}

\begin{figure}[h]
\begin{minipage}[t]{7.5cm}
\centering
{\includegraphics[width=7.5cm,height=7cm]{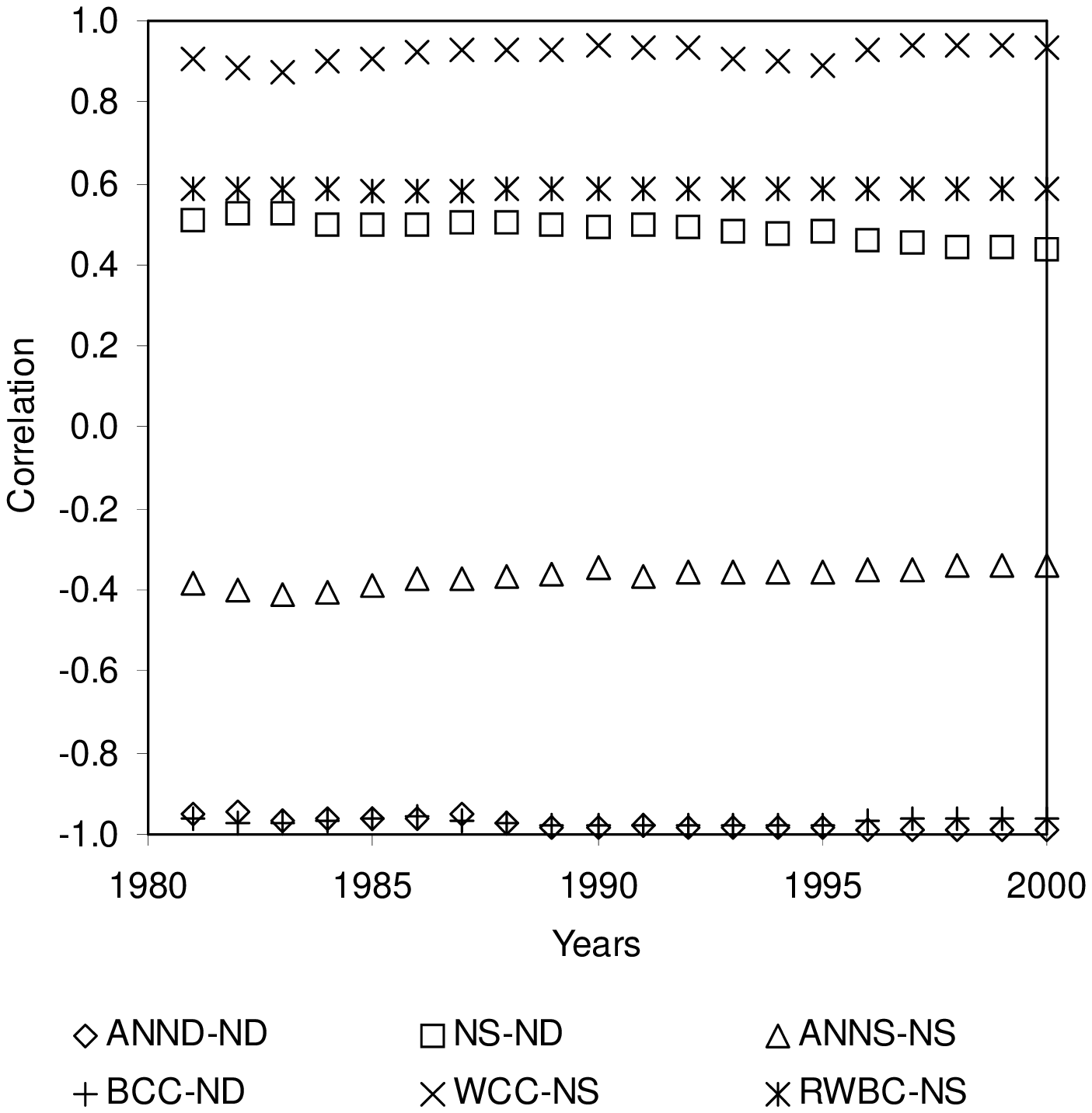}}
\caption{Correlation between node statistics vs. years.}
\label{Fig:corr_stats}
\end{minipage}\hfill
\begin{minipage}[t]{7.5cm}
\centering
{\includegraphics[width=7.5cm,height=7cm]{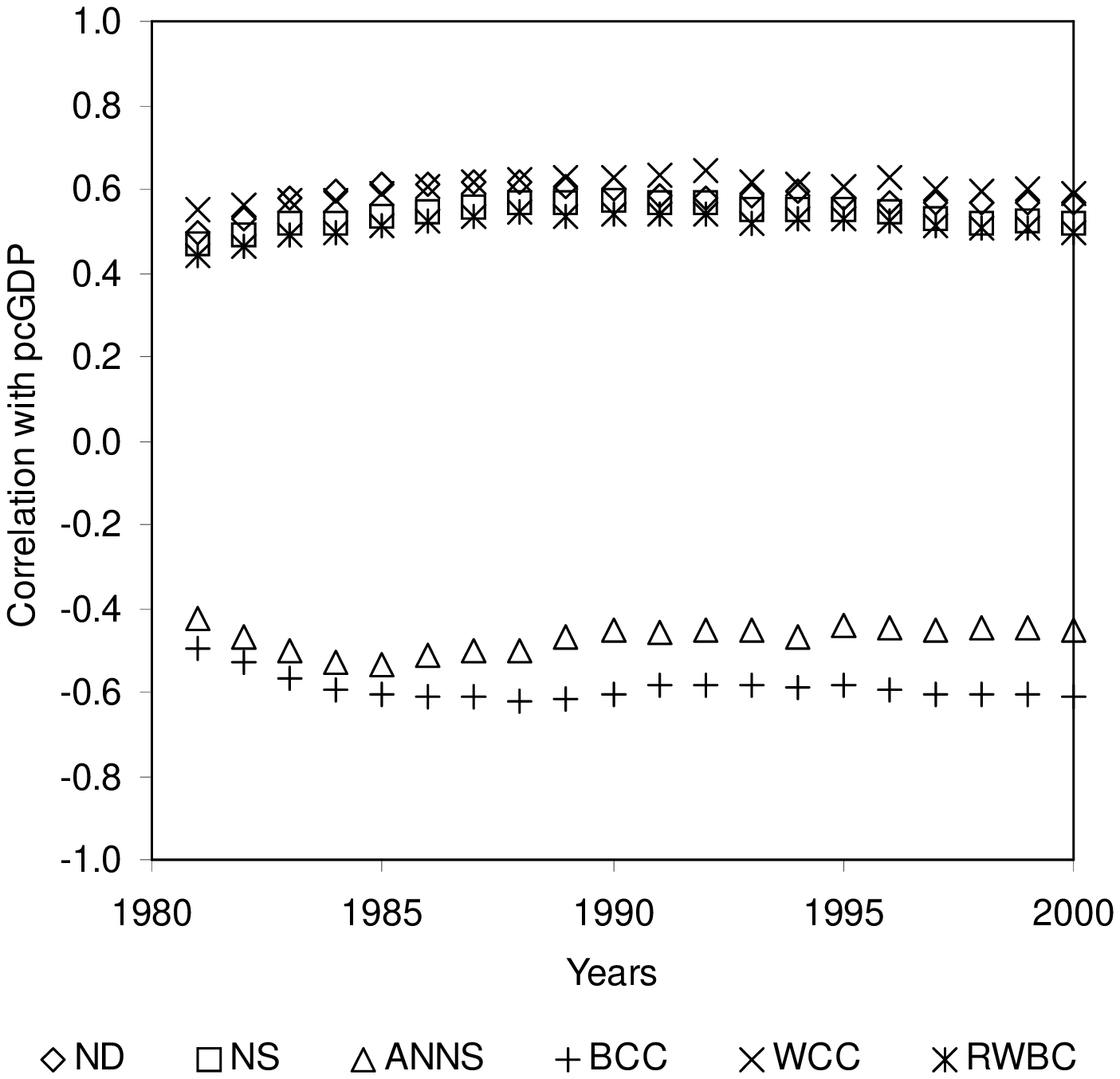}}
\caption{Correlation between node statistics and country per-capita
gross-domestic product (GDP) vs. years.} \label{Fig:corr_pcgdp}
\end{minipage}
\end{figure}

To further explore the topological properties of the WTW, we turn
now to examine the correlation structure existing between binary-
and weighted-network statistics \footnote{More precisely, the
correlation coefficient between two variables $X$ and $Y$ is defined
here as the product-moment (Pearson) sample correlation, i.e.
$\sum_{i}{(x_i-\overline{x})(y_i-\overline{y})}/[(N-1)s_X s_Y]$,
where $\overline{x}$ and $\overline{y}$ are sample averages and
$s_X$ and $s_Y$ are sample standard deviations.}. As expected
\cite{SeBo03,Garla2004,Garla2005}, Figure \ref{Fig:corr_stats} shows
that the binary version of the WTW is strongly disassortative in the
entire sample period. Furthermore, countries holding many trade
partners do not typically form trade triangles. Conversely, the
weighted WTW turns out to be a weakly disassortative network.
Moreover, countries that are intensively connected (high NS) are
also more clustered (high WCC). This mismatch between binary and
weighted representations can be partly rationalized by noticing that
the correlation between NS and ND is positive but not very large (on
average about 0.45), thus hinting to a topological structure where
having more trade connections does not automatically imply to be
more intensively connected to other countries in terms of total
trade controlled. As to centrality, RWBC appears to be positively
correlated with NS, signalling that in the WTW there is little
distinction between global and local centrality \footnote{These
results can be made more statistically sound by comparing
empirically-observed correlation coefficients with their expected
counterparts under random schemes B-RS and W-RS (see above for their
definitions). Simulation results (not shown in the paper for the
sake of brevity) indicate that almost all empirical correlation
coefficients (in every year) are in absolute value larger than the
absolute value of their expected counterpart under either B-RS or
W-RS. This means that the magnitude of almost all observed
correlations are bigger than expected. The only exception is the
ANNS-NS correlation that, albeit positive, is not significantly
larger than in W-RS. This indicates that whereas the binary WTW is
strongly disassortative, the weak-disassortative nature of the
weighted WTW is not statistically distinguishable from what we would
have observed in comparable random graphs.}.

Another interesting issue to explore concerns the extent to which
country specific characteristics relate to network properties. We
focus here on the correlation patterns between network statistics
and country per capita GDP (pcGDP) in order to see whether countries
with a higher income are more/less connected, central and clustered.
The outcomes are very clear and tend to mimic those obtained above
for the correlation structure among network statistics. Figure
\ref{Fig:corr_pcgdp} shows that high-income countries tend to hold
more, and more intense, trade relationships and to occupy a more
central position. However, they trade with few and weakly-connected
partners, a pattern suggesting the presence of a sort of ``rich-club
phenomenon'' \footnote{Again, simulation results (not shown here)
suggest that all empirical correlations are in absolute values
larger than their expected level under reshuffling schemes B-RS and
W-RS, except for the ANNS-pcGDP correlation.}.

\begin{figure}[h]
\begin{minipage}[t]{7.5cm}
\centering
{\includegraphics[width=7.5cm,height=7cm]{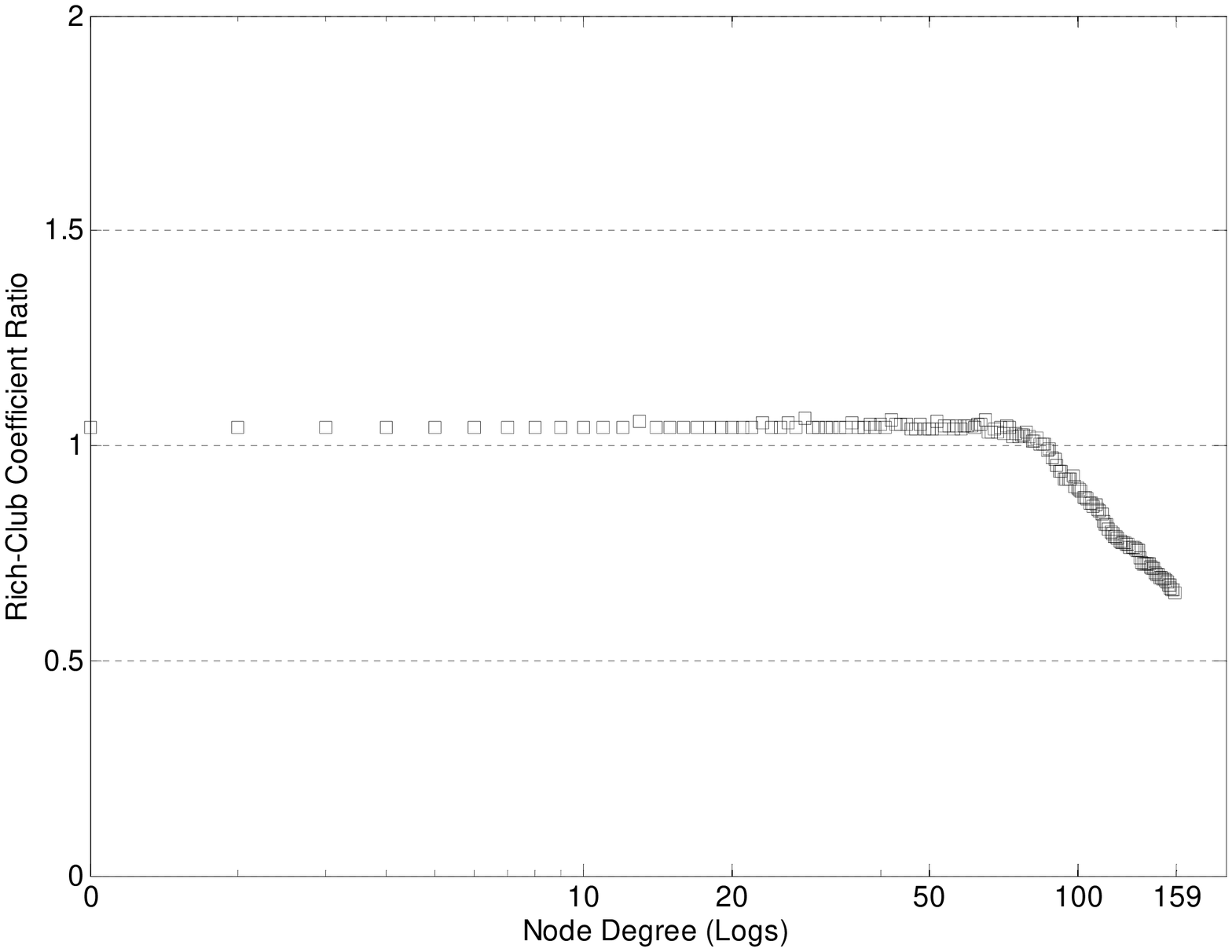}}
\caption{Rich-club behavior in the binary WTW (year=2000). Rich-club
coefficient ratio vs. logged degree. The rich-club coefficient ratio
is obtained dividing the standard rich-club coefficient by its value
in random uncorrelated networks \citep{RichClub2}. Values larger
than one indicate rich-club ordering.} \label{Fig:richclub_binary}
\end{minipage}\hfill
\begin{minipage}[t]{7.5cm}
\centering
{\includegraphics[width=7.5cm,height=7cm]{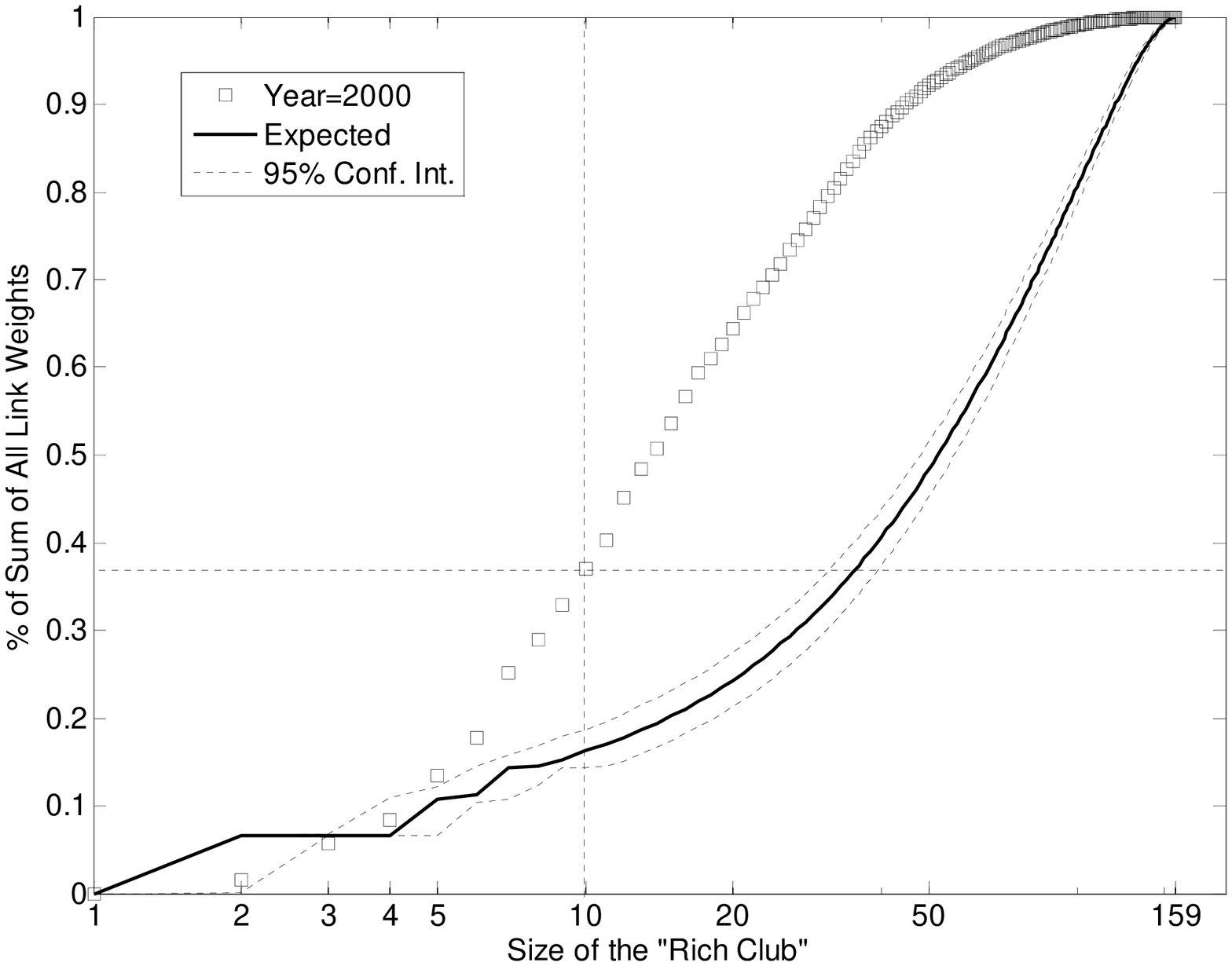}}
\caption{Rich-club behavior in the weighted WTW (year=2000).
Percentage of total link-weight sum explained by the richest $k$
countries in terms of NS. Expected values and 95\% confidence
intervals computed using a W-RS reshuffling scheme (10000
replications).} \label{Fig:richclub_weighted}
\end{minipage}
\end{figure}

To further explore this evidence, we have firstly considered the
binary version of the WTW and we have computed the rich-club
coefficient $R^t(k)$, defined, for each time period $t$ and degree
$k$, as the percentage of edges in place among the nodes having
degree higher than $k$ (see, e.g., Ref. \cite{RichClub1}). Since a
monotonic relation between $k$ and $R^t(k)$ is to be expected in
many networks, due to the intrinsic tendency of hubs to exhibit a
larger probability of being more interconnected than low-degree
nodes, $R^t(k)$ must be corrected for its version in random
uncorrelated networks (see Ref. \cite{RichClub2} for details). If
the resulting (corrected) rich-club index $\tilde{R}^t(k)>1$,
especially for large values of $k$, then the corresponding graph
will exhibit statistically-significant evidence for rich-club
behavior. In our case, the binary WTW does not seem to show any
clear rich-club ordering, as Figure \ref{Fig:richclub_binary} shows
for year 2000. This contrasts with, e.g., the case of scientific
collaborations networks, but is well in line with the absence of any
rich-club pattern in protein interaction and Internet networks
\citep{RichClub2}. This may be intuitively due to the very high
density of the underlying binary network, but also to the fact that,
as suggested by Ref. \cite{RichClub2} and discussed above, any
binary structure underestimates the importance of intensity of
interactions carried by the edges. In fact, if studied from a
weighted-network perspective, the WTW exhibits indeed much more
rich-club ordering than its binary version. To see why, for any
given year we have sorted in a descending order the nodes
(countries) according to their strength, taken as another measure of
richness. We have then computed the percentage of the total trade
flows in the network that can be imputed to the trade exchanges
occurring (only) among the first $k$ nodes of the NS year-$t$
ranking, i.e. a $k$-sized rich club. More precisely, let
$\{i_1^t,i_2^t,...,i_N^t\}$ the labels of the $N$ nodes sorted in a
descending order according to their year-$t$ NS. The coefficient for
a given rich-club size $k>1$ is computed as the ratio between
$\sum_{j=1}^{k}\sum_{h=1}^{k}{w_{i_j^t i_k^t}}$ and the sum of all
entries of the matrix $W^t$. Figure \ref{Fig:richclub_weighted}
shows how our crude ``weighted rich-club coefficient'' behaves in
year 2000 for an increasing rich-club size. It is easy to see that
the 10 richest countries in terms of NS are responsible of about
40\% of the total trade flows (see dotted vertical and horizontal
lines), a quite strong indication in favor of the existence of a
rich club in the weighted WTW. This is further confirmed if one
compares the empirical observations (very high) with their expected
values under a W-RS reshuffling scheme (much lower, also after 95\%
confidence intervals have been considered) \footnote{Additional
research on the rich-club phenomenon in the WTW could entail a
backbone-extraction analysis like the one performed in Ref.
\citep{serrano-2008}.}.

In summary, the overall picture that our correlation analysis
suggests is one where countries holding many trade partners and/or
very intense trade relationships are also the richest and most
(globally) central; typically trade with many countries, but very
intensively with only a few very-connected ones; and form few, but
intensive, trade clusters (triangles).

Furthermore, our correlation analysis provides further evidence to
the distributional stability argument discussed above. Indeed, we
have already noticed that the first four moments of the
distributions of statistics under study (ND, NS, ANNS, WCC, RWBC)
display a marked stability over time. Figure \ref{Fig:corr_stats}
shows that also their correlation structure is only weakly changing
during the sample period. This suggests that the whole architecture
of the WTW has remained fairly stable between 1981 and 2000. To
further explore the implications of this result, also in the light
of the ongoing processes of internationalization and globalization,
we turn now to a more in-depth analysis of the in-sample dynamics
and out-of-sample evolution of WTW topological structure.

\subsection{Within-Sample Distribution Dynamics}
\label{SubSec:DistrDyn}

The foregoing evidence suggests that the shape of the distributions
concerning the most important topological properties of WTW displays
a rather strong stability in the 1981-2000 period. However,
distributional stability does not automatically rule out the
possibility that between any two consecutive time steps, say
$t-\tau$ and $t$, a lot of shape-preserving turbulence was actually
going on at the node and link level, with many countries and/or link
weights moving back and forth across the quantiles of the
distributions. In order to check whether this is the case or not, we
have computed stochastic-kernel estimates
\citep{ChungStocKern1960,Futia1982} for the distribution dynamics
concerning node and link statistics. More formally, consider a
real-valued node or link statistic $X$. Let
$\phi^{\tau}(\cdot,\cdot)$ be the joint distribution of $(X^t,
X^{t-\tau})$ and $\psi^{\tau}(\cdot)$ be the marginal distribution
of $X^{t-\tau}$. We estimate the $\tau$-year stochastic kernel,
defined as the conditional density
$s^{\tau}(x|y)=\phi^{\tau}(x,y)/\psi^{\tau}(y)$ \footnote{Cf. Refs.
\cite{df1999,FiaschiLavezzi2007} for economic applications. Here and
in what follows, Markovianity of the statistics under analysis has
been assumed without performing more rigorous statistical tests
\cite{Markovianity}. This is actually one of the next points in our
agenda.}.

\begin{figure}[h]
\begin{minipage}[t]{7.5cm}
\centering {\includegraphics[width=7.5cm,height=6cm]{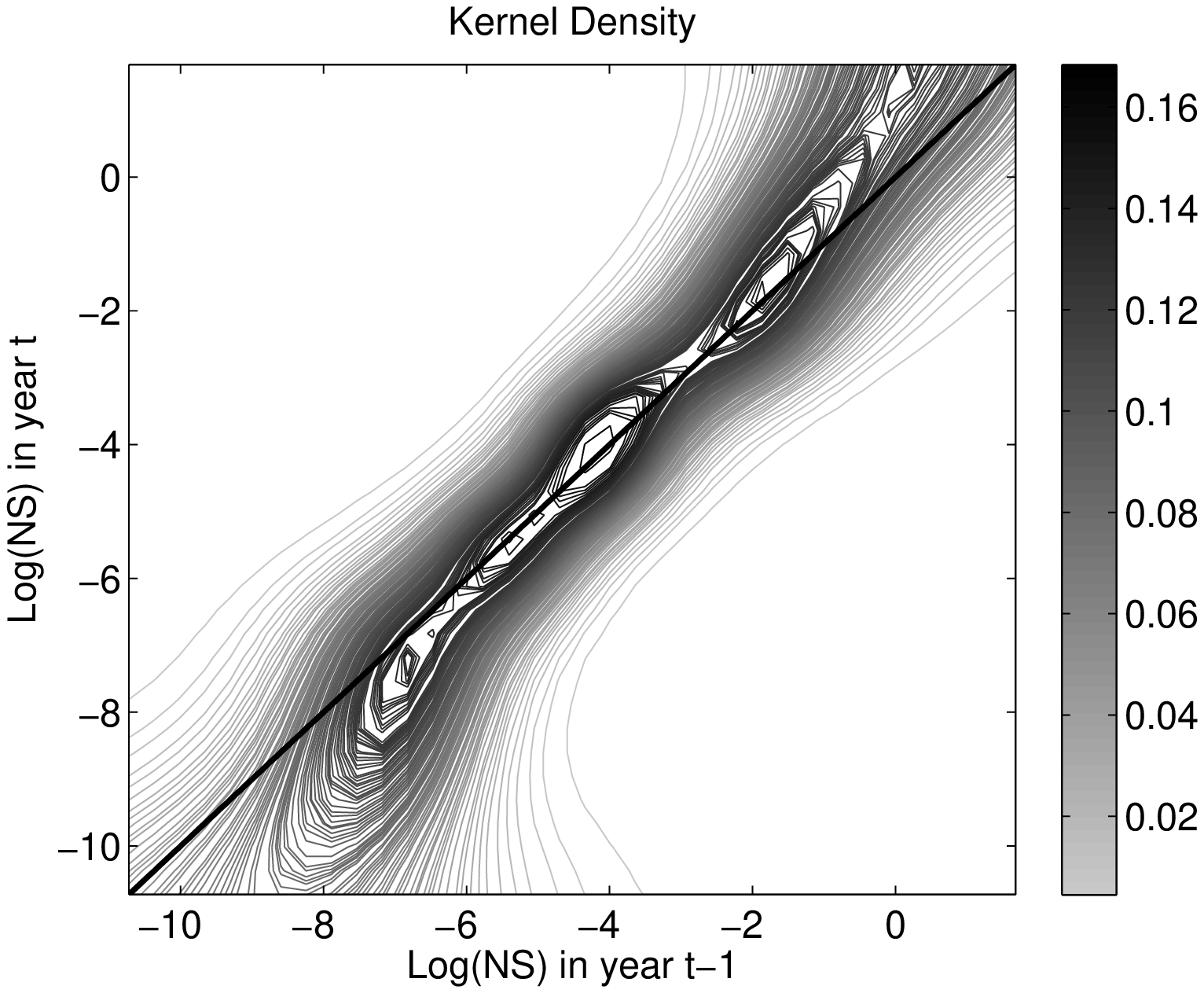}}
\caption{Contour plot of stochastic-kernel estimates for logged node
strength (NS). Solid line: Main $45^\circ$ diagonal.}
\label{Fig:kernel_ns}
\end{minipage}\hfill
\begin{minipage}[t]{7.5cm}
\centering
{\includegraphics[width=7.5cm,height=6cm]{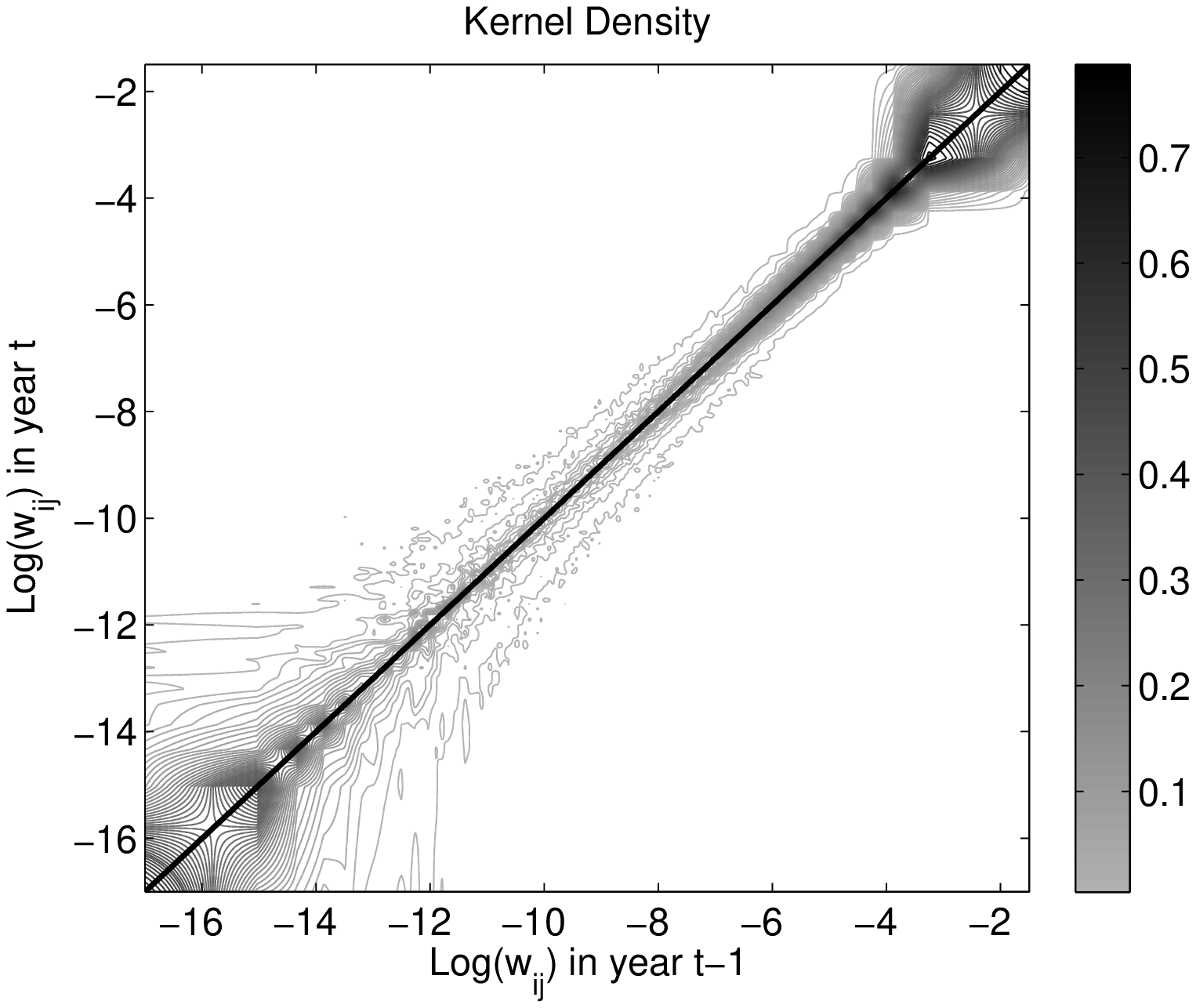}}
\caption{Contour plot of stochastic-kernel estimates for logged
positive link-weights (\textsf{log}$(w_{ij}^t), w_{ij}^t>0$). Solid
line: Main $45^\circ$ diagonal.} \label{Fig:kernel_linkweights}
\end{minipage}
\end{figure}

Figures \ref{Fig:kernel_ns} and \ref{Fig:kernel_linkweights} present
the contour plots of the estimates of the 1-year kernel density of
logged NS and logged positive link-weights. Notice that the bulk of
the probability mass is concentrated close to the main diagonal
(displayed as a solid $45^\circ$ line). Similar results are found
for all other real-valued node statistics (ANNS, WCC and RWBC) also
at larger time lags. The kernel density of logged positive link
weights, contrary to the logged NS one, is instead extremely
polarized towards the extremes of the distribution range, whereas in
the middle of the range it is somewhat flatter (Figure
\ref{Fig:kernel_linkweights}). We will go back to the implications
that this feature has on out-of-sample distributional evolution
below.

This graphical evidence hints to a weak turbulence for the
distributional dynamics of all node and link statistics under
analysis. To better appreciate this point, we have estimated, for
all five node statistics employed above (ND, NS, ANNS, WCC, RWBC),
as well as link-weight distributions, the entries of $\tau$-step
Markov transition matrices \cite{StuartOrd1994}, where
$\tau=1,2,\dots,T-1$ is the time lag. More formally, suppose that
the distribution dynamics of the statistic $X$ can be well described
using $K$ quantile classes (QCs) in every year $t$ (see footnote
72). Given the above stability results, we can assume that the
process driving the distribution dynamics of $X$ is stationary and
can be well represented by a discrete-state Markov process defined
over such $K$ QCs. Let $n_{i,j}^{t-\tau,t}$ be the number of
countries whose statistic $X$ was in QC $i$ in year $t-\tau$ and
moved to QC $j$ in year $t$. Then the statistic:
\begin{equation}
\hat{p}_{ij}^{\tau}=\frac{\sum_{t=\tau+1}^{T}{n_{i,j}^{t-\tau,t}}}{\sum_{t=2}^{T}\sum_{h=1}^{K}{n_{i,h}^{t-\tau,t}}}
\end{equation}
can be shown to be the maximum likelihood estimators of the true,
unobservable, $\tau$-step transition probability $p_{ij}^{\tau}$,
i.e. the probability that $X$ belongs to QC $i$ at $t-\tau$ and to
QC $j$ at $t$ \cite{AndersonGoodman1957}.

In order to build a measure of persistence of distribution dynamics,
we have computed for each node or link statistic $X$ the percentage
mass of probability that lies within a window of $\omega$ quantiles
from the main diagonal of the estimated $\tau$-step
transition-probability matrix
$\hat{P}^{\tau}=\{\hat{p}_{ij}^{\tau}\}$, defined as:

\begin{equation}
M_{\omega,K}^{\tau}(X)=\frac{1}{K}\sum_{h=1}^{K}\sum_{l:|l-h|\leq
w}^{K}{\hat{p}_{hl}^{\tau}},
\end{equation}
where the window $\omega=0,1,\dots,K-1$. For example, when
$\omega=0$, $M_{0,K}^{\tau}(X)$ boils down to the trace of
$\hat{P}^{\tau}$ divided by $K$, whereas if $\omega=1$,
$M_{1,K}^{\tau}(X)$ is the average of all the entries in the main
diagonal and those lying one entry to the right and one entry to the
left of the main diagonal itself. The statistic
$M_{\omega,K}^{\tau}(X)\in [0,1]$ and increases the larger the
probability that a country remains in the same (or nearby) QC
between $t-\tau$ and $t$ (for any given choice of $\tau$, $w$ and
$K$).

Table \ref{Tab:distr_dynamics_nodes} shows for our main statistics
(ND, NS, ANNS, WCC, RWBC) and $K=10$, the values of
$M_{\omega,K}^{\tau}(X)$ as $\tau \in \{1,4,7,10\}$ and $\omega=0,1$
\footnote{Parameters outside these ranges and choices do not change
the main implications of the analysis. Similar results also are
obtained if one computes the statistic $M_{\omega,K}^{\tau}$ on
logged distributions of $X$ (i.e., logs of node statistics and
positive link weights) and/or one preliminary re-scales the data by
removing the time-averages of the distributions in order to wash
away possible trends.}. The figures strongly supports the result
obtained by looking at the estimated stochastic kernels. Indeed, the
entries of $\hat{P}^{\tau}$ close to the main diagonal always
represent a large mass of probability, thus hinting to a
distribution dynamics that in the period 1981-2000 is characterized
by a rather low turbulence. For example, more than 96\% of countries
are characterized by node statistics that either stick to the same
QC between $t-\tau$ and $t$, or just move to a nearby QC of the
distribution. This share is often close to 99\%. To better
statistically evaluate the figures in Table
\ref{Tab:distr_dynamics_nodes}, we have also estimated the
distribution of $M_{\omega,K}^{\tau}(X)$ under reshuffling scheme
W-RS, i.e. in random graphs where we keep fixed the observed
adjacency matrix $A^t$ and we re-distribute weights at random by
reshuffling the observed link-weight distribution \footnote{The
distributions of $M_{\omega,K}^{\tau}(X)$ turn out to be
well-proxied by Gaussian densities.}. This allows us to compute
confidence intervals (at 95\%) for $M_{\omega,K}^{\tau}(X)$. As
reported in Table \ref{Tab:distr_dynamics_nodes}, the empirical
values are always larger than the upper bound of these confidence
intervals, thus confirming the relatively strong persistence found
in WTW node-statistic dynamics.

The same analysis can be also applied to the link-weight
distribution $w^t=\{w_{ij}^t, i\neq j=1,\dots,N\}$. In order not to
treat the same way existing links (with strictly positive weight)
and absent links (with a zero weight), we first define the two link
sets $L_0^t=\{(i,j),i\neq j=1,\dots,N:w_{ij}^t=0\}$ and
$L_{+}^t=\{(i,j),i\neq j=1,\dots,N:w_{ij}^t>0\}$ and we then
separately study the within-sample dynamics of the associated link
distributions $w_0^t=\{w_{ij}^t\in w^t: (i,j)\in L_0^t\}$ and
$w_{+}^t=\{w_{ij}^t\in w^t: (i,j)\in L_{+}^t\}$. To begin with,
notice that a strong persistence also characterizes the dynamics of
transition from an absent link to an existing one (and back).
Indeed, the estimated probability of remaining an absent link (zero
weight) is 0.9191, while that of remaining a present link (positive
weight) is 0.9496. Thus, the link birth-rate is on average about
8\%, while the death-rate is around 5\% \footnote{Standard
deviations of such estimates are quite small. Let $\hat{p}_{00}$ and
$\hat{p}_{++}$, respectively, be the probability of remaining a zero
and positive link weight. We find that $\sigma(\hat{p}_{00})=0.0034$
and $\sigma(\hat{p}_{++})=0.0023$. Similarly, let $\hat{p}_{0+}$ and
$\hat{p}_{+0}$ be the probability of becoming a positive
(respectively, zero) link weight. Since
$\hat{p}_{00}=1-\hat{p}_{0+}$ and $\hat{p}_{++}=1-\hat{p}_{+0}$ by
construction, then $\sigma(\hat{p}_{0+})=\sigma(\hat{p}_{00})$ and
$\sigma(\hat{p}_{+0})=\sigma(\hat{p}_{++})$.}. This means that in
the period 1981-2000, the WTW has shown a slight tendency toward an
increase in trade relationships. This is remarkable for two reasons.
First, our panel of countries has been balanced in order to focus on
a fixed number of nodes. Second, the density of the network was
already very high at the beginning. Table
\ref{Tab:distr_dynamics_linkweights} shows instead the persistence
measure $M_{\omega,K}^{\tau}(X)$ where $X$ are the distributions of
positive link-weights $w_{+}^t$ and the number of QCs is set to
$K=20$. Again, most of transitions occur within the same or nearby
QCs, signaling that also the dynamics of weight distributions of
existing links is rather persistent. Furthermore, as happens for
node statistics, also in this case confidence intervals (at 95\%)
for randomly-reshuffled weights always lie to the left of the
observed value of $M$. Very similar results are obtained computing
the persistence measure $M$ to logged link-weight distributions.

\subsection{Country-Ranking Dynamics}

The distributional-stability results obtained in the foregoing
sections naturally hint to the emergence of a lot of stickiness in
country rankings (in terms of node statistics) as well. To explore
this issue, for each year $t=1981,\dots\,2000$ we have ranked our
$N=159$ countries according to any of the five main statistics
employed so far (ND, NS, ANNS, WCC, RWBC) in a descending order. The
first question we are interested in is assessing to which extent
also these rankings are sticky across time. We check stability of
rankings by computing the time-average of Spearman rank-correlation
coefficients (SRCC) \cite{Spearman,HollanderWolfe1973} between
consecutive years \footnote{The SRCC between two variables
$(Z^1_i,Z^2_i)$, $i=1,\dots,N$ is simply the Pearson product-moment
correlation coefficient defined above, now computed between
$(z^1_i,z^2_i)$, where $z^j_i$ are the ranks of each observation $i$
according to the original variables $Z^j_i$, $j=1,2$. Therefore, the
SRCC equals one if the $N$ observations are ranked the same under
$Z^1$ and $Z^2$; it equals $-1$ if the ranks according to the two
variables are completely reversed; and it is zero if there is no
correlation whatsoever between the ranking of the observations
according to $Z^1$ and $Z^2$. We focus here only on one-year lags
between rankings. An interesting extension to the present analysis
would be to check for stability of rankings across time lags of
length $\tau>1$.}. More formally, let $r_{(i)}^t (X)$ be the rank of
country $i=1,\dots,N$ in year $t$ according to statistic $X$, and
$\rho_{t-1,t}(X)$ be the SRCC between rankings at two consecutive
years $t-1$ and $t$, for $t=2,\dots,T$. Our ranking-stability index
(RSI) for the statistic $X$ is defined as
\begin{equation}
RSI(X)=\frac{1}{T-1}\sum_{t=2}^{T}{\rho_{t-1,t}(X)}.
\end{equation}
Of course $RSI(X)\in [-1,1]$, where $RSI(X)=-1$ implies the highest
ranking turbulence, whereas $RSI(X)=1$ indicates complete stability.
The results for the WTW suggest that even rankings are very stable
over time. Indeed, one has that $RSI(ND)=0.9833$, $RSI(NS)=0.9964$,
$RSI(ANNS)=0.9781$, $RSI(WCC)=0.9851$ and $RSI(RWBC)=0.9920$. Notice
that, since $\rho_{t-1,t}(X)\rightarrow N(0,N^{-1})$, our $RSI(X)$
should tend to a $N(0,[N(T-1)]^{-1})\cong N(0,3.3102E-4)$.
Therefore, our empirical values are more than 50 standard deviations
to the right of 0 (no average rank correlation).

The second issue that deserves a closer look concerns detecting
which countries rank high according to different node statistics.
Table \ref{Tab:rankings} displays the top-20 countries in each given
node-statistic ranking in 2000, which, given the stability results
above, well represents the entire sample period. First note that,
apart from ANNS, all ``usual suspects'' occupy the top-ten
positions. Germany scores very high for all statistics but ANNS,
while the U.S. and Japan are characterized by a very high rank for
weighted statistics but not for ND. This implies that they have
relatively less trade partners but the share of trade that they
control, the capacity to cluster, and their centrality is very high.
Conversely, countries like Switzerland, Italy and Australia have a
more diversified portfolio of trade partners with which they
maintain less intense trade relationships. Furthermore, it is worth
noting that China was already very central in the WTW in 2000,
despite its clustering level was relatively lower. India was instead
not present among the top-20 countries as far as NS and WCC were
concerned; it was only 14th according to centrality and 11th in the
ND ranking. Notice how all top-20 countries in the ANNS are micro
economies: they typically feature a very low NS and ND, but only
tend connect to the hubs of the WTW. Table \ref{Tab:rankings} also
presents country rankings in 2000 according to GDP and per-capita
GDP (expressed in US dollars per person). As expected, the group of
countries topping the rankings based on weighted statistics (except
ANNS) are also among those having highest GDP levels. This is due to
the fact that link weights are expressed in terms of total trade,
which is typically positively correlated with GDP, and
node-statistics like NS, WCC and RWBC partly reflect this ex-ante
correlation. This is not the case, however, if one compares GDP with
ND rankings, meaning that top countries in terms of number of trade
relations are not also those at the top of GDP rankings. A similar
mismatch occurs between node-statistic and per-capita GDP rankings.
Hence, after one washes away country-size effects (e.g., population
size), top countries in terms of income ---as measured by per-capita
GDP--- do not necessarily occupy the first positions of the rankings
according to their intensity of connections, centrality, and
clustering.

Notwithstanding the presence of a relatively high ranking stability,
there are indeed examples of countries moving up or falling behind
over the period 1981-2000. For example, as far as centrality is
concerned, Russia has steadily fallen in the RWBC ranking from the
6th to the 22th position. A similar downward pattern has been
followed by Indonesia (from 17th to 36th). South Africa has instead
fallen from 23th (in 1981) to 32th (in 1990) and then has become
gradually more central (16th in 2000). On the contrary, the majority
of high-performing Asian economies (HPAE), have been gaining
positions in the RWBC ranking. For example, South Korea went from
the 24th to the 8th position; Malaysia from the 43th to the 21th;
Thailand started from the 42th position in 1981 and managed to
become the 18th best central country in 2000. This evidence strongly
contrast with the recent experience of Latin American (LATAM)
economies (e.g., Mexico and Venezuela) that have -- at best --
maintained their position in the ranking of centrality
\cite{RFS2008wp}.

\subsection{Within-Sample Autocorrelation Structure and Growth
Dynamics} \label{SubSec:Growth} To further explore the properties of
within-sample distribution dynamics, we now investigate
autocorrelation structure and growth dynamics of node and link
statistics. More precisely, let $X_i^t$ the observation of statistic
$X$ for node or link $i$ at time $t$, where $i=1,\dots,I$,
$t=1,\dots,T$, and $I$ stands either for $N$ (in case of a node
statistic) or for $N(N-1)/2$ (in case of link weights). We first
compute the (node or link) distribution of first-order
autocorrelation coefficients (ACC) defined as:

\begin{equation}
\hat{r}_i(X)=\frac{\sum_{t=2}^{T}{(X_i^t-\bar{X}_i^0)(X_i^{t-1}-\bar{X}_i^1)}}{\sqrt{\sum_{t=2}^{T}{(X_i^t-\bar{X}_i^0)^2}\sum_{t=2}^{T}{(X_i^{t-1}-\bar{X}_i^1)^2}}}
\label{Eq:ACC}
\end{equation}
where $\bar{X}_i^j=(T-1)^{-1}\sum_{t=2}^{T}{X_i^{t-j}}$, $j=0,1$.

Second, we compute the first-order ACC $\hat{r}(X)$ on the (node or
link) distribution of $X$ pooled across years. To do so, we
preliminary standardize the distributions $\{X_i^t,i=1,\dots,I\}$
for each $t$, so as to have zero-mean and unitary standard deviation
in each year, and then we pool all $T$ year-distributions together
\footnote{We stop at first-order autocorrelation coefficients
because of the few time observations available. Notwithstanding
their low statistical significance, also second-order ACCs turn out
to be positive albeit much smaller than first-order ones.}.

The left part of Table \ref{Tab:autocorr} shows the values of
$\hat{r}(X)$ together with the population mean and standard
deviation of $\hat{r}_i(X)$ for our five node statistics and link
weights. We also report the percentage of observations (nodes or
links) for which the ACC $\hat{r}_i(X)$ turns out to be larger than
zero. Both $\hat{r}(X)$ and the percentage of positive-ACF
observations indicate a relatively strong persistence in the
dynamics of both node and link statistics.

Given that the pooled ACC figures are very close to unity, we
further check whether autoregressive dynamics governing the
evolution of logged network statistics is close to a random-walk. In
particular, we test whether a Gibrat dynamics (i.e., a
multiplicative process on the levels $X_i^t$, where rates of growth
of $X_i^t$ are independent on $X_i^t$) applies to our variables or
not \cite{Gib31,Sut97}. Notice that, under a Gibrat dynamics,
$X_i^t$ should be in the limit log-normally distributed, which is
what we actually observe in our sample for the majority of node
statistics (see Section \ref{SubSec:Shape}). More formally, we begin
by fitting the simple model:

\begin{equation}
\Delta log(X_i^t)=\beta_i log(X_i^{t-1})+\epsilon_i^{t},
\label{Eq:Gibrat}
\end{equation}
where $\Delta log(X_i^t)=log(X_i^t)-log(X_i^{t-1})$ is the rate of
growth of $X_i^t$ and $\epsilon_i^{t}$ are white-noise errors
orthogonal to $log(X_i^{t-1})$. If a Gibrat dynamics applies for a
given node or link, then $\beta_i=0$. We also fit the model in
\eqref{Eq:Gibrat} to the time-pooled sample by setting
$\beta_i=\beta$, where again we first standardize in each year our
variables in order to wash away trends and spurious dynamics.

As the right part of Table \ref{Tab:autocorr} shows, our data reject
the hypothesis that network statistics follow a Gibrat dynamics.
Indeed, both the population average of $\hat{\beta}_i$ and the
pooled-sample estimate $\hat{\beta}$ are significantly smaller than
zero, thus implying a process where small-valued entities (i.e.,
nodes and links characterized by small values of any given
statistic) tend to grow relatively more than large-valued ones. This
is further confirmed by the percentage of nodes or links for which
$\hat{\beta}_i$ turns out to be significantly smaller than zero.

\begin{figure}[h]
\begin{minipage}[t]{7.5cm}
\centering
{\includegraphics[width=7.5cm,height=6cm]{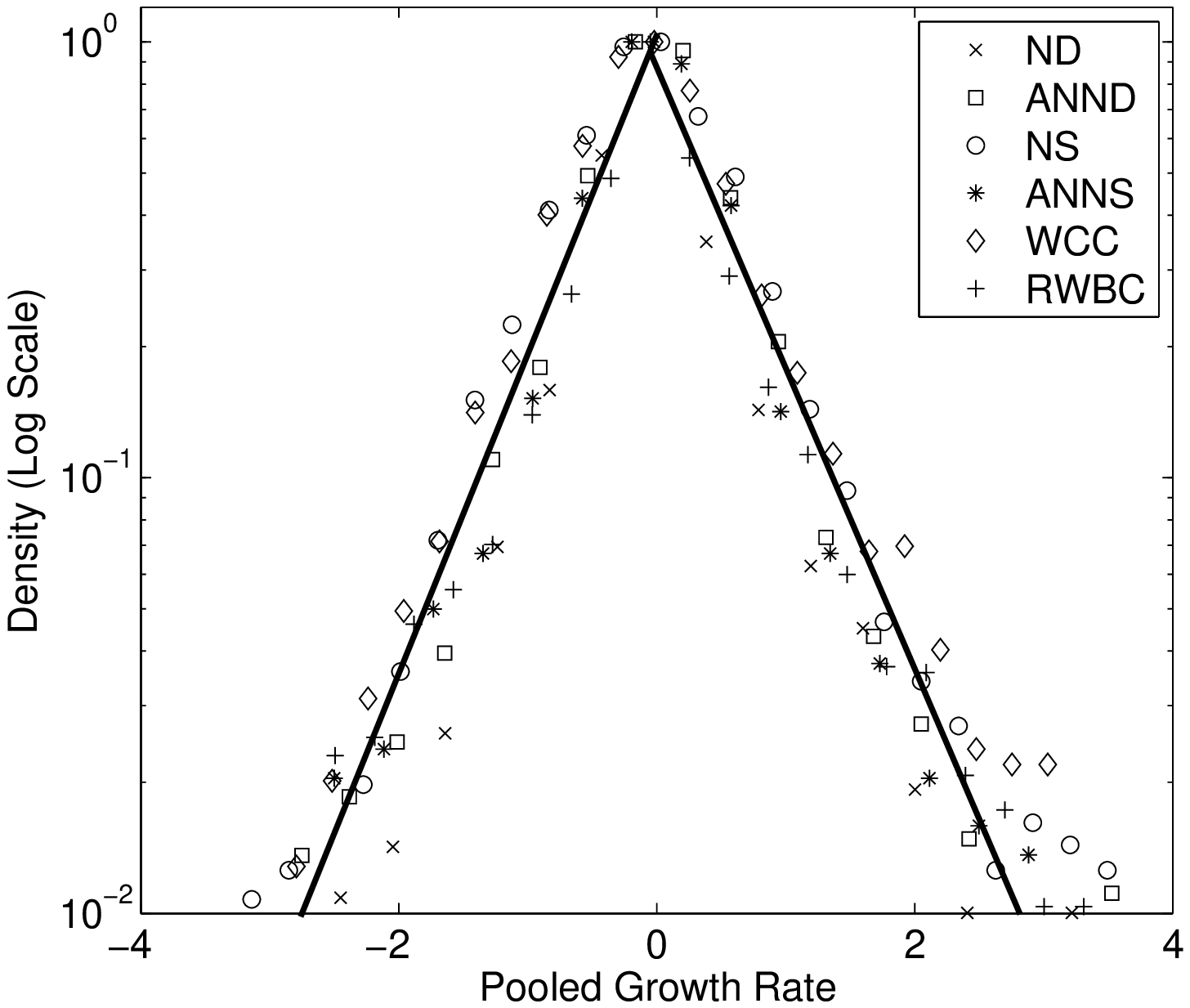}}
\caption{Pooled growth-rate distributions for node statistics.
Y-axis: Log Scale. Solid lines: Laplace fit.}
\label{Fig:gr_distr_node}
\end{minipage}\hfill
\begin{minipage}[t]{7.5cm}
\centering
{\includegraphics[width=7.5cm,height=6cm]{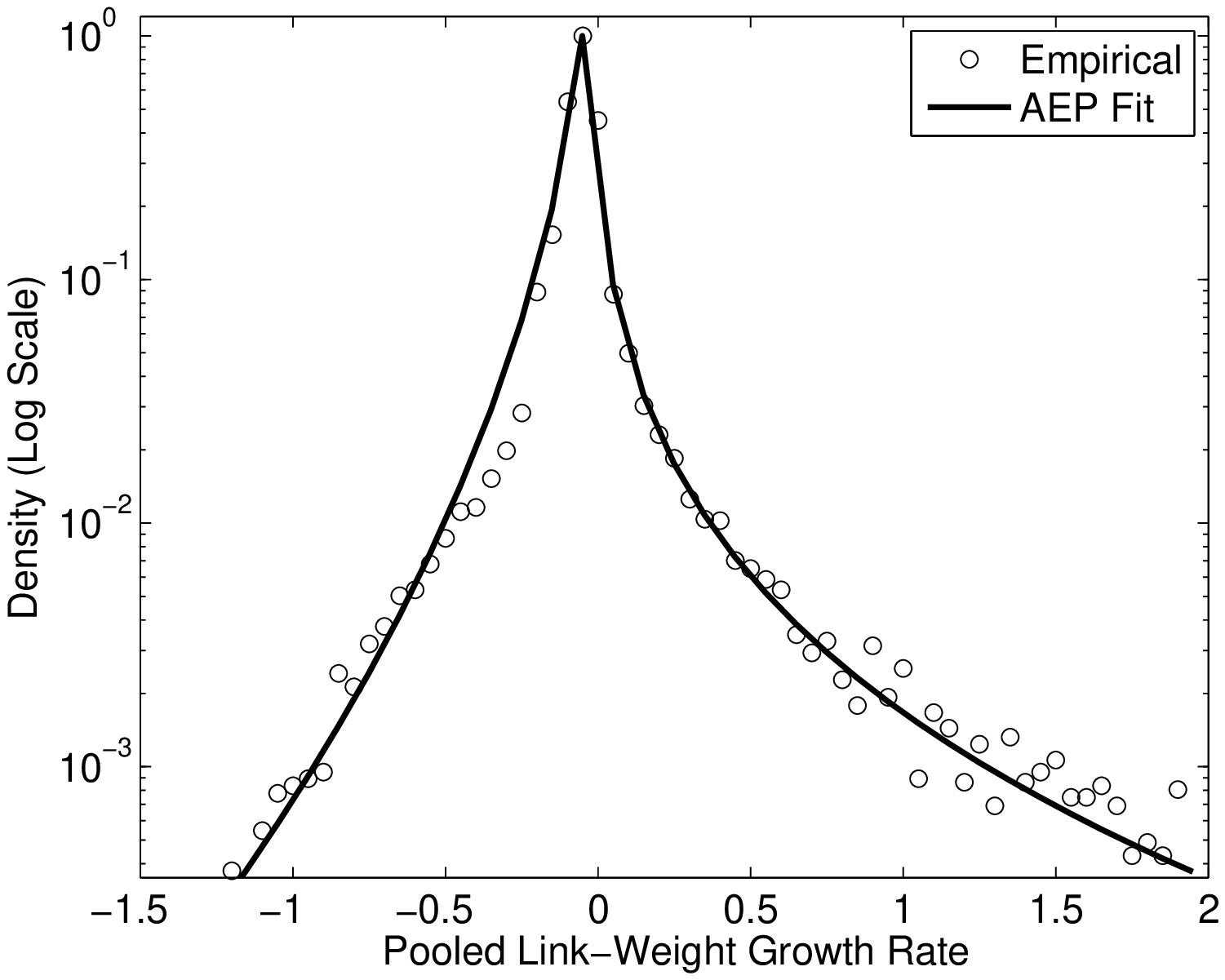}}
\caption{Pooled growth-rate distribution of link weights. Y-axis:
Log Scale. Solid lines: Asymmetric exponential-power (AEP) fit
\eqref{Eq:subboasym}. Parameter estimates: $\hat{b}_l=0.5026$,
$\hat{b}_r=0.2636$, $\hat{a}_l=0.0511$, $\hat{a}_r=0.0615$,
$\hat{m}=-0.0202$.} \label{Fig:gr_distr_link}
\end{minipage}
\end{figure}

Rejection of a Gibrat dynamics also implies that the distributions
of growth rates $\Delta log(X_i^t)$ should depart from Gaussian ones
\cite{BnS06}. This is confirmed by all our pooled fits. Indeed, as
Figure \ref{Fig:gr_distr_node} shows for node statistics, pooled
growth-rate distributions are well proxied by Laplace (fat-tailed,
symmetric) densities. Furthermore, the pooled distribution of growth
rates $g$ for positive link weights is nicely described by an
\textit{asymmetric exponential power} (AEP) density \cite{BnS06b}:

\begin{equation}
d(g;a_{l},a_{r},b_{l},b_{r},m)=\left\{
\begin{array}{cc}
\Upsilon^{-1}e^{-\frac{1}{b_{l}}|\frac{g-m}{a_{l}}|^{b_{l}}},& g<m \\
\Upsilon^{-1}e^{-\frac{1}{b_{r}}|\frac{g-m}{a_{r}}|^{b_{r}}},& g\geq m \\
\end{array}
\right. \label{Eq:subboasym},
\end{equation}
where $\Upsilon=a_l b_l^{1/b_l}\Gamma(1+1/b_l)+a_r b_r^{1/b_r}
\Gamma(1+1/b_r)$, and $\Gamma$ is the Gamma function \footnote{The
AEP features five parameters. The parameter $m$ controls for
location. The two $a$'s parameters control for scale to the left
($a_{l}$) and to the right ($a_{r}$) of $m$. Larger values for $a$'s
imply -- \textit{coeteris paribus} -- a larger variability. Finally,
the two $b$'s parameters govern the left ($b_{l}$) and right
($b_{l}$) tail behavior of the distribution. To illustrate this
point, let us start with the case of a symmetric exponential power
(EP), i.e. when $a_{l}=a_{r}=a$ and $b_{l}=b_{r}=b$. It is easy to
check that if $b=2$, the EP boils down to the normal distribution.
In that case, the correspondent HCE distribution would be
log-normal. If $b<2$, the EP displays tails thicker than a normal
one, but still not heavy. In fact, for $b<2$, the EP configures
itself as a medium-tailed distribution, for which all moments exist.
In the case $b=1$ we recover the Laplace distribution. Finally, for
$b>2$ the EP features tails thinner than a normal one and still
exponential.}. Maximum-likelihood estimation of tail parameters
indicate that link-weight growth rates display tails much fatter
than Laplace ones. Moreover, the right tail is remarkably thicker
than the left one (as $\hat{b}_l=0.5026>0.2636=\hat{b}_r$), see
Figure \ref{Fig:gr_distr_link}. Therefore, link weights are
characterized by a relatively much higher likelihood of large
positive growth events than of negative ones. This result brings
further evidence in favor or the widespread emergence of fat-tailed
growth-rate distributions in economics. In fact, recent studies have
discovered that Laplace (and more generally AEP) densities seem to
characterize the growth processes of many economic entities, from
business companies \cite{Sea96,Aea97,Cea98,Fuetall05} to
world-country GDP and industrial production \cite{FagNapRov2007}.

\subsection{Out-of-Sample Evolution} \label{SubSec:OutofSample}

In the preceding sections, we have investigated the within-sample
dynamics of the distributions of node and link statistics. Now we
turn out attention to the out-of-sample (long-run) evolution of such
distributions by estimating their limiting behavior. To do so, we
employ kernel density estimates obtained above to compute ergodic
densities, which represent the long-run tendency of the
distributions under study \footnote{Given the real-valued statistic
$X$, its ergodic distribution $\phi_{\infty}(\cdot)$ is implicitly
defined for any given $\tau$ as
$\phi_{\infty}(x)=\int{s^{\tau}(x|z)\phi_{\infty}(z)dz}$, where
$s^{\tau}(x|z)$ is the stochastic kernel defined in Section
\ref{SubSec:DistrDyn}. See also Ref. \cite{df1999}.}.

\begin{figure}[h]
\begin{minipage}[t]{7.5cm}
\centering
{\includegraphics[width=7.5cm,height=6cm]{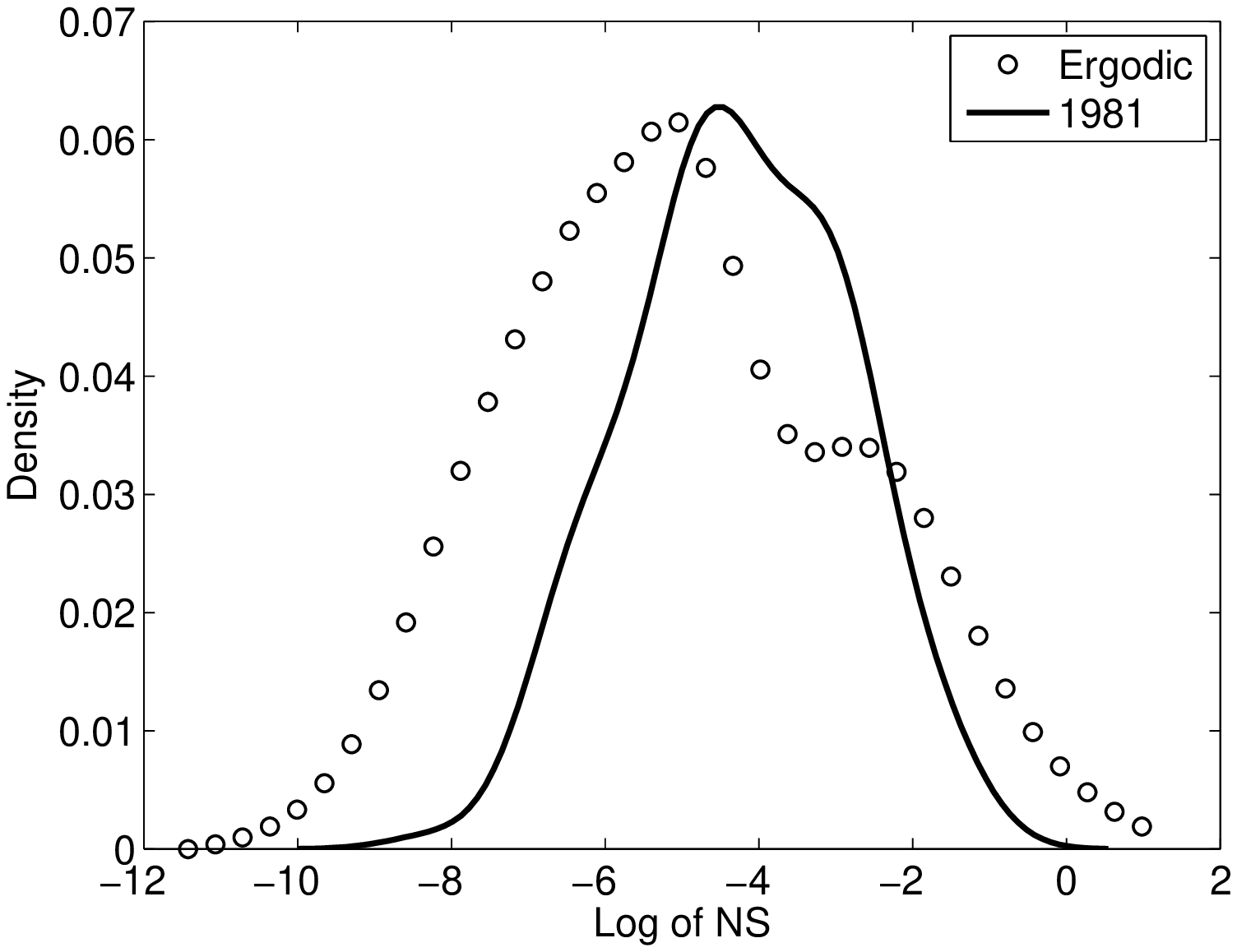}}
\caption{Node Strength (NS): Kernel density of initial distribution
(year=1981) vs. estimate of ergodic (limiting) distribution.}
\label{Fig:ergodic_ns}
\end{minipage}\hfill
\begin{minipage}[t]{7.5cm}
\centering
{\includegraphics[width=7.5cm,height=6cm]{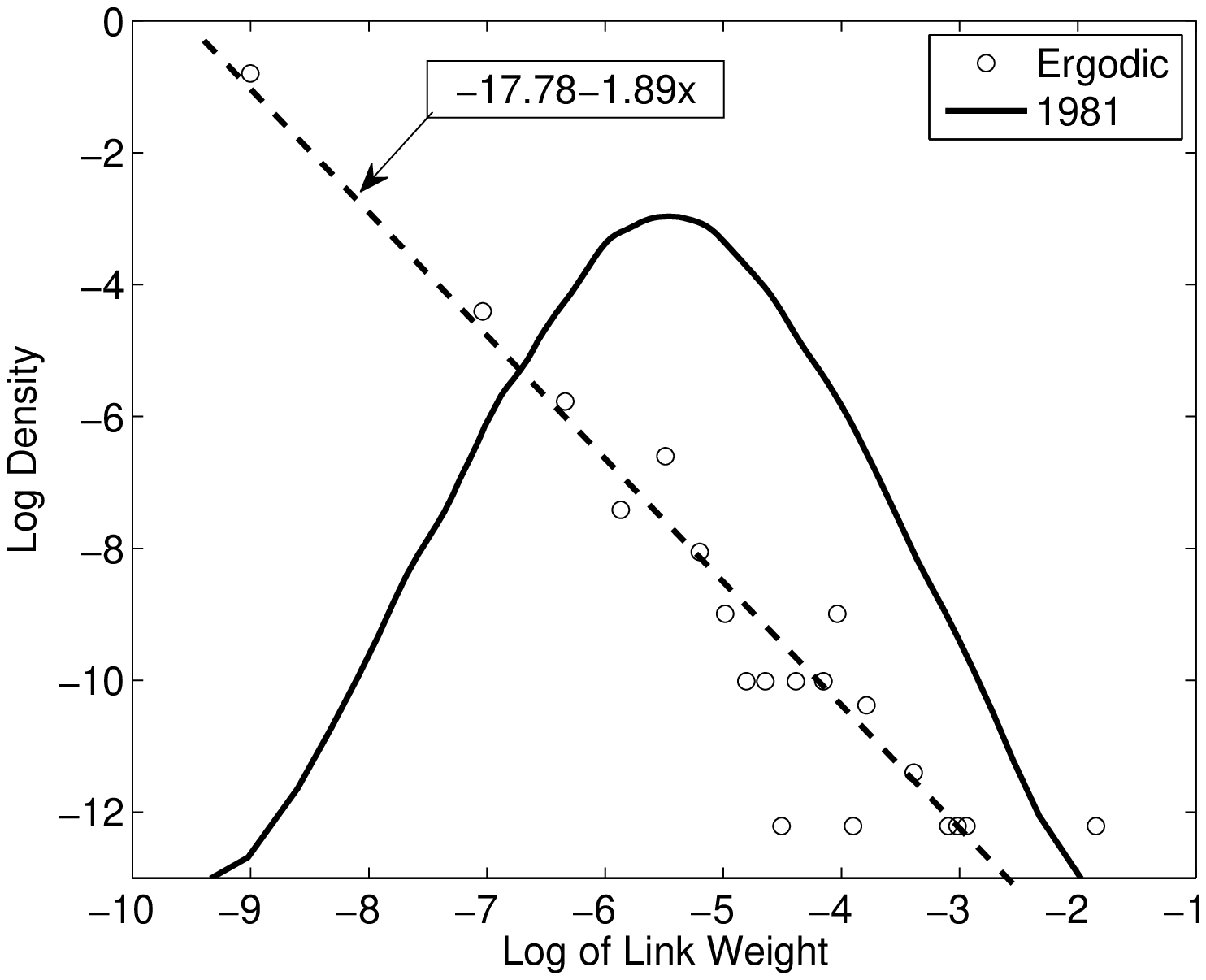}}
\caption{Positive link weights: Kernel density of initial
distribution (year=1981) vs. estimate of ergodic (limiting)
distribution. Dotted line: Power-law fit (equation shown in inset).
Note: Log scale on y-axis.} \label{Fig:ergodic_linkweights}
\end{minipage}
\end{figure}

As already noted above, stochastic kernels of all node statistics
are quite concentrated and evenly distributed along the $45^\circ$
line. Therefore, it is no surprising that also their limiting
distributions look quite similar to the ones in year 1981. This can
be seen in Figure \ref{Fig:ergodic_ns}, where we exemplify this
point by plotting initial vs. estimates of the ergodic distribution
for the logs of NS. Both distributions present a similar shape. If
any, the ergodic one exhibits a larger variability, a shift to the
left of the lower tail and a shift to the right of the upper tail.
This can be explained by noticing that the kernel density estimate
(Figure \ref{Fig:kernel_ns}) shows a relatively larger probability
mass under the main diagonal in the bottom-left part of the plot,
whereas in the top-right part this mass was shifted above the main
diagonal. Such shape-preserving shifts hold also for the other node
statistics under analysis. In particular, the ergodic distribution
for node RWBC roughly preserves its power-law shape, as well as its
scale exponent.

On the contrary, the shape of the stochastic kernel for logged link
weights hinted at a concentration of transition densities at the
extremes of the range. Middle-range values presented instead a
flatter and more dispersed landscape. This partly explains why we
observe a radical difference between initial and ergodic
distributions of logged link weights. Whereas the initial one is
close to a Gaussian (i.e., link weights are well-proxied by a
log-normal density), the ergodic distribution displays a power-law
shape with very small exponent. This can be seen in Figure
\ref{Fig:ergodic_linkweights}, where the two plots have been
superimposed.

These findings imply that the architecture of the WTW will probably
evolve in such a way to undergo a re-organization of link weights
(i.e., country total trade volumes) that is nevertheless able to
keep unchanged the most important node topological properties. Such
a re-organization seems to imply a polarization of link weights into
a large majority of links carrying moderate trade flows and a small
bulk of very intense trade linkages. The power-law shape of the
ergodic distribution suggests that such a polarization is much more
marked than at the beginning of the sample period, when the
distribution of link weights was well proxied by a log-normal
density. Furthermore, it must be noted that results on Gibrat
dynamics in Section \ref{SubSec:Growth} indicate that some
catching-up between low- and high-intensity links is going on within
our sample period. The findings on out-of-sample evolution discussed
here, on the contrary, seem to imply that such a catching-up
dynamics is not so strong to lead to some convergence between e.g.
low-intensity and high-intensity link weights.

\subsection{Robustness to Alternative Weighting Schemes}
\label{SubSec:Robustness}

All results obtained so far refer to a particular weighting
procedure. To recall, the weight of a link from $i$ to $j$ is, after
symmetrization, proportional to the total trade (imports plus
exports) flowing through that link in a given year. This baseline
weighting scheme is very common in the literature
\cite{LiC03,Bhatta2007a,Bhatta2007b,Garla2007}, but treats the same
way all countries irrespective of their economic importance. Are our
findings robust to alternative weighting schemes? To address this
issue, we have considered here two alternative
economically-meaningful setups, where we wash away size effects by
scaling directed link weights with the GDP of either the exporter or
the importer country.

More formally, in the first alternative setup, each directed link
from node $i$ to $j$ is now weighted by total exports of country $i$
to country $j$ and then divided by the country $i$'s GDP (i.e., the
\textit{exporter} country). Such a weighting setup allows one to
measure how much economy $i$ depends on economy $j$ as a buyer. In
the second setup, we still remove size effects from trade flows, but
we now divide by the GDP of the \textit{importer} country ($j$'s
GDP). This allows us to appreciate how much economy $i$ depends on
$j$ as a seller \footnote{We have also experimented with the
weighting scheme where trade is scaled \textit{by the sum} of
importer's and exporter's GDPs without detecting any significant
difference.}.

\begin{figure}[h]
\begin{minipage}[t]{7.5cm}
\centering
{\includegraphics[width=7.5cm,height=6cm]{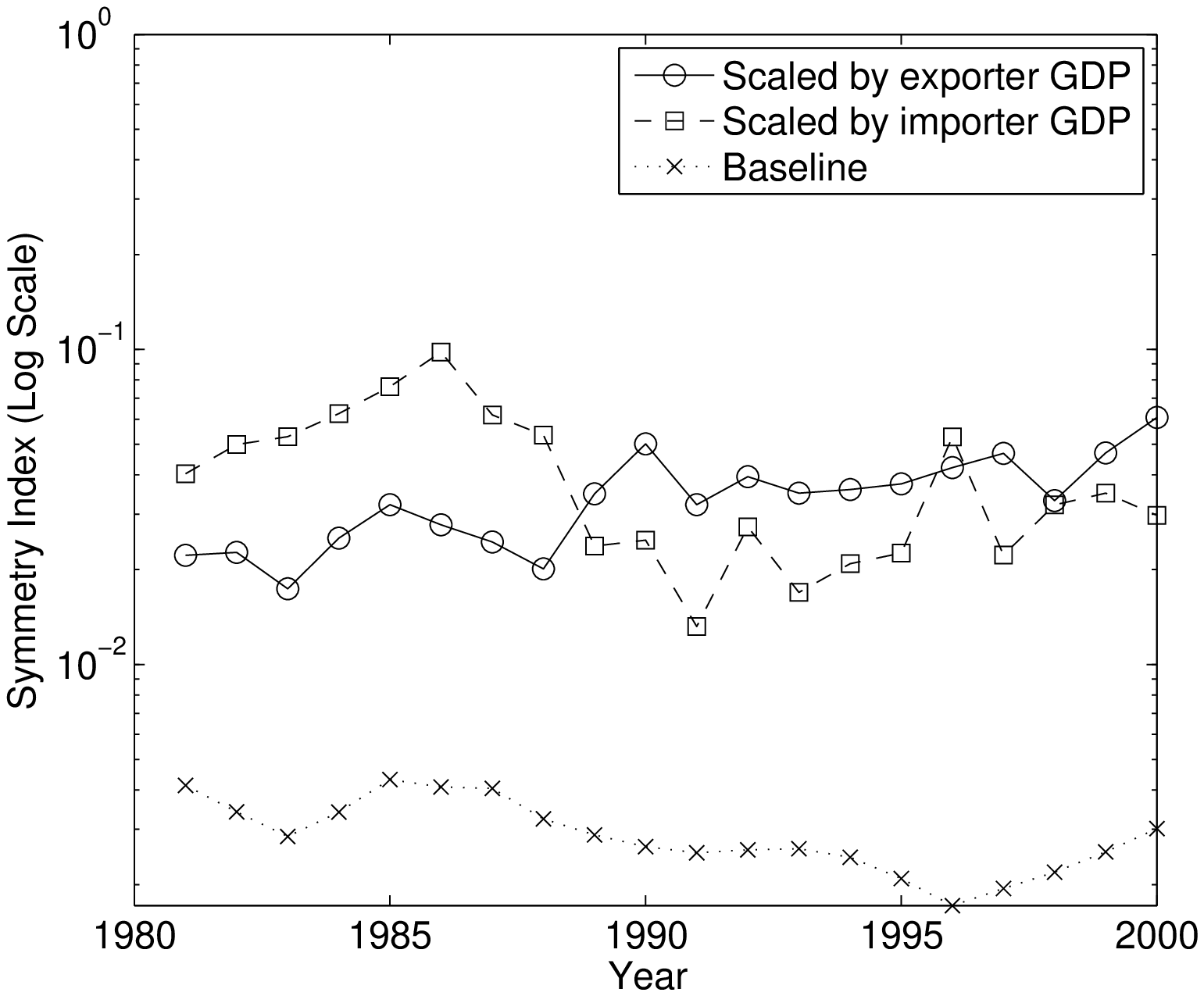}}
\caption{Symmetry index applied to the baseline weighting scheme vs.
symmetry index in the two alternative weighting schemes analyzed.
First scheme: Exports scaled by exporter GDP. Second scheme: Exports
scaled by importer GDP.} \label{Fig:robustness_symmetry}
\end{minipage}\hfill
\begin{minipage}[t]{7.5cm}
\centering
{\includegraphics[width=7.5cm,height=6cm]{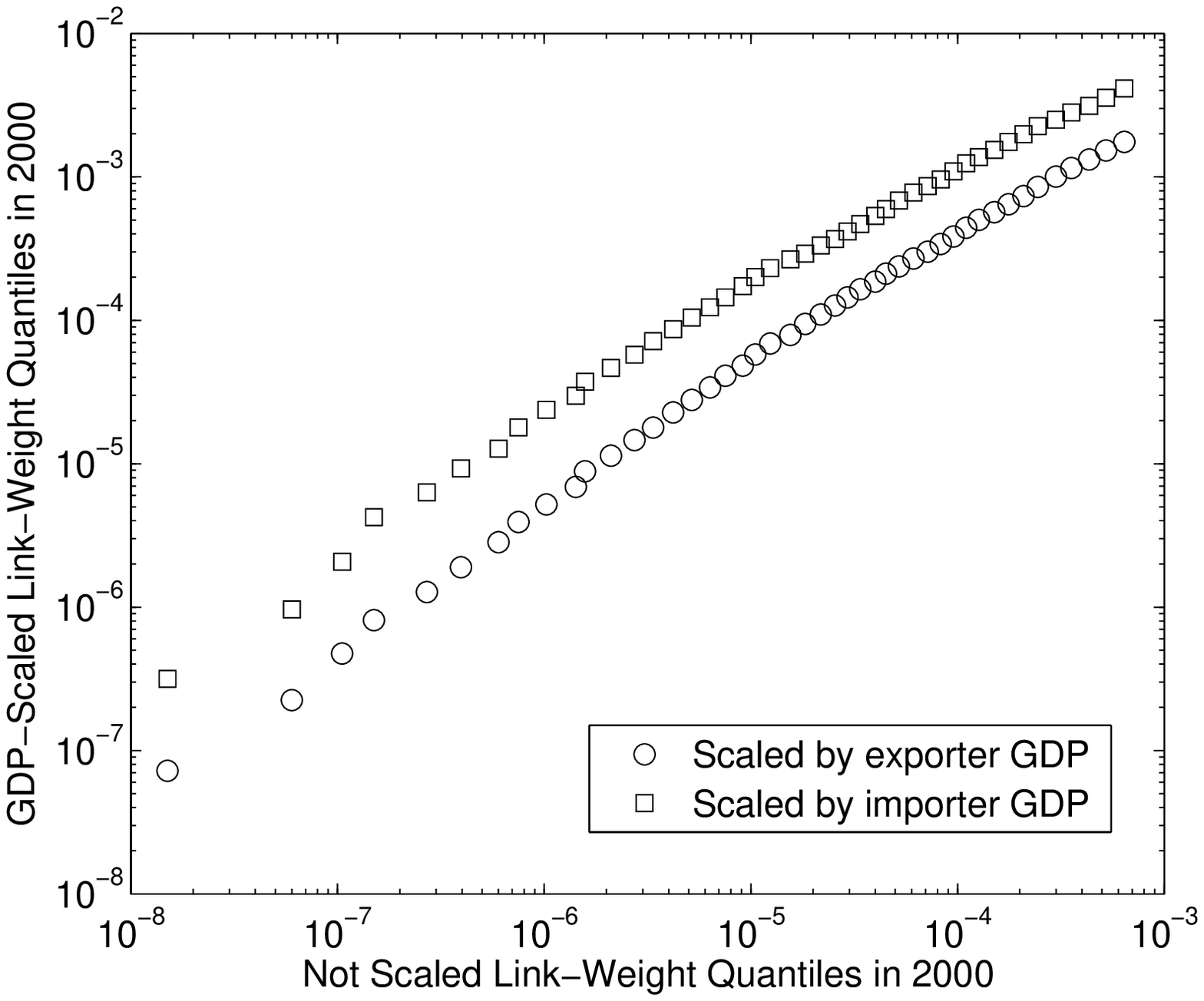}}
\caption{Quantile-quantile plots of logged link-weight distributions
in year 2000. X-axis: Quantiles of logged link-weight distribution
for the baseline weighting scheme. Y-axis: Quantiles of logged
link-weight distributions in the two alternative weighting schemes
analyzed. First scheme: Exports scaled by exporter GDP. Second
scheme: Exports scaled by importer GDP.}
\label{Fig:robustness_qqplot}
\end{minipage}
\end{figure}

\begin{figure}[h]
\begin{minipage}[t]{7.5cm}
\centering {\includegraphics[width=7.5cm,height=6cm]{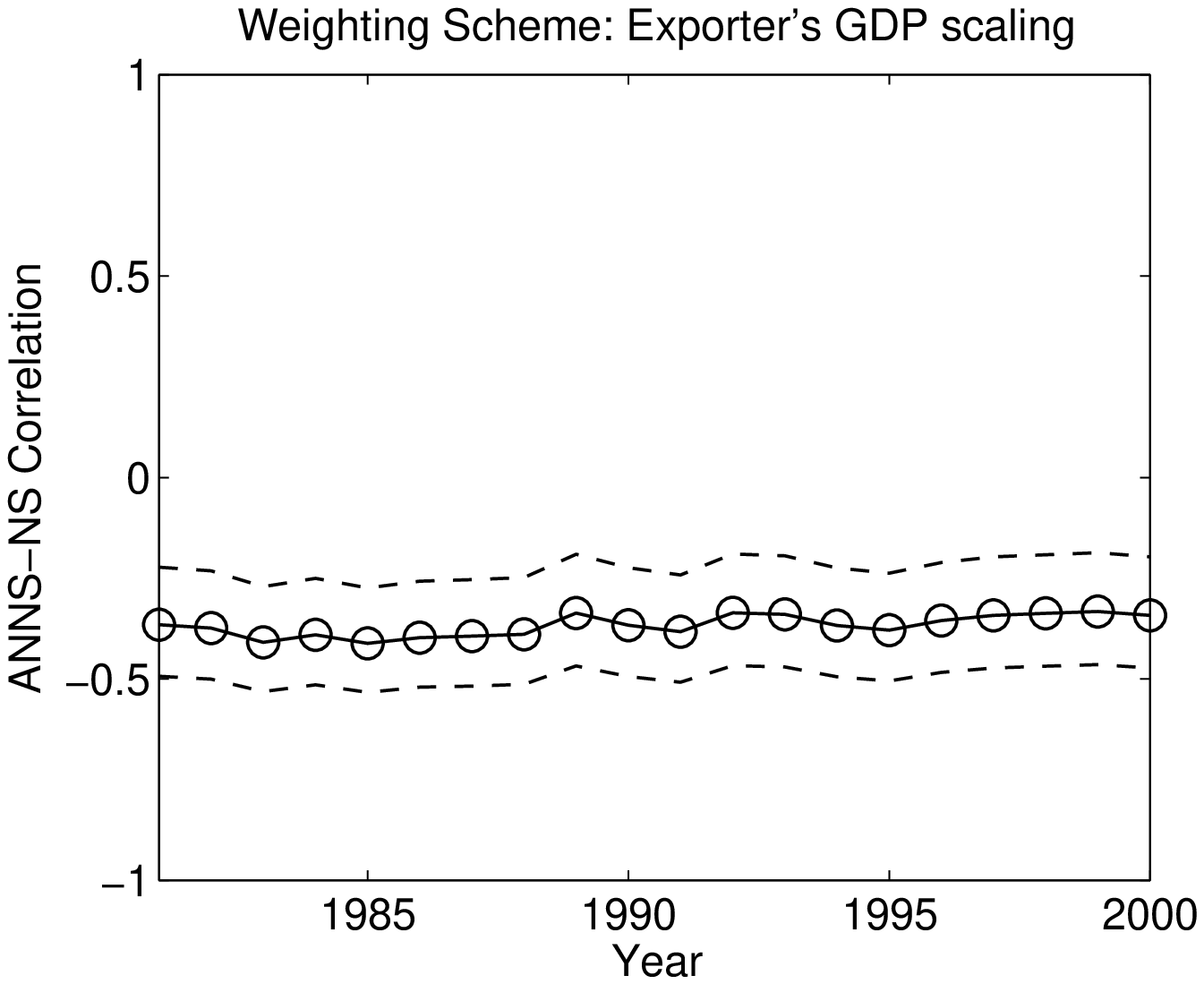}}
\end{minipage}\hfill
\begin{minipage}[t]{7.5cm}
\centering {\includegraphics[width=7.5cm,height=6cm]{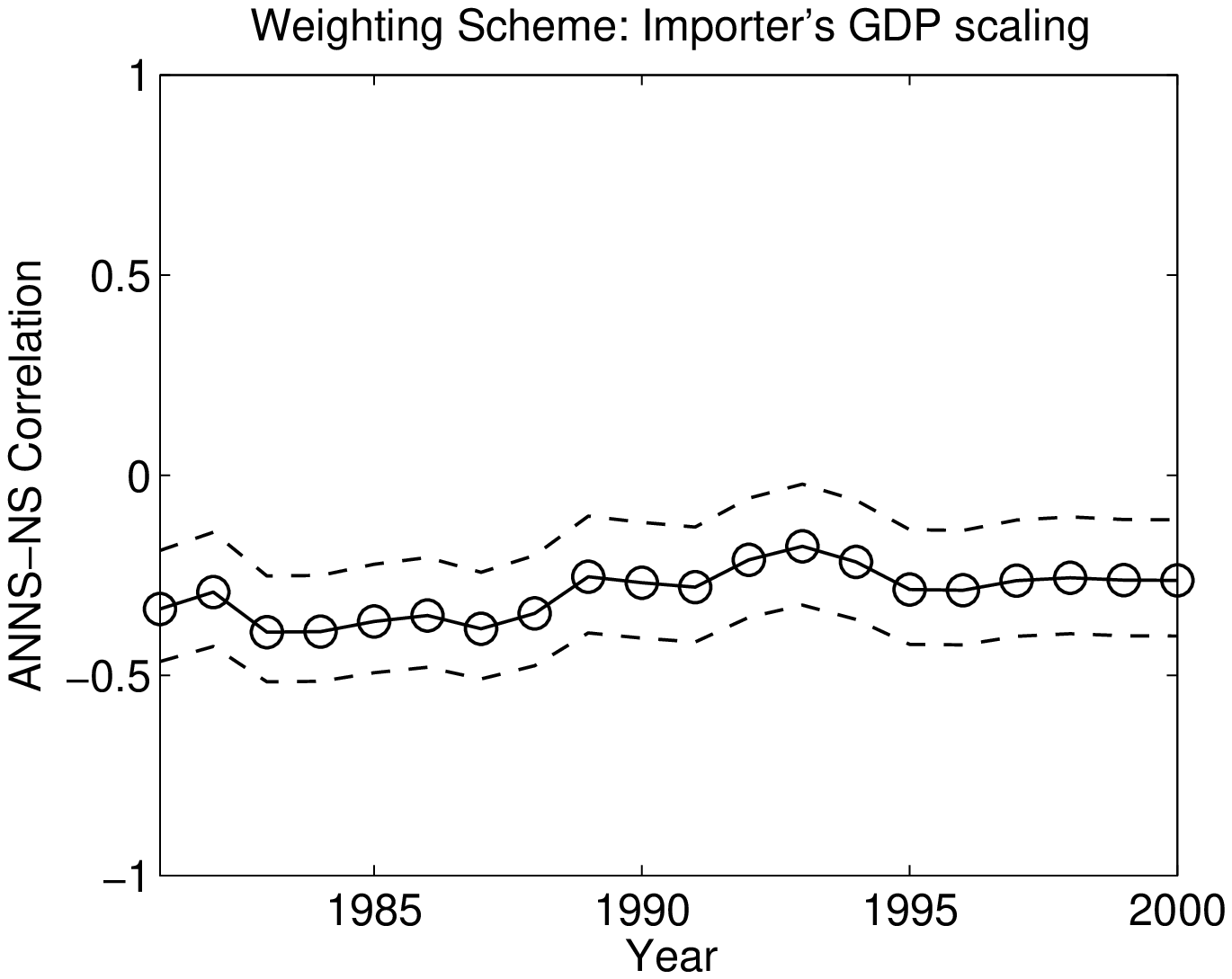}}
\end{minipage}
\begin{minipage}[t]{7.5cm}
\centering {\includegraphics[width=7.5cm,height=6cm]{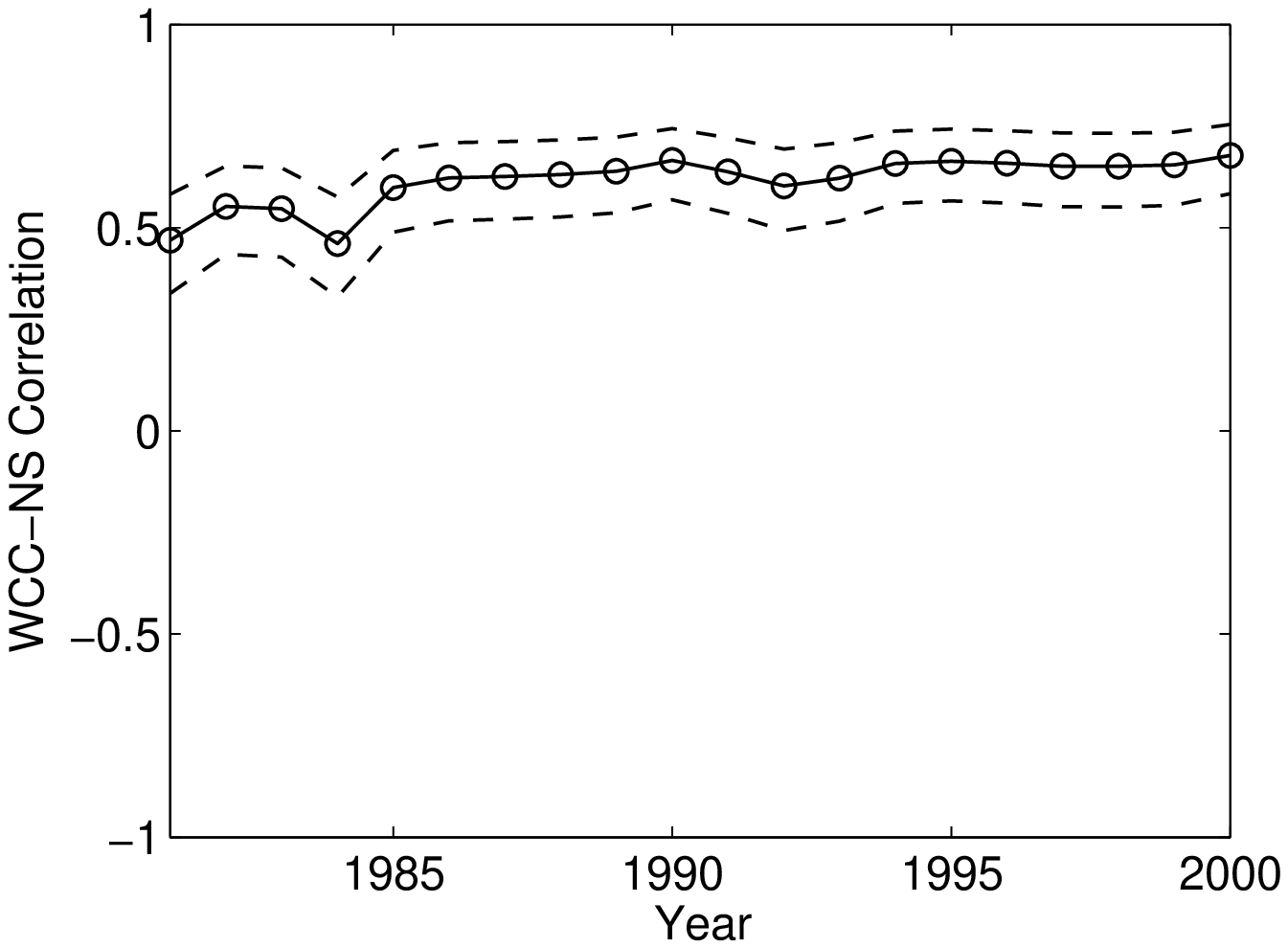}}
\end{minipage}\hfill
\begin{minipage}[t]{7.5cm}
\centering {\includegraphics[width=7.5cm,height=6cm]{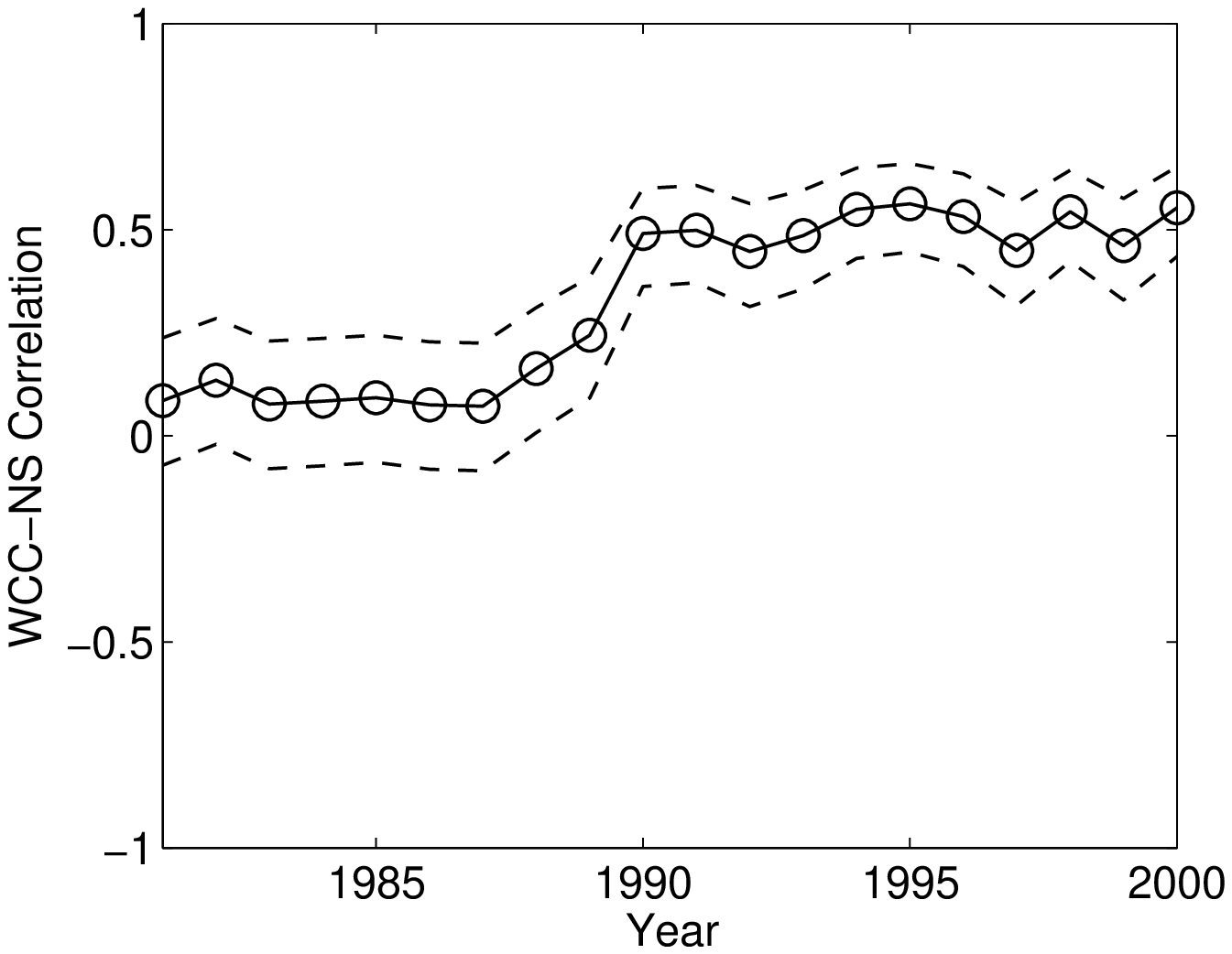}}
\end{minipage}
\begin{minipage}[t]{7.5cm}
\centering {\includegraphics[width=7.5cm,height=6cm]{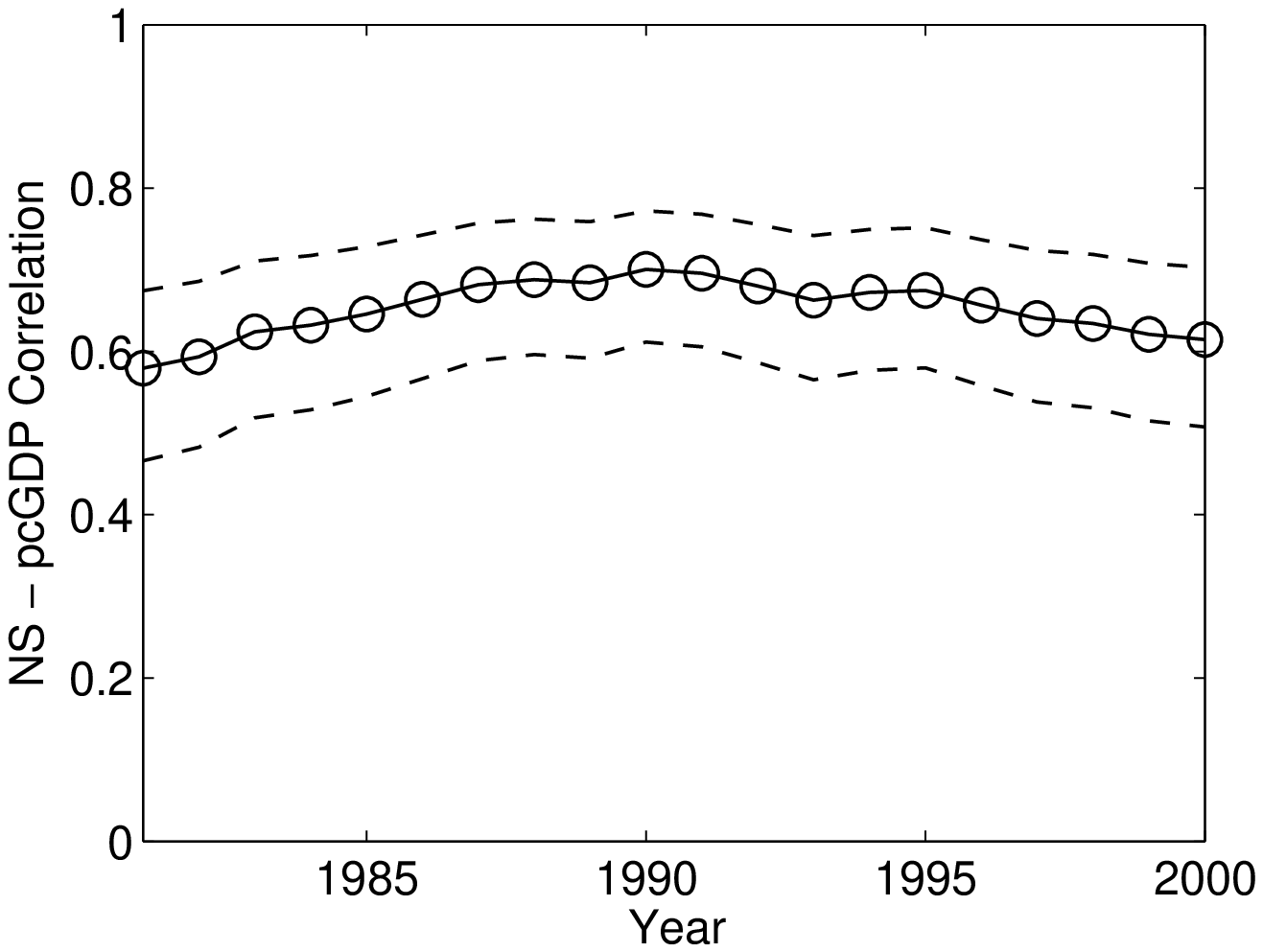}}
\end{minipage}\hfill
\begin{minipage}[t]{7.5cm}
\centering {\includegraphics[width=7.5cm,height=6cm]{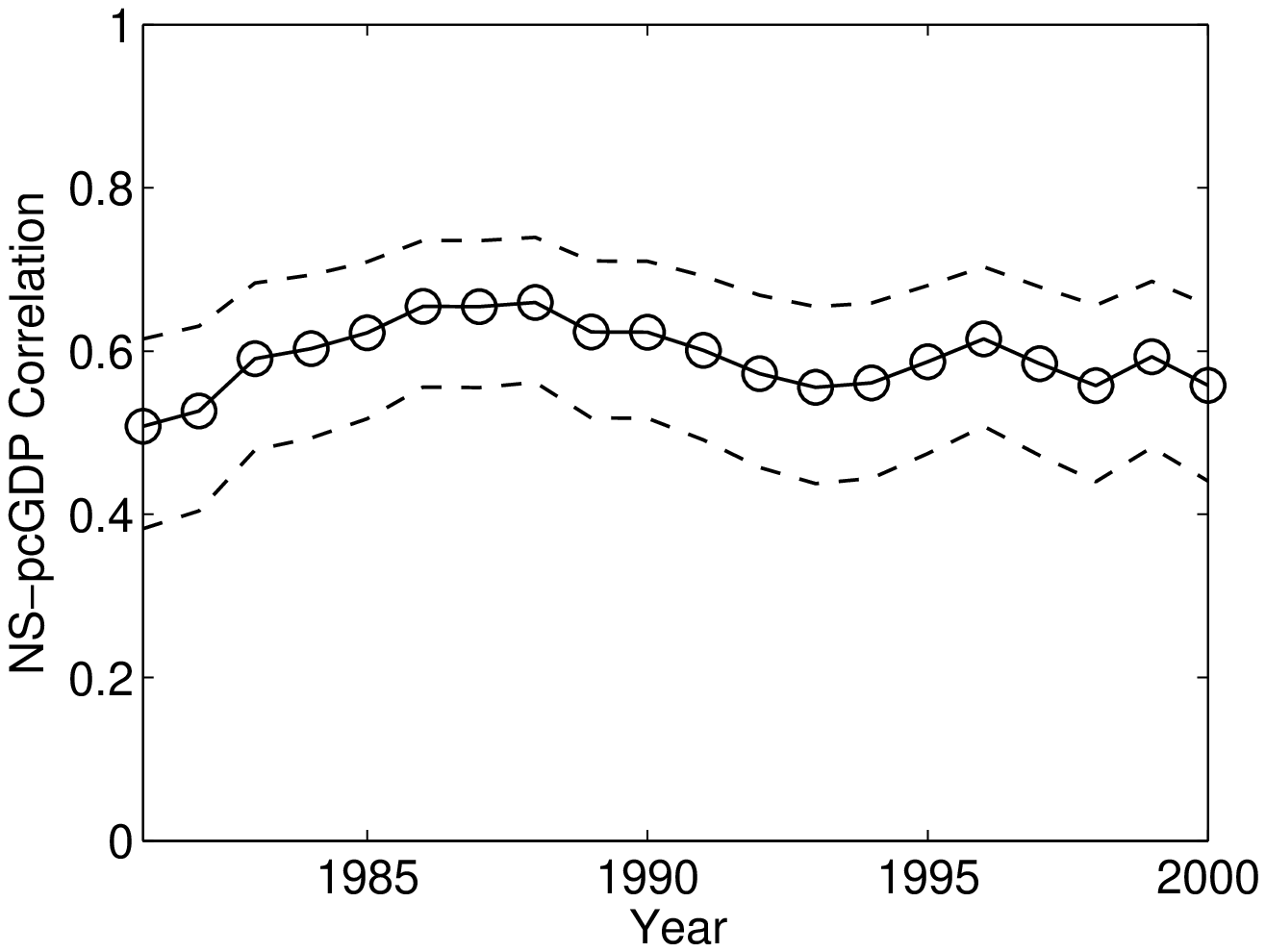}}
\end{minipage}
\caption{Correlation structure among node statistics in the two
alternative weighting schemes analyzed. Left figures: Exports scaled
by exporter GDP. Right Figures: Exports scaled by importer GDP.
Dotted lines: 95\% confidence intervals for population averages.}
\label{Fig:robustness}
\end{figure}

All our main results turn out to be quite robust to these two
alternatives. This is an important point, as a weighted network
analysis might in principle be sensible to the particular choice of
the weighting procedure. To illustrate this point, we first compare
the symmetry index for the three weighting schemes across the years,
cf. Figure \ref{Fig:robustness_symmetry}. If one scales exports by
exporter's or importer's GDP the symmetry index still remains very
low and close to the one found in the baseline weighting scheme.
This indicates that under all three schemes an undirected-network
analysis is appropriate. As a further illustration, Figure
\ref{Fig:robustness_qqplot} reports the quantile-quantile plots of
logged link-weight, GDP-scaled, distributions vs. baseline logged
link weights in year 2000. It is easy to see that both alternative
link-weight distributions are very similar to the baseline one. This
results holds also for pooled distributions, as well as for node
statistics ones. Finally, Figure \ref{Fig:robustness} depicts some
examples of the across-time correlation patterns between node
statistics and pcGDP. Left panels refer to the first alternative
weighting scheme (exports scaled by exporter GDP) whereas right
panels shows what happens under the second alternative setup
(exports scaled by importer GDP). All previous results (see Figures
\ref{Fig:corr_stats} and \ref{Fig:corr_pcgdp}) are confirmed. Notice
that GDP scaling results in weaker but still significantly different
from zero correlation coefficients (especially for WCC-NS). Of
course, we do not expect our results to hold irrespective of
\textit{any} weighting scheme to be adopted. In fact, the binary
characterization of the WTW, where some of the weighted-network
results are reversed, is itself a particular weighting scheme, one
that assigns to each existing link the same weight \footnote{In this
respect, an interesting exercise would imply to find (if any) a
proper re-scaling or manipulation of original trade flows that makes
weighted and binary results looking the same.}.

\section{Conclusions} \label{Sec:Conclusions}
In this paper we have explored, from a purely descriptive
perspective, the within-sample dynamics and out-of-sample evolution
of some key node and link statistic distributions characterizing the
topological properties of the web of import-export relationships
among world countries (WTW). By employing a weighted-network
approach, we have shown that WTW countries holding many trade
partners (and/or very intense trade relationships) are also the
richest and most (globally) central; typically trade with many
partners, but very intensively with only a few of them (which turn
out to be themselves very connected); and form few but
intensive-trade clusters. All the distributions and country rankings
of network statistics display a rather strong within-sample
stationarity. Our econometric tests show that node and link
statistics are strongly persistent. However, Gibrat-like dynamics
are rejected. This is confirmed also by the fact that the
growth-rate distributions of our statistics can be well approximated
by fat-tailed Laplace or asymmetric exponential-power densities.
Furthermore, whereas the estimated ergodic distributions of all
node-statistics are quite similar to the initial ones, the
(positive) link-weight distribution is shifting from a log-normal to
a power law. This suggests that a polarization between a large
majority of weak-trade links and a minority of very intense-trade
ones is gradually emerging in the WTW. Interestingly, such a process
is likely to take place without dramatically changing the
topological properties of the network.

Many extensions to the present work can be conceived. First,
building on Refs. \cite{Garla2004,Bhatta2007a}, one may try to
explore simple but economically-meaningful models of WTW dynamics
that are able to reproduce the main stylized facts put forth by our
purely empirical analysis. To do that, one may attempt to deduce the
probability distributions of link and node statistics of interest by
postulating some given growth model for link weights. For example,
it is well-known that Gibrat-like multiplicative, statistically
independent, processes at the level of links can easily generate
---as their limiting distribution--- log-normal densities such as the
observed ones. Notice, however, that the discussion in Sections
\ref{SubSec:DistrDyn} , \ref{SubSec:Growth} and
\ref{SubSec:OutofSample}, points towards patterns of within-sample
growth dynamics persistently deviating from the standard Gibrat
model, and indicates that also out-of-sample evolution does not seem
to be driven by such a simple mechanics. Furthermore, statistical
independence among growth processes for different link weights (both
belonging to a given country and among different countries) can be
easily dismissed on simple economic arguments. This suggests that,
in order to single out simple stochastic growth models accounting at
the same time both for the observed link-weight dynamics, and for
the ensuing statistical properties of node-statistics computed from
such a dynamics, more complicated models might be conceived. A
possibility, indeed the very next point in our agenda, would be to
adapt existing models of weighted-network evolution \cite{Barr05} in
such a way to allow for more plausible rules ---gathered e.g. from
international-trade literature--- governing the emergence of trade
relationships and the subsequent evolution of their intensity.

Second, one would like to explore in more details the topological
properties of the WTW, both cross-sectionally and over time.
Interesting questions here concern the role of geographical
proximity in shaping the structure of international trade, the
degree of fragility of the network, and so on. More specifically,
trade flows could be disaggregated across product classes to explore
how trade composition affects network properties.

Third, one could abstract from aggregate statistical properties and
analyze at a finer level the role of single countries in the network
structure. For instance, how does the dynamics of degree, strength,
clustering, etc. behave for single relevant countries in different
regions? Do country-specific network indicators display the same
time-stationarity of their aggregate counterparts?

Finally, in line with Ref. \cite{KaRe07}, one can ask whether node
statistics characterizing connectivity, clustering, centrality and
so on, can be employed as explanatory variables for the dynamics of
country growth rates and development patterns.


\newpage \clearpage


\begin{table}[h]
\begin{center}
\begin{tabular}{rcccc}
\hline \hline
           &         \multicolumn{ 4}{c}{Average Absolute Growth Rates} \\
           &       Mean &    Std Dev &   Skewness &   Kurtosis \\
\hline
        ND &     0.0143 &     0.0047 &     0.0375 &     0.0086 \\
      ANND &     0.0079 &     0.0279 &     0.0116 &     0.0260 \\
       BCC &     0.0043 &     0.0197 &     0.0615 &     0.0205 \\
        NS &     0.0379 &     0.0412 &     0.0263 &     0.0317 \\
      ANNS &     0.0479 &     0.0512 &     0.0223 &     0.0452 \\
       WCC &     0.0544 &     0.0097 &     0.0274 &     0.0543 \\
      RWBC &     0.0049 &     0.0107 &     0.0251 &     0.0556 \\
\hline \hline
\end{tabular}
\end{center}
\caption{Average over time of absolute-valued 1-year growth rates of
the first four moments of node statistics. Given the value of the
node statistic X at time $t$ for country $i$ ($X_i^t$) and
$M^k(\cdot)$ the moment operator that for $k=1,2,3,4$ returns
respectively the mean, standard deviation, skewness and kurtosis,
the time-average of absolute-valued 1-year growth rates of the
$k$-th moment-statistic is defined as
$\frac{1}{T-1}\sum_{t=2}^{T}{|(E^k(X_i^t)/E^k(X_i^{t-1})-1)|}$.}
\label{Tab:moments_gr}
\end{table}

\newpage \clearpage


\begin{sidewaystable}[h]
\begin{center}
\begin{tabular}{lllllllllll}
\hline \hline
           &       1981 &       1982 &       1983 &       1984 &       1985 &       1986 &       1987 &       1988 &       1989 &       1990 \\
\hline
        ND & $0.0000^{***}$ & $0.0000^{***}$ & $0.0010^{***}$ & $0.0000^{***}$ & $0.0000^{***}$ & $0.0000^{***}$ & $0.0010^{***}$ & $0.0000^{***}$ & $0.0010^{***}$ & $0.0000^{***}$ \\
      ANND & $0.0277^{**}$ & $0.0261^{**}$ & $0.0400^{**}$ & $0.0403^{**}$ & $0.0525^{*}$ & $0.0309^{**}$ & $0.0333^{**}$ & $0.0197^{**}$ & $0.1254^{}$ & $0.1020^{}$ \\
       BCC & $0.0060^{***}$ & $0.0040^{***}$ & $0.0040^{***}$ & $0.0040^{***}$ & $0.0050^{***}$ & $0.0020^{***}$ & $0.0050^{***}$ & $0.0030^{***}$ & $0.0080^{***}$ & $0.0040^{***}$ \\
        NS & $0.2925^{}$ & $0.2046^{}$ & $0.4021^{}$ & $0.2870^{}$ & $0.4344^{}$ & $0.6804^{}$ & $0.6238^{}$ & $0.5300^{}$ & $0.3496^{}$ & $0.5343^{}$ \\
      ANNS & $0.1118^{}$ & $0.2500^{}$ & $0.2724^{}$ & $0.2463^{}$ & $0.2816^{}$ & $0.2532^{}$ & $0.3243^{}$ & $0.1633^{}$ & $0.1666^{}$ & $0.1065^{}$ \\
       WCC & $0.5673^{}$ & $0.2525^{}$ & $0.2821^{}$ & $0.2874^{}$ & $0.2867^{}$ & $0.2601^{}$ & $0.3564^{}$ & $0.2035^{}$ & $0.2005^{}$ & $0.1202^{}$ \\
      RWBC & $0.0000^{***}$ & $0.0000^{***}$ & $0.0000^{***}$ & $0.0000^{***}$ & $0.0000^{***}$ & $0.0000^{***}$ & $0.0000^{***}$ & $0.0000^{***}$ & $0.0000^{***}$ & $0.0000^{***}$ \\
           &            &            &            &            &            &            &            &            &            &            \\
           &       1991 &       1992 &       1993 &       1994 &       1995 &       1996 &       1997 &       1998 &       1999 &       2000 \\
\hline
        ND & $0.0010^{***}$ & $0.0020^{***}$ & $0.0010^{***}$ & $0.0010^{***}$ & $0.0020^{***}$ & $0.0000^{***}$ & $0.0000^{***}$ & $0.0000^{***}$ & $0.0000^{***}$ & $0.0000^{***}$ \\
      ANND & $0.0810^{*}$ & $0.0630^{*}$ & $0.0900^{*}$ & $0.0700^{*}$ & $0.0580^{*}$ & $0.0400^{**}$ & $0.0480^{**}$ & $0.0460^{**}$ & $0.0400^{**}$ & $0.0260^{**}$ \\
       BCC & $0.0090^{***}$ & $0.0040^{***}$ & $0.0040^{***}$ & $0.0020^{***}$ & $0.0020^{***}$ & $0.0060^{***}$ & $0.0050^{***}$ & $0.0050^{***}$ & $0.0050^{***}$ & $0.0050^{***}$ \\
        NS & $0.5367^{}$ & $0.2398^{}$ & $0.2917^{}$ & $0.2016^{}$ & $0.3685^{}$ & $0.4693^{}$ & $0.6000^{}$ & $0.6312^{}$ & $0.5918^{}$ & $0.5260^{}$ \\
      ANNS & $0.2450^{}$ & $0.0905^{*}$ & $0.1402^{}$ & $0.1133^{}$ & $0.1269^{}$ & $0.0574^{*}$ & $0.0734^{*}$ & $0.0899^{*}$ & $0.0668^{*}$ & $0.1385^{}$ \\
       WCC & $0.2661^{}$ & $0.1166^{}$ & $0.1562^{}$ & $0.1356^{}$ & $0.1358^{}$ & $0.0638^{*}$ & $0.1206^{}$ & $0.1095^{}$ & $0.1072^{}$ & $0.1583^{}$ \\
      RWBC & $0.0000^{***}$ & $0.0000^{***}$ & $0.0000^{***}$ & $0.0000^{***}$ & $0.0000^{***}$ & $0.0000^{***}$ & $0.0000^{***}$ & $0.0000^{***}$ & $0.0000^{***}$ & $0.0000^{***}$ \\
\hline \hline
\end{tabular}
\end{center}
\caption{P-values for Jarque-Bera normality test
\citep{jbtest1,jbtest2}. Null hypothesis: Logs of (positive-valued)
distribution are normally-distributed with unknown parameters.
Asterisks: (*) null hypothesis rejected at 10\%; (**) null
hypothesis rejected at 5\%; (***) null hypothesis rejected at 1\%.}
\label{Tab:normality}
\end{sidewaystable}

\newpage \clearpage

\begin{table}[h]
\begin{center}
\begin{tabular}{lllllll}
\hline \hline
           & \multicolumn{ 2}{c}{Log Normal} & \multicolumn{ 2}{c}{Stretched Exponential} & \multicolumn{ 2}{c}{Power Law} \\

   KS Test &  Statistic &    p-Value &  Statistic &    p-Value &  Statistic &    p-Value \\
\hline
     $ND$ &     0.1262 & $0.0114^{**}$ &     0.5326 & $0.0000^{***}$ &     0.4185 & $0.0000^{***}$ \\
     $NS$ &     0.0647 & $0.5059^{}$ &     0.3563 & $0.0000^{***}$ &     0.2862 & $0.0000^{***}$ \\
   $ANNS$ &     0.0911 & $0.1351^{}$ &     0.5141 & $0.0000^{***}$ &     0.1336 & $0.0061^{***}$ \\
    $WCC$ &     0.0528 & $0.7658^{}$ &     0.6493 & $0.0000^{***}$ &     0.2569 & $0.0000^{***}$ \\
   $RWBC$ &     0.2103 & $0.0000^{***}$ &     0.4830 & $0.0000^{***}$ &     0.1065 & $0.0564^{}$ \\
\hline \hline
\end{tabular}
\end{center}
\caption{Kolmogorov-Smirnov goodness-of-fit test results for year
2000 distributions. The three null hypotheses tested are that the
observed distributions come from, respectively, log-normal,
stretched exponential, or power-law pdfs. Asterisks: (*) null
hypothesis rejected at 10\%; (**) null hypothesis rejected at 5\%;
(***) null hypothesis rejected at 1\%.}
\label{Tab:kstest}
\end{table}

\newpage \clearpage


\begin{table}[h]
\begin{center}
\begin{tabular}{rcccc}
\hline \hline
$\omega=0$ &                      \multicolumn{ 4}{c}{$\tau$} \\
\hline
 Statistic &          1 &          4 &          7 &         10 \\
\hline
        $M_{\omega,K}^{\tau}(ND)$ &     0.8674 &     0.7794 &     0.7282 &     0.6943 \\
        $M_{\omega,K}^{\tau}(NS)$ &     0.9346 &     0.8612 &     0.8234 &     0.7874 \\
      $M_{\omega,K}^{\tau}(ANNS)$ &     0.8541 &     0.7577 &     0.7145 &     0.6656 \\
       $M_{\omega,K}^{\tau}(WCC)$ &     0.8553 &     0.7490 &     0.6776 &     0.6377 \\
      $M_{\omega,K}^{\tau}(RWBC)$ &     0.9004 &     0.8280 &     0.7899 &     0.7515 \\
\hline
C.I. (Reshuffled) & [0.1854,0.2146] & [0.1842,0.2159] & [0.1823,0.2176] & [0.1800,0.2200] \\
\hline
           &            &            &            &            \\
$\omega=1$ &                      \multicolumn{ 4}{c}{$\tau$} \\
\hline
 Statistic &          1 &          4 &          7 &         10 \\
\hline
        $M_{\omega,K}^{\tau}(ND)$ &     0.9950 &     0.9881 &     0.9753 &     0.9640 \\
        $M_{\omega,K}^{\tau}(NS)$ &     0.9997 &     0.9965 &     0.9923 &     0.9875 \\
      $M_{\omega,K}^{\tau}(ANNS)$ &     0.9930 &     0.9835 &     0.9710 &     0.9579 \\
       $M_{\omega,K}^{\tau}(WCC)$ &     0.9980 &     0.9918 &     0.9787 &     0.9686 \\
      $M_{\omega,K}^{\tau}(RWBC)$ &     0.9990 &     0.9965 &     0.9952 &     0.9906 \\
\hline
C.I. (Reshuffled) & [0.5020,0.5370] & [0.5004,0.5385] & [0.4981,0.5406] & [0.4954,0.5436] \\
\hline \hline
\end{tabular}
\end{center}
\caption{Distribution dynamics. Persistence measure
$M_{\omega,K}^{\tau}(X)$ for the distributions of node statistics
and for alternative choices of the window $\omega\in \{0,1\}$ and
the time lag $\tau\in \{1,4,7,10\}$. All statistics refer to $K=10$
quantile classes. The lines labeled as ``C.I. (Reshuffled)'' contain
confidence intervals (at 95\%) for the mean of the distribution of
the statistic $M$ in random graphs where the observed adjacency
matrices $A^t$ are kept fixed and weights are re-distributed at
random by reshuffling the observed link-weight distributions
$w^t=\{w_{ij}^t, i\neq j=1,\dots,N\}$. Values of
$M_{\omega,K}^{\tau}(X)$ close to one and to the right of confidence
intervals indicate a strong persistence of the associated
with-sample distribution dynamics.} \label{Tab:distr_dynamics_nodes}
\end{table}


\begin{table}[h]
\begin{center}
\begin{tabular}{rcccc}
\hline \hline
           &                      \multicolumn{ 4}{c}{$\tau$} \\
           &          1 &          4 &          7 &         10 \\
\hline
$\omega=0$ &     0.8116 &     0.7073 &     0.6464 &     0.6012 \\
C.I. (Reshuffled) & [0.1974,0.2026] & [0.1971,0.2029] & [0.1969,0.2032] & [0.1962,0.2037] \\
\hline
$\omega=1$ &     0.9910 &     0.9733 &     0.9562 &     0.9397 \\
C.I. (Reshuffled) & [0.5175,0.5238] & [0.5172,0.5241] & [0.5170,0.5248] & [0.5167,0.5253] \\
\hline \hline
\end{tabular}
\end{center}
\caption{Distribution dynamics. Persistence measure
$M_{\omega,K}^{\tau}(X)$ for the distributions of positive link
weights and for alternative choices of the window $\omega\in
\{0,1\}$ and the time lag $\tau\in \{1,4,7,10\}$. All statistics
refer to $K=20$ quantile classes. The lines labeled as ``C.I.
(Reshuffled)'' contain confidence intervals (at 95\%) for the mean
of the distribution of the statistic $M$ in random graphs where the
observed adjacency matrices $A^t$ are kept fixed and weights are
re-distributed at random by reshuffling the observed link-weight
distributions $w^t=\{w_{ij}^t, i\neq j=1,\dots,N\}$. Values of
$M_{\omega,K}^{\tau}(X)$ close to one and to the right of confidence
intervals indicate a strong persistence of the associated
with-sample distribution dynamics.}
\label{Tab:distr_dynamics_linkweights}
\end{table}

\newpage \clearpage


\begin{sidewaystable}[h]
\begin{center}
\begin{tabular}{cccccccc}
\hline \hline
      Rank &         ND &         NS &       ANNS &        WCC &       RWBC &   Real GDP &      pcGDP \\
\hline
         1 &    Germany &        USA & Sao Tome-Principe &        USA &        USA &        USA & Luxembourg \\
         2 &      Italy &    Germany &   Kiribati &    Germany &    Germany &      Japan & Switzerland \\
         3 &         UK &      Japan &      Nauru &      Japan &      Japan &    Germany &      Japan \\
         4 &     France &     France &      Tonga &         UK &     France &         UK &     Norway \\
         5 & Switzerland &      China &    Vanuatu &      China &         UK &     France &        USA \\
         6 &  Australia &         UK &     Tuvalu &     France &      China &      China &    Denmark \\
         7 &    Belgium &     Canada &    Burundi &      Italy &      Italy &      Italy &    Iceland \\
         8 & Netherlands &      Italy &   Botswana & Netherlands &   S. Korea &     Canada &     Sweden \\
         9 &    Denmark & Netherlands &    Lesotho &   S. Korea & Netherlands &     Brazil &    Austria \\
        10 &     Sweden &    Belgium &   Maldives &  Singapore &    Belgium &     Mexico &    Germany\\
\hline
        11 &      India &   S. Korea & Solomon Islands &     Mexico &      Spain &      Spain &    Ireland \\
        12 &      Spain &     Mexico &     Bhutan &    Belgium &  Australia &      India & Netherlands \\
        13 &        USA &     Taiwan &    Comoros &      Spain &  Singapore &   S. Korea &    Finland \\
        14 &      China &  Singapore & Seychelles &     Taiwan &      India &  Australia &      Qatar \\
        15 &     Norway &      Spain & Saint Lucia &     Canada &     Taiwan & Netherlands &    Belgium \\
        16 &      Japan & Switzerland & Guinea-Bissau & United Arab Emirates & South Africa &     Taiwan &  Singapore \\
        17 &     Taiwan &   Malaysia &   Mongolia & Saudi Arabia &     Brazil &  Argentina &     France \\
        18 &   Malaysia &     Sweden & Cape Verde &       Iraq &   Thailand &     Russia &         UK \\
        19 &    Ireland &   Thailand &    Grenada & Switzerland & Saudi Arabia & Switzerland &      China \\
        20 &     Canada &  Australia &       Fiji &     Russia &     Canada &     Sweden & United Arab Emirates \\
\hline \hline
\end{tabular}
\end{center}
\caption{Country rankings in year 2000 according to node statistics,
real gross-domestic product (GDP, in current US dollars) and
per-capita GDP (US dollars per person).} \label{Tab:rankings}
\end{sidewaystable}


\begin{table}[h]
\begin{center}
\begin{tabular}{r|cccc|cccc}
\hline \hline
           & \multicolumn{ 4}{c}{First-Order Autocorrelation} &          \multicolumn{ 4}{|c}{Gibrat-Regression Parameter} \\
\hline
           &       Mean &    SD &    $\%(>0)$ &     Pooled &       Mean &     StdDev &    $\%(<0)$ &     Pooled \\
\hline
        ND &     0.6438 &     0.2222 &     0.8428 &     0.9859 &    -0.3636 &     0.2235 &     0.6667 &    -0.1330 \\
        NS &     0.6795 &     0.1170 &     0.9623 &     0.9949 &    -0.3186 &     0.1176 &     0.7799 &    -0.1910 \\
      ANNS &     0.6353 &     0.0686 &     0.9811 &     0.9539 &    -0.3609 &     0.0670 &     0.9811 &    -0.2235 \\
       WCC &     0.6404 &     0.1414 &     0.9119 &     0.9855 &    -0.3596 &     0.1414 &     0.8302 &    -0.2866 \\
      RWBC &     0.6203 &     0.2007 &     0.8050 &     0.9983 &    -0.3795 &     0.2007 &     0.7862 &    -0.3651 \\
Pos Link Weights &     0.4330 &     0.3069 &     0.4171 &     0.9940 &    -0.4368 &     0.2299 &     0.9196 &    -0.1422 \\
\hline \hline
\end{tabular}
\end{center}
\caption{First-order autocorrelation coefficient $\hat{r}_i(X)$
\eqref{Eq:ACC} and $\hat{\beta}_i$ parameter in Gibrat regressions
\eqref{Eq:Gibrat} for node and link statistics. Mean and SD columns:
Population average and standard deviation computed across node or
links. Columns labeled by $\%(>0)$ or $\%(<0)$ report the percentage
of nodes or links whose estimate is larger or smaller than zero.
Columns labeled by ``Pooled'' report estimates for the time-pooled
normalized sample (i.e., the sample obtained by first standardizing
each observation by the mean and standard deviation of the year, and
then stacking all years in a column vector).} \label{Tab:autocorr}
\end{table}


\newpage \clearpage


\end{document}